\documentclass[
11pt,
fleqn
]
{classjch}

\begin{document}

\begin{titlepage}

\vskip2.5cm

 \begin{center}
   {\Huge \textsc{Primordial Black Holes in \\
                  \vskip0.3cm
                  Non-linear perturbation theory} }
 \end{center}

\vskip3cm

\begin{center}
  {\Large \textsc{Juan~Carlos~Hidalgo~Cuellar} 
          \vskip0.5cm
    \large
    \textsc{Astronomy Unit}\\
    \textsc{School of Mathematical Sciences}\\
    \textsc{Queen Mary College}\\
    \textsc{University Of London}\\
    \textsc{United Kingdom}
  }
\end{center}

\vskip5cm

\begin{center}
  \textsc{%
    A dissertation submitted in candidature for the degree of Doctor of
    Philosophy in the University of London
}
\end{center}

\vskip2cm
\centerline{\textsc{Friday 1st May 2009}}

\end{titlepage}



\vskip2.5cm

\begin{abstract}

 The thesis begins with a study of the origin of
 non-linear cosmological fluctuations. In particular, a class of
 models of multiple field inflation are considered, with specific
 reference to those cases in which the non-Gaussian correlation
 functions are large. The analysis shows that perturbations
 from an almost massless auxiliary field generically produce large
 values of the non-linear parameter $\fnl$. \\
 
 Next, the effects of including
 non-Gaussian correlation functions in the statistics of cosmological
 structure are explored. For this purpose, a non-Gaussian probability
 distribution function (PDF) for the curvature perturbation $\R$ is
 required. Such a PDF is derived  from first principles in the
 context of quantum field theory, with n-point correlation
 functions as the only input. Under reasonable power-spectrum
 conditions, an explicit expression for the PDF is presented, with
 corrections to the Gaussian distribution from the three-point
 correlation function $\langle \R \R \R\rangle$. \\

 The method developed for the derivation of the non-Gaussian PDF is then
used to explore two important problems in the physics of primordial
black holes (PBHs). First, the non-Gaussian probability is used to compute
corrections to the number of PBHs generated from the primordial curvature
fluctuations. Particular characteristics of such corrections are
explored for a variety of inflationary models. The non-Gaussian
corrections explored consist exclusively of non-vanishing three-point
correlation functions.  \\

The second application concerns new cosmological observables. The
formation of PBHs is known to depend on two main physical
characteristics:  the strength of the gravitational field 
produced by the initial curvature inhomogeneity and the pressure
gradient at the edge of the curvature configuration. The latter has
so far been ignored in the estimation of the 
probability of PBH formation. We account for this by using two
parameters to describe the profile: The amplitude of the inhomogeneity
and its second radial derivative, both evaluated at the centre of the
configuration. The method developed to derive the non-Gaussian PDF is
modified to find the joint probability of these two parameters. We
discuss the implications of the derived probability for the fraction
of mass in the universe in the form of PBHs.

 \end{abstract}

\clearemptydoublepage
\thispagestyle{empty}
{\large
$~$

\vspace{2cm}

 I hereby certify that this thesis, which is
approximately $45,000$ words in length, has been written by me, that
it is the record of the work carried out by me at the Astronomy Unit, Queen Mary, University of London,  and that it has not
been submitted in any previous application for a higher degree. \\

Some of the work contained in Chapter \ref{chaptertwo} was
carried out in collaboration with Dr David Seery and Dr Filippo Vernizzi
and is unpublished. Chapter \ref{chapterthree}
presents a project developed in collaboration with Dr David Seery,
published as an article in the Journal of Cosmology and 
Astroparticle Physics \citep{sh06}. The work in Chapter
\ref{chapterfour} was done by me alone and it is described in
an article available on-line \citep{hidalgo}. The material in
Chapter \ref{chapterfive} was done in collaboration with Dr
Alexander Polnarev and is published in  Physical
Review D \citep{hidalgo-polnarev}.  I made a major contribution to all
the original research presented in this thesis. \\

\vspace{1.23cm}
\begin{align*}
\qquad\qquad&\mbox{Juan Carlos Hidalgo}\\
\qquad\qquad&\mbox{Queen Mary, University of London}\\
\qquad\qquad&\mbox{London, United Kingdom}\\
\qquad\qquad&\mbox{April, 2009}
\end{align*}

}

\clearemptydoublepage
\thispagestyle{empty}

  \vskip3cm
\begin{center}
\addcontentsline{toc}{chapter}{Acknowledgements}
{\large \bf Acknowledgements }
\end{center}
\vskip1cm

{\large

I would first like to thank my parents Jes\'us and L\'{\i}dice, and my
sister Aura for their continual support, encouragement and inspiration
throughout my studies.   

I am grateful to my supervisor Prof Bernard Carr for his invaluable
support and guidance  during my PhD. 

I am also very grateful to my collaborators, Dr David Seery and Dr
Alexander Polnarev, for their assistance through several
challenges. Special thanks also to Dr Karim Malik for his
encouragement and sense of humour. (Apologies for the `Oscar winning'
speech). 

I would like to thank all those who have helped me from the School of
Mathematical Sciences and in particular, Mr William White and Prof
Malcolm MacCallum.

Dr Sergio Mendoza in the Instituto de Astronom\'{\i}a, UNAM, has
always been supportive and attentive to my academic development. I am
grateful to him for many years of inspiration.    

I would like to express my gratitude to my entire  family and
especially my grandfather Alfonso Cu\'ellar, my cousin L\'{\i}dice
Cuellar Quintero and my aunt Carmen Watson for their resolute encouragement.  

Finally, I would like to thank Christine Rooks for being brave enough
to join me on this journey. All friends who have stood by me: Rodrigo,
Adri\'an, Juli\'an, Rogelio, Gustavo and Mauricio, also to my
flatmates No\`elia and Periklis. Last but not least, I would like to
include Julio, Gian Paolo, Guillermo, Cesar, Rub\'en, Mariana, Paola
and Pedro. To all of them a `big' thanks.\\

This work was fully funded by the Mexican council for Science and
Technology (CONACYT scholarship No. 179026), with complementary
support from the School of Mathematical Sciences at Queen Mary,
University of London.  I gratefully acknowledge this support.

}

\clearemptydoublepage
\thispagestyle{empty}

  \vskip3cm
\begin{center}
{\large \bf Publications resulting from the work in this thesis}
\end{center}
\vskip1cm
{\large
 
 \begin{enumerate}
 
 \item   D.~Seery and J.~C.~Hidalgo,\\
  \emph{``Non-Gaussian corrections to the probability distribution of
  the  curvature perturbation from inflation,''}\\
  JCAP {\bf 0607} (2006) 008\\
  $[$arXiv:astro-ph/0604579$]$.

 \item   J.~C.~Hidalgo,\\
\emph{  ``The effect of non-Gaussian curvature perturbations on the
formation of primordial black holes,''}\\
  arXiv:0708.3875 [astro-ph].

 \item   J.~C.~Hidalgo and A.~G.~Polnarev,\\
 \emph{ ``Probability of primordial black hole formation and its
 dependence on the radial profile of initial configurations,''}\\ 
  Phys.\ Rev.\  D {\bf 79} (2009) 044006\\
    arXiv:0806.2752 $[$astro-ph$]$.

 \end{enumerate}
}

\clearemptydoublepage
\thispagestyle{empty}

  \vskip3cm
\begin{center}
\end{center}
\vskip5cm

\begin{flushright}
\textsc{To my parents, }\\
\textsc{and to the loving memory of my grandmother Juana.}
\end{flushright}
\clearemptydoublepage
\tableofcontents
\clearemptydoublepage
\listoffigures
\clearemptydoublepage
\begin{onehalfspace}
{\large
\chapter{Introduction}
\label{chapterone}

Cosmology is at the forefront of modern physics. Over the last two
decades, it has moved from a predominantly theoretical discipline to a sound
observational science.  
Today's experiments are capable of observing tiny fluctuations of a
faint signal coming from the Big Bang, emitted  about thirteen billion
years ago. The observations of primordial inhomogeneities are a unique
probe of the physical conditions in the early universe. The
inflationary paradigm indicates that the inhomogeneities are the
result of quantum fluctuations of the matter dominating the universe
in its first moments. In this widely accepted picture  the
observed inhomogeneities fix the normalisation of an inflationary
potential setting the energy scales for inflation to about
$10^{16}\mathrm{ GeV}$, the GUT scale. This is $10^6$ times more than
the energy of particles released by supernovae. A similar ratio arises
for the energy scales to be tested by the large hadron collider
(LHC). These numbers show how the geometry of the universe and its
inhomogeneities constitute a unique probe of high energy physics.  

Several observational parameters have been defined in cosmology in
order to determine the physical conditions of the early universe. The
density and nature of matter observed today, the distribution and mean
amplitude of initial inhomogeneities, and most recently the
non-Gaussianity of primordial fluctuations are among these
parameters. The latter has received considerable attention from
cosmologists but the analysis of the latest observations has not yet
provided conclusive evidence for departures from Gaussian
statistics. A great deal of effort is under way to reduce the 
detection thresholds of the non-Gaussian parameters. Even if
non-Gaussianity remains undetected by future experiments,  we can still
constrain theoretical models that are known to develop large
non-Gaussianity. 
  
The main objective of the present work is to study how non-Gaussian
statistics, inherited from inflation, can modify the probability of
primordial black hole formation. The class of models of inflation
that motivate this study and the development of statistical tools to
address this question are complementary projects, and both are
included in the present thesis.
In the rest of this chapter we provide a brief description of the
state of the art in cosmology, with special attention to the open
questions that motivate this thesis. 
 
\section{Cosmological observations and the Big Bang}
\label{intro-obs}

It has been more than four decades since \cite{penzias-wilson} managed
to identify, for the first time, the  cosmic microwave background
(CMB) radiation. This was detected, almost by accident, while
calibrating a large reflector at the Bell Laboratories. The uniform
and isotropic radiation observed corresponds to the most perfect
black-body radiation ever measured, peaking at $\lambda =
1.9~\mathrm{mm}$, with a red-shifted temperature of $T_{\rm CMB}\,=
2.725~\mathrm{Kelvin}$ \citep{jaffe-param}.  

The detection of the CMB gave  decisive support to the Big Bang
theory. The standard Big Bang model considers a universe dominated by
uniform and isotropic matter. Its  dynamics is governed by gravity,
with equations prescribed by the theory of general
relativity. (Gravity is the only long-range force to be considered
since the universe is electrically neutral.) The conditions of
isotropy and homogeneity, in this context, imply that the spacetime
admitting these properties is necessarily a Friedmann-Robertson-Walker
universe (FRW) (see e.g. \cite{wald}).   

\subsection{Basic dynamics of the universe}

We write the FRW metric in the form of the line-element in spherical
coordinates.  
\beq
	ds^2 = - dt^2 +\frac{ a(t)}{1 - \kappa r^2 }  \left(dr^2 + r^2
	\left[d\theta^2 + \sin^2  
\theta d\phi^2\right] \right),
\eeq

\noindent where $t$ and $\kappa$ are the coordinate time and the
	uniform curvature of the spatial sections
	respectively. The usual spherical coordinates in the spatial
	hypersurfaces are $r, \theta$ and $\phi$. Finally, $a(t)$
	is the scale factor, with present value $a_0 = 1$. The
	Einstein equations of general relativity provide the dynamical
	relation  between the matter and spacetime variables. Assuming 
	homogeneous and isotropic matter, with density $\rho$ and
	isotropic pressure $p$, the Einstein equations show that the
	evolution of the scale factor is given by
\beq
	H^2 \equiv \, \left(\frac{\dot{a}}{a}\right)^2 =\,
	\frac{1}{3}\rho  - \frac{\kappa}{a^2},
	\label{friedmann:kappa}
\eeq

\noindent where an over-dot is the coordinate time derivative and $H$ is the Hubble 
parameter, a measure of the expansion rate. Its present value is $H_0 =
100 h~\mathrm{km s^{-1} Mpc^{-1}}$, with $ h = 0.71~\pm~0.08 $
\citep{freedman-h}. This last equation is known as the Friedmann
equation. We use throughout units where $c  =  \hbar = 8\pi G  = 1$.  

The matter contents of our universe has several components and the
fraction of each  component  relative to the critical density is
called the density parameter $\Omega_{i} =  \rho^{(i)} / 3 H^2$. If we
denote the sum of all matter components as $\Omega_{\rm T}$, the
Friedmann equation can be  written simply as 
\beq
	\Omega_\kappa(t) + \Omega_{\rm T}(t) = 1, 
        \label{friedmann:omega}
\eeq

\noindent where $\Omega_\kappa = \kappa / (a H )^2$ is the curvature
density parameter. When the matter density is equal to the critical
density $3 H^2$, then $\Omega_{\rm T} = 1 $ and the universe is
flat at all times. Observations tell us that we live in a nearly flat
universe ($|\Omega_{\kappa}| < 10^{-2} $), so we assume 
$\Omega_{\kappa} = 0$ hereafter. The energy density is dominated by
two main components, a cold dark matter component ($\Omega_{\rm CDM}
\simeq 0.23$) and another component referred as dark energy
($\Omega_{\Lambda} \simeq 0.72$).  The nature of both these components
is a crucial question in cosmology and has  motivated a lot of
research. We will return to this point and
to an analysis of the Einstein equations later in this work.  

 \subsection{The Big Bang model}
The hot Big Bang model is now accepted as the standard model
describing the evolution of the universe. This model characterises, with
impressive accuracy, the evolution after the first second. At this
time, the universe was a primordial fireball with high enough
temperature and pressure to dissociate any nuclei. The formation of
nuclei was only possible once the cosmic expansion reduced the average  
kinetic energy sufficiently.  The formation of the first 
elements took place at temperatures of around $T \simeq  0.1~{\rm
MeV}$, when the universe was around $1\,{\rm s}$ old.  This process
involves conditions that cannot be replicated elsewhere (cf. stellar
nucleosynthesis). Within the current observational limitations, the
Big Bang prediction for the present abundance of light elements is
confirmed remarkably  by the present measurements.   

Big Bang nucleosynthesis halted once matter had cooled down enough, due 
to the cosmic expansion. The electrical neutrality of the matter was
reached at a more recent event: the so called `recombination' process
refers to the time when each electron was captured by a nucleus forming
the first neutral atoms. Subsequently, at a temperature of around
$T \approx 0.1~\mathrm{eV}\, (\approx 10^3~\mathrm{Kelvin})$, CMB photons
decoupled from ordinary matter and have since travelled freely. These
same photons reach us in the form of microwave radiation. The surface
of emission of these primordial photons is called the last-scattering
surface. CMB observations constitute irrefutable proof that the
universe was homogeneous at early epochs and dominated by radiation
when  $T>10^3~\mathrm{Kelvin}$.   

The current temperature of the CMB radiation ($T_{\rm CMB} =
2.725~\mathrm{Kelvin}$) 
is measured  with such precision because its fluctuations are
tiny. The first observational evidence for the CMB anisotropies came
from the COBE satellite \citep{smoot92,bennett-banday}. The results of
this experiment showed that the temperature fluctuations have a mean
amplitude  $\delta T /T \sim 10^{-5}$. The amplitude of such
deviations was predicted by \cite{peebles70} and \cite{zeldovich72} in
terms of the matter density perturbation $ \delta
\rho / \rho  \sim 10^{-5}$. These inhomogeneities are related through
the Sachs-Wolfe formula \citep{sachs-wolfe}. This prescribes that for
inhomogeneities of comoving size $\lambda$, 
\beq
	\frac{\delta T}{T} \approx -  
        \frac{1}{2} \left({a_{\rm LS}H_{\rm LS}}{\lambda}\right)^2\, 
        \delta_{\rho}
\eeq

\noindent 
where we have defined $\delta_{\rho} \equiv \delta \rho / \rho$, and 
where a subscript ${\rm LS}$ indicates an evaluation at the
last-scattering surface. 

More recent experiments, such as BOOMERANG \citep{boomerang},
MAXIMA \citep{maxima-1} and
WMAP \citep{wmap2006-temp,wmap2008-params}, managed to measure the  
acoustic oscillations in the radiation plasma due to the small-scale
density variations in the early universe. Measurements of acoustic
oscillations in the CMB demonstrated the flatness of the universe to
$1\%$ precision (i.e. $|\Omega_{\kappa}| < 10^{-2}$). They were also
used to rule out cosmic strings as a significant contributor to
structure formation and suggested `cosmological inflation' as the
theory of structure formation \citep{jaffe-param}.

\section{Cosmological inflation}
\label{intro-inflation}

\subsection{Motivation and achievements} 

The observations mentioned above provided strong arguments in
favor of the  Big Bang model but also showed the necessity of a
larger theoretical framework due to the following problems: 
\begin{enumerate}
	\item \textbf{Horizon problem}. In the Big Bang model, the
	distance light could have traveled up to the time of
	last-scattering $d_{\rm LS}$ is of order $ 180 \Mpc$. This is
	called the particle horizon and determines the radius of causally
	connected regions at that time. The particle horizon today is much
	larger, with radius $d_{0} \sim 6000 \Mpc$. Therefore, the
	measurements of CMB radiation at angular scales larger than
	one degree include regions that were causally disconnected at
	the time of the photon decoupling. The temperature at such
	scales is observed to be uniform up to one part in
	$10^5$. This means that causally disjoint patches of the
	universe in the past had the same thermal history. In the
	context of the hot Big Bang model there is no plausible
	explanation for this fact.    

	\item \textbf{Flatness problem}. The density of matter components in
	the universe is diluted with time due to the cosmic
	expansion. Conversely, if there was an initial curvature
	component $\kappa$, then this would rapidly dominate the matter
	contents. This is easily derived from
	Eq.~\eqref{friedmann:omega}, which can be written in the form 
        \beq
        \Omega_{\rm T}-1 = \frac{\kappa}{a^2H^2} \equiv \Omega_{\kappa}.
        \eeq

        \noindent The product $aH$ decreases with time in a radiation or matter 
        dominated universe. If the universe is initially flat,
        then it remains flat for subsequent times, but observations
        show that $|\Omega_{\kappa}| \lesssim 10^{-2}$ today, and the
        Friedmann evolution demands an even smaller curvature in the 
        past. For example, at nucleosynthesis, when the universe was
        around $1~{\rm s}$ old, we require $|\Omega_{\kappa}| \lesssim
        10^{-16}$ to be consistent with the present value. Such a
        small value requires an extreme fine-tuning of initial
        conditions $\Omega_{\rm T}$, for which a causal explanation
        would be desirable.      

\end{enumerate}
	
        A solution to these problems is provided by the inflationary
	paradigm, which we will study in detail in
	Chapter \ref{chaptertwo}. The main feature of this theory is
	that it  changes the behaviour of the comoving cosmological
	horizon by considering an accelerated expansion of the
	universe at early times, i.e., at times prior to
	nucleosynthesis. In terms of the scale factor, this condition
	demands  
\beq
	\ddot{a} > 0 \quad\Rightarrow \quad \dtotal{}{t}
	\left[\frac{1}{aH}\right] < 0.
	\label{inf:cond}
\eeq
	
	\noindent The shrinking of the cosmological horizon represents
	a `reverse' evolution of spacetime which avoids the
	fine-tuning of initial conditions demanding homogeneity and
	flatness. If we consider an inhomogeneous patch of the
	universe when inflation starts, at an initial time $t_{i}$,
	the cosmic accelerated expansion brings all initial
	inhomogeneities out of the comoving cosmological horizon. If
	inflation lasts long enough, then after the inflationary
	period we are left with a much larger region composed of small
	patches of size of the cosmological horizon which are out of
	causal contact but with common physical characteristics. The number of
	e-folds of expansion required for the listed problems to be
	solved is  
	\beq
		N = \ln\left(\frac{a(t_{\rm end})}{a(t_{\rm i})}\right) 
                \gtrsim 60. \quad{\rm } 
	\eeq 
	
\noindent This number is required to guarantee that the comoving 
	scale of the current size of the universe exited the horizon
        at the beginning of inflation \citep{liddle-lyth}. This
        indicates that inflation must last longer than 60
        e-folds. Arguably, it was \cite{guth-inflation} who first
        brought  these ideas together.  
 
 The theory of inflation has received important contributions from
 particle physics. In particular, the theory of particle creation from
 vacuum fluctuations \citep{hawking-fluct,starobinsky-fluct} gave
 inflation its strongest argument: the vacuum fluctuations generated
 during inflation are redshifted to superhorizon scales by the action
 of the inflationary mechanism. At the end of inflation, the
 thermalisation of the inflaton false vacuum reheats the universe and
 the standard hot Big Bang phase begins. In this transition, the
 vacuum fluctuations of the inflaton field are transformed into matter
 density perturbations with a prescribed amplitude. From this
 transition onwards, the modes re-enter the expanding comoving
 horizon. Thus, initial conditions of cosmological perturbations in
 the hot Big Bang are set by inflation. 
 The observed mean amplitude of the temperature
 inhomogeneities \citep{smoot92,boomerang,wmap2006-params} sets the
 energy scale at which the initial vacuum fluctuations were generated
 by tracing back the evolution of fluctuations described above.  
 This simple explanation of the origin of the temperature fluctuations
 constitutes a decisive argument in favour of the inflationary
 scenario. It represents the greatest advantage of inflation over many
 other alternative extensions of the standard Big Bang scenario.  

 In summary, the requirements for a period of inflation are:
 (1) a mechanism to generate an accelerated expansion maintained  for
 at least 60 e-folds of expansion;
 (2) a way of accounting for the transition to the subsequent FRW
 stages of evolution, thereby providing the suitable initial conditions for
 the Big Bang scenario; 
 (3) quantum fluctuations of the inflationary field, generated at
 observable scales such that the matter density fluctuations of size
 $\lambda$ meet the relation $(aH \lambda) \delta_\rho \simeq
 10^{-5}$ and this product is almost invariant over the observed scales.   
 
 In practice, measurements of CMB anisotropies, combined with measurements of 
background parameters inferred from supernovae surveys
\citep{astier05,riess06}, indicate that the root-mean-square 
(${\rm RMS}$) amplitude of temperature fluctuations is   
 \beq
 	\left(\frac{\delta T}{T}\right)_{{\rm RMS}} \approx 2 \times10^{-5},
 \eeq
 
 \noindent at the pivot scale with comoving size  $\lambda_{\rm CMB} =
 150\,{\rm Mpc}$ customarily used in CMB studies.   
  Observations also indicate that this value does not vary
 significantly over the range of observed scales. In other words the
 mean amplitude is almost scale-invariant for angular scales larger
 than one degree. In
 Chapter \ref{chaptertwo} we show how this relates to the curvature
 perturbation $\zeta$ and discuss its basic properties. In particular,
 we will show that, in the cases which concern us, $\zeta$ is
 constant for scales larger than the particle horizon.  

\subsection{An embarrassment of richness}

  The required amount of inflation and the corresponding amplitude of
 the curvature perturbations determine the kind of matter and energy scale
 necessary to satisfy the conditions for accelerated expansion. 
  These  prerequisites have been met by several models of inflation which
 may or may not be motivated by more fundamental theories of
 physics. One of the main problems faced by the inflationary paradigm
 is that of richness. There are many models that meet the dynamical
 requirements. Most of them invoke one or more scalar fields
 $\{\phi_i\}$ with dynamics governed by a potential $V(\phi_i)$.
 There are a plethora of models, each of which corresponds to
 particular realisation of this potential, which satisfy the
 observational constraints up to the level of the observed
 inhomogeneities. Consequently, many of the models cannot  
 be distinguished at the level of linear perturbation theory. This
 demands the formulation and experimental determination of new
 parameters that provide complementary information about the early
 universe. An important constraint on the inflationary models can be
 obtained by considering the statistical deviations from a Gaussian
 field of fluctuations.  
This idea has opened a new window in the study of the early universe,
 namely the nonlinear extension of perturbation theory and its
 non-Gaussian statistics. 
    
\section{Non-Gaussianity}
  \label{intro-ng}
  By non-Gaussianity in cosmology we refer to the small
  deviations of observed fluctuations from the random field of linear,
  Gaussian, curvature perturbations $\zeta_1(t,\bx)$.  $\zeta(t,\bx)$
  is the curvature perturbation in the comoving gauge, that is, as
  measured by an observer which sees no net-momentum
  flux. The mathematical expression for $\zeta(t,\bx)$ in terms of the
  matter density perturbation is provided in Chapter \ref{chaptertwo}. 

  Among the parameters of nonlinearity, the nonlinear coupling
  $\fnl$ is the most useful observable for describing non-Gaussianity.
  Its definition comes from the second order expansion of curvature
  perturbations in real space, which can be written as 
   \beq
   \zeta(\bx) = \zeta_{\rm 1} + \frac{1}{2}\zeta_{\rm  2}, 
  \label{ng:exp}
  \eeq
  	
  \noindent where $\zeta_{\rm 1}$ refers to the Gaussian perturbation
  with variance $\Sigma^2_\zeta (x)= \zeta_{\rm RMS}^2 (x)$ and
  $\zeta_{\rm 2}$ is the second order perturbation parametrised by the
  nonlinear parameter
  $\fnl$ in the following way
  \beq
  \qquad \zeta_{\rm 2}(\bx) =  - \frac{6}{5} \fnl 
  (\zeta_{\rm 1}(\bx)^2  - \zeta_{\rm RMS}(x)^2).
  \eeq

 Note that the perturbative expansion of $\zeta$ implies also the
  rough definition  
  \beq	
  	\fnl = - \frac{5}{6}\frac{\zeta_2(\bx)}{\zeta_1^2(\bx)}, 
  \eeq
  
  \noindent which gives an intuitive notion of this
  parameter. Historically, non-Gaussianity as a test of the accuracy
  of perturbation theory was first suggested
  by \cite{Allen-Grinstein-Wise}.  
   The definition of $\fnl$ used here was first introduced
  by \cite{salopek-bond} in terms of the Newtonian or Bardeen
  potential $\Phi_B$ (defined in Chapter \ref{chaptertwo}). Their
  initial definition has been preserved by
  convention \citep{gangui-lucchin,verde-ngI,komatsu-spergel}, which
  is why the  transformation to the curvature perturbation $\zeta_2$
  involves the numerical factor $-5/6$.    
  In the context of perturbation theory, the study of dynamical
  equations at second order  yields important information independent
  of the parameters of linear perturbations.  
Thus, in the nonlinear regime, we can discriminate different models
  of inflation which are degenerate at linear order. This fact has
  motivated the search for non-Gaussianity in the CMB and large-scale
  structure.   
  
  Statistically, the lowest order effect of including a non-Gaussian
  contribution  is a non-vanishing correlator of three copies of the
  curvature field $\zeta$. The three-point function in Fourier space
  is given by the bispectrum $B$, defined by 
    \beq
 	\langle  \zeta(\bk_1) \zeta(\bk_2) \zeta(\bk_3)\rangle =
 	(2\pi)^3 B_{\zeta} (k_1,k_2,k_3) 
        \delta^{(3)} (\bk_1 + \bk_2 + \bk_3), 
        \label{ng:bisp}
\eeq
         
   \noindent where $\delta^{(3)}$ is the three-dimensional Dirac
   delta function. 

The bispectrum is directly related to the parameter $\fnl$ and for
each mode $k = |\vect{k}|$. Moreover, being a function of three
momenta, the $k$-dependence of the bispectrum also provides valuable
information which could help us to understand the physics of the early
universe.  
   
  The nonlinear parameters have been investigated through the analysis
  of higher order correlations in the CMB anisotropies observed mostly
  by the WMAP satellite \citep{wmap2006-params}. After five years of
  collecting data, WMAP observations give the limits $ -151
  < \fnl^{\rm equil} < 253$ \citep{wmap2008-params} for an equilateral
  triangulation of the momenta and  $ -4 < \fnl^{\rm local} <
  80$ \citep{smith09} for a local triangulation. The triangulation
  of the bispectrum is a characteristic which arises
  due to the following: The momentum conservation in the three point
  correlation is guaranteed by the delta function in
  Eq.~\eqref{ng:bisp}, which demands that the sum of the three vectors
  is zero. In consequence the three momenta represent the sides of a
  triangle in $k$-space. Two main triangulations can be distinguished:
  the equilateral triangulation and the isosceles or local triangulation,
  which are characteristic shapes of different models of inflation
  (see e.g. \cite{babich-creminelli}). An experimental detection of
  $\fnl$ would greatly narrow the range of cosmological models which
  meet the observational bounds. In the near future, space telescopes,
  and in particular the PLANCK satellite, are expected to tighten
  these bounds considerably. Specifically, any signal with $|\fnl
  | \gsim 5$ should be observed by
  PLANCK \citep{komatsu-spergel,liguori-hansen}. This raises the 
  exciting possibility of looking for particular signatures of
  inflationary models.   

  Another attractive observational prospect for non-Gaussianity is to
  look at the implications of considering primordial non-Gaussian
  fluctuations in the study of the statistics of galaxies and other
  large-scale structures
  (LSS) \citep{verde-ngI,verde-ngII,loverde-shandera}.  Such
  observations  probe inhomogeneities at scales smaller than those
  observed in the CMB. 
  
  The effects of non-Gaussianity in the LSS can be classified into two
 categories, which provide distinct observational methods for
 detecting non-Gaussianity.  The first is  the bispectrum of galaxies,
 potentially determined by computing the three-point correlation
 function from redshift
 catalogues \citep{verde-jimenez,scoccimarro-ng}. The second is  the
 non-Gaussian correlations in the probability distribution function
 (PDF) which leads to modifications in the number of galaxies and
 other structures with respect to the Gaussian
 case \citep{verde-ngI,verde-ngII}.  
  
  Both methods involve delicate issues, crucial for the  correct
  interpretation of observations. 
Most important is the fact that the inhomogeneities that collapse to
  form galaxies evolve in a nonlinear fashion at late times. This is
  because the primordial fluctuations enter the horizon much before
  they form virialised structures. Consequently, the nonlinear
  evolution of fluctuations may blur the primordial non-Gaussianity of
  the initial statistics.  
    
  Another important problem is that there is no single way of
  constructing a non-Gaussian PDF from theoretical models, i.e.,
  several non-Gaussian PDFs can be constructed with a common  variance
  and skewness. This well known problem has been expressed pithily
  by \cite{heavens-dog}: ``We know what a dog is, but, what is a
  no-dog? A no-dog can be anything''.  The effects on, say, the
  integrated number of galaxies may change substantially with every
  realisation of the PDF. This complicates the interpretation of
  non-Gaussian signatures.  
  
  In Chapter \ref{chapterthree}, a formalism is presented to attack this 
problem. We construct the PDF of the curvature perturbations with a
direct input from its higher-order correlations. The formalism is then
applied to compute the modification which a non-Gaussian distribution
of fluctuations brings to the abundance of primordial black holes.

  \section{Primordial black holes}
  \label{ngpbhs}

\subsection{Standard picture}

The idea that large amplitude matter overdensities in the universe could have
collapsed through self-gravity to form  primordial black holes (PBHs)
was first put forward by \cite{zeldovich-novikov-I} and then 
independently by \cite{hawking-pbhsI} and \cite{carr-hawking}
more than three decades ago. They suggested that at early times
large-amplitude overdensities would overcome
internal pressure forces and collapse to form black holes. The
standard picture of PBH formation from initial inhomogeneities
prescribes that an overdense region with size $r_{\rm i}$ will overcome
pressure and collapse to form a black hole if its size is bigger than
the associated Jeans length 
\beq
r_{\rm J} = 4\pi\frac{\sqrt{w}}{5 + 9w}d_{H},
\eeq

\noindent where the particle horizon $d_H$ is of order of the Hubble
radius $r_{ H} = 1 /H$. Here we assume an equation of state$ p =
w\rho$, where $w$ is constant. For the case of
radiation-domination, for example, $w = 1/3 $. 

The size of the initial inhomogeneity must also be smaller than the
separate universe scale    
\beq
 r_{\rm U} =  \frac{1}{H} f(w),
\eeq

\noindent where the function $f(w)$ has been derived by
\cite{harada04}, and is of order unity. Thus, $r_{\rm J} < r_{\rm i} <
r_{\rm U}$, both limits being of order the Hubble radius. Consequently
the mass of a PBH is close to the Hubble horizon mass. This gives a
simple formula for the mass of a PBH 
forming at time $t$ during radiation domination \citep{carr}: 
\beq
	M_{\rm PBH} \simeq M_H = \frac{4}{3}\pi r_{H}^3\rho = \ten{15}
	\left(\frac{t}{\ten{-23}~\mathrm{s}}\right)~\gr. 
\eeq

The PBH mass spectrum depends mainly on two characteristics of the
early universe:  
the equation of state $w$, which determines how large the amplitude of
initial inhomogeneities should be to halt the background expansion and
recollapse, and the nature of the initial density fluctuations,
which determines how likely such amplitudes are. 
\cite{carr} determined the threshold amplitude 
$\delta_{\rm th} \equiv (\delta_\rho)_{\rm th}$
required for the density perturbation to collapse to a PBH to be 
$\delta_{\rm th} \sim w$. In this case, one needs
perturbations to the FRW metric with mean amplitude of order
unity to form a significant number of PBHs.  

 The special characteristic of PBHs is that they can form
at very early epochs and have very small masses. The smallest PBHs
  would have formed at the end of the inflationary 
  expansion \citep{carr-lidsey-constraints},
  even from field fluctuations that never exited the horizon 
  \citep{lyth-malik-zaballa,zaballa-green}. The mass of the horizon at
  the end of inflation is \citep{zaballa-green}
\beq
	M_{H}  \simeq 10^{17} \gr \left(\frac{\ten{7}\gev}{T_{\rm RH}}\right)^2,
	\label{intro:horizonmass}
\eeq
  
\noindent where the reheating temperature $T_{\rm RH}$ depends
	sensitively on the model of inflation considered. In the
	canonical slow-roll inflationary model this temperature can be
	well above  $\ten{10}~\gev$ \citep{kolb-turner}. Taking on
	account the production of dark matter candidate particles in
	supersymetric models, this temperature could be dropped by
	several orders of magnitude, however, leptogenesis does not
	alow the reheating scale to be smaller than
	$\ten{9}~\gev$ \citep{buchmuller}. This in turn means that
	PBHs could have been produced with masses much smaller than
	$10^{11}~\gr$. On the other hand, PBHs that formed at $1~{\rm s}$
	have masses of order $10^{5}M_{\odot}$ which is already in the
	range of masses of black holes at the centre of galaxies.

  The small masses of PBHs prompted the investigation of their quantum
  properties. The well known result of \cite{hawking-pbhs} shows that
  black holes radiate with a temperature 
  \beq
  	T
	\simeq \ten{-7}\left(\frac{M}{M_{\odot}}\right)^{-1}\mathrm{Kelvin}
  \eeq

  \noindent  and evaporate entirely on a time scale
  \beq
  	t_{\rm evap}  
	\simeq \ten{64}\left(\frac{M}{M_{\odot}}\right)^3~\mathrm{y}, 
  \eeq
 
 \noindent where $M_{\odot}$ is the solar mass. 
 With the age of the universe estimated as $1.37 \pm 0.015 \times
 10^{10}~\mathrm{y}$ \citep{wmap2006-params}, we can predict that PBHs
 with mass $M_{\rm crit} = 5\times10^{14}~\gr$ are evaporating
 now. PBHs are also the only type of black holes for which 
 the effect of Hawking evaporation  could be observed. Indeed,
 the black holes evaporating now would be producing photons with
 energy $100~\mathrm{MeV} $ \citep{page-hawking}. The observed
 $\gamma$-ray background radiation at this energy implies that the
 density parameter of such PBHs must satisfy \citep{page-hawking}  
 \beq
 \Omega_{\rm PBH}(M \sim \ten{15}~\gr) \,\lesssim\,10^{-8}.
\label{pbh:bound}
 \eeq
 
 \noindent   This bound remains the tightest constraint to the
 abundance of PBHs.  Additional cosmological
 bounds to the mass fraction of PBHs are reviewed in Chapter
 \ref{chapterfour}.

The  mass fraction of the universe turning into PBHs of mass $M$ at
the time of their  formation is denoted by
$\beta_{\rm{PBH}}(M)$. This is equivalent to the  probability of
formation of PBHs of mass $M$. In a rough calculation,
$\beta_{\rm{PBH}}(M)$ is given by the Press-Schechter formalism  
\citep{press-schechter,carr} as the integral of the PDF over all
amplitudes $\delta_{\rho}$  above the threshold $\delta_{\rm{th}}$:
\beq
	\beta_{\rm{PBH}}(M) = 2 \int_{\delta_{\rm th}}^{\infty}
	\Prob(\delta_{\rho})\,  
	\d\delta_{\rho},
	\label{intro:press-schechter}
\eeq

\noindent where the factor two has been added to account for the half
	volume of the universe that is necessarily underdense. With
	this factor the Press-Schechter formula gives a good fit to
	the results of N-body simulations for the case of galactic
	haloes \citep{peebles80}. For the case of PBHs, an upper limit
	of integration is formally required. This is the amplitude of
	an inhomogeneity for which the total mass would form a
	separate closed universe. However, the contribution of higher
	values to the probability is almost negligible and we do not
	include an upper limit here. For the case of a Gaussian PDF
	with variance $\Sigma_{\rho}(M)$ this integral is approximated
	by \citep{carr}  
\beq
	\beta_{\rm{PBH}}(M) \approx \delta_{\rm th} \exp\left(-
	\frac{\delta_{\rm th}^2}{2\Sigma_{\rho}^2(M)}\right). 
\label{intro:beta}
\eeq

\noindent This equation demonstrates the sensitive dependence of the
probability of PBH formation with $\delta_{\rm th}$.  The above
integral is expected to be small due to the exponential dependence on
the threshold value $\delta_{\rm th}$. $\beta_{\rm{PBH}}$ is also
known to be small because it is related to the current density
parameter $\Omega_{\rm PBH}$ of PBHs formed at time $t$ and with mass
$M$ by 
 \beq
 	\Omega_{\rm PBH} =  \beta_{\rm PBH}~\Omega_{\rm R} 
	\left(\frac{a_0}{a(t)}\right)\simeq  10^6~\beta_{\rm PBH}
	\left( \frac{t}{1~{\mathrm{s}}}\right)^{-1/2}
	\simeq  10^{18}~\beta_{\rm PBH}
	\left(\frac{M}{10^{15}\gr}\right)^{-1/2},
 \eeq

\noindent where $\Omega_{\rm R} = 8\times\ten{-5}$. The factor
$a^{-1}$ arises because PBHs form mostly during the
radiation-dominated era but PBH density scales as $a^{-3}$, while
radiation scales as $a^{-4}$. From this relation we see that any limit on
$\Omega_{\rm PBH}$ places a direct constraint on  $\beta_{\rm PBH}$. For example,
from the bound in Eq.~\eqref{pbh:bound}, we infer that $\beta_{\rm
PBH}(M=10^{15}\gr)$ can only have a small value of order $10^{-26}$. 

\subsection{Shortcomings}

The simple picture of PBH formation described above has several shortcomings
\begin{enumerate}
\item In the radiation era the inhomogeneities forming PBHs must have a
large amplitude when they enter the horizon and they must be bigger than
the horizon for a considerable period of their evolution. As we will
show in Chapter \ref{chaptertwo}, 
the inhomogeneities at superhorizon scales are best described in terms
of curvature perturbations because they are constant in this regime. The
curvature perturbation has already been used in the more recent
numerical simulations of PBH formation
\citep{shibata-sasaki,niemeyer-dynamics,musco-polnarev}. Here, as in
several other recent works on the subject
\citep{yokoyama99,green-liddle,zaballa-green,josan09}, we
 compute the probability of formation of PBHs from the statistics of
 the curvature perturbations. This has the  
advantage of relating the formation of PBHs directly to the
initial perturbation spectrum. Additionally, it avoids the gauge anomaly
 associated to the matter density fluctuation.  
\item In the calculation of the probability of PBH formation, one
could argue that the Press-Schechter formula in
Eq.~\eqref{intro:press-schechter} 
is only an empirical approximation. Alternative approaches have
therefore considered the theory of
peaks \citep{green-liddle}. However, this does not render significant
corrections to the Press-Schechter  result. Moreover, the
Press-Schechter formula can be used to calculate the probabilities of
large-scale structure formation from non-Gaussian
PDFs \citep{verde-ngII}. Indeed, the latest numerical simulations
confirm that it is a good approximation even 
in this case \citep{grossi09}. This justifies our choice of
the Press-Schechter formalism to explore new aspects of the
probability of PBH formation.
\item A severe oversimplification of the usual calculation of the
probability of PBH formation is the
assumption of Gaussianity. The exponential decay of the
Gaussian PDF is preserved after its integration in the Press-Schechter
formula \eqref{intro:beta}. The fact that the
mass fraction involves an integration over the tail of the normal
distribution, where the probability density is small, leads us to
consider that a slight variation on the profile of the PDF might
modify this picture significantly. Indeed,  non-Gaussian probability
distributions have been considered in studies of the probability of
PBH formation by \cite{bullock-primack} and \cite{ivanov}. The
discrepancy in their results and the large departures from the
Gaussian case make this problem worth revisiting. One main objective
of this thesis is to derive the modifications that non-Gaussian PDFs
bring to the probability of PBH formation in the most general cases.  
We explore for the first time the modifications that a non-Gaussian
PDF may bring for the bounds on the amplitude of fluctuations and the
higher order statistics parameter $\fnl$ on the cosmological scales
relevant to PBH formation.

\item The last important problem in the calculation of
$\beta_{\textrm{PBH}}$ is the determination of the precise value of the
threshold amplitude $\delta_{\rm th}$ or $\zeta_{\rm th}$ for the
density or the curvature inhomogeneity. This approximation of
$\beta_{\textrm{PBH}}$ prompted several studies of PBH formation to
determine the precise value of the threshold amplitude. Early
 numerical simulations of gravitational collapse, however, already
 showed that this value depends sensitively on the shape and profile of
 the initial configuration $\delta_{\rho}(\bx)$
 \citep{nadezhin-novikov}. This dependence indicates that the lower
 limit of the  integral \eqref{intro:press-schechter} is not uniquely
 prescribed for all  configurations collapsing to form PBHs. The
 problem then is how to 
 differentiate profiles of initial inhomogeneities in the calculation
 of the probability of PBH formation. This is another problem we
 address in this thesis. We calculate the
probability of PBH formation by taking into account the radial profiles of
initial curvature  inhomogeneities. This represents a first attempt to
incorporate profiles into the calculation of $\beta_{\textrm{PBH}}$
and allow for a more precise estimation of the probability of PBH formation.    

\end{enumerate}

\subsection{Alternative mechanisms of PBH formation}

The formation of PBHs is not limited to the collapse of overdensities. PBHs 
may also form at the phase transitions expected in the early
universe. Let us here briefly review other known mechanisms of PBH
formation.   
 
\begin{enumerate}
\item[$\bullet$]PBHs may form at early phase transitions where the
equation of state is soft for a small period of time. In such
transitions, the effective pressure in the universe is reduced due to
the the formation of non-relativistic particles. Hydrodynamical
simulations show that at such a phase transition the value of
$\delta_{\rm th}$ is reduced below the value pertaining to the
radiation era. This mechanism enhances the probability of PBH
formation at a mass scale of the order of the horizon mass at that
time \citep{khlopov-polnarev,jedamzik-qcd}.

\item[$\bullet$] Loops of cosmic strings can collapse to form
PBHs. Cosmic strings are topological defects formed at the phase
transitions in the very early universe.
Closed loops can be formed from string self-intersection. The scale of
a loop will be larger than the Schwarzschild radius by a factor $(G\mu
)^{-1}$, where $\mu$ is the string mass per unit length, a free
parameter in the theory. In the cosmic
string scenario, these loops are responsible for the formation of 
cosmological structures if $(G\mu )$ is of order $\ten{-6}$. In this
scenario, there is always a small probability that particular
configurations, in which all the loop dimensions lie within its
Schwarzschild radius, can collapse to form black holes.   This
mechanism has been discussed by many authors
(see e.g. \cite{hawking-pbhsIII,polnarev-strings,garriga-sakellariadou}).  
However, WMAP and observations of galaxy distributions show that
cosmic strings can at most contribute to $10\%$ of the temperature
anisotropy in the CMB \citep{wyman-pogosian}. The mass per unit length
is less constrained by the observational limits on primordial black
holes \citep{caldwell-strings}. Because the $\mu$ parameter is
scale-invariant and its most stringent limit comes from CMB
observations, we can say that the formation of PBHs from cosmic string
loops is subdominant with respect to the standard picture of collapse
of overdensities.

\item[$\bullet$] One can also consider closed domain walls which form
black holes. Domain walls are hypothetical topological defects of
higher order. In a phase transition of second order, such as might be
associated with inflation, sufficiently large domain walls may be
produced \citep{domain-wallsI}. This leads to the formation of PBHs in
the  lower end of the range of masses  \citep{domain-wallsII}. 
 
\item[$\bullet$] Recently, a mechanism to form PBHs as the result of
warping cosmic necklaces has  been suggested. These topological
defects arise in the process of symmetry breaking in the framework of
quantum strings \citep{matsuda}.   
\end{enumerate}

 \noindent In all these  mechanisms the PBHs have mass of order the horizon mass
 at phase transitions in the early universe. They are also expected to
 produce PBHs with a Gaussian distribution. Here we are interested
 mostly in PBHs with a non-Gaussian distribution in order to produce
 constraints on models of inflation, so we do not study these
 alternative formation mechanisms.

\section{Thesis outline}
\label{outline}
Chapter \ref{chaptertwo} presents a study of non-Gaussianity from
inflationary scalar perturbations. It first introduces the relevant
definitions and the main tools used in the study of inflationary
perturbations. It then focuses on the derivation of non-Gaussian
correlation functions. Specifically, the three-point correlation is
studied in models where an auxiliary scalar field during inflation is
responsible for the generation of non-adiabatic fluctuations. The
cases in which the non-adiabatic fluctuations may generate large
values of $\fnl$ is considered in detail. 
 
The method used to derive the non-Gaussian correlators requires the
solution of the Klein-Gordon equation beyond linear order. This
equation is solved considering a perturbative expansion of the
nonlinear terms without taking on account the metric back-reaction. For the
cases in which analytic solutions are possible, the derivation of the
three-point correlation is presented. Finally, the observational
limits on $\fnl$ are used to constrain models of inflation which
include a curvaton field, a special case of an isocurvature field.

Chapter \ref{chapterthree} discusses the decomposition of the
curvature perturbation $\R$ into harmonics. This is a technical step,
which is necessary in order to write down a path integral for the PDF
$\Prob(\R)$. We present the calculation for the Gaussian case first,
in order to clearly explain our  method with a minimum of technical
details. This is followed  by the equivalent calculation including
non-Gaussian corrections which follow from a non-zero three-point
function. Finally we calculate the probability $\Prob[\R(k)]$, which
will be used to derive a non-Gaussian probability of PBH formation.

In Chapter~\ref{chapterfour} we compute the mass fraction $\beta_{\rm
PBH}$ resulting from a non-Gaussian PDF of primordial curvature
fluctuations $\R$. We restrict ourselves to the case in which the
non-Gaussian PDF corresponds to a constant value of $\fnl$. It is
first shown how to reconcile the discrepancy between two previous
studies of non-Gaussian PBH formation \citep{bullock-primack,ivanov}.
We then calculate the modifications to the observational bounds to
$\beta_{\rm PBH}$ when  a large value of $\fnl$ is included.

Chapter \ref{chapterfive} explores the probability of finding
non-trivial spatial profiles for the perturbations that form PBHs. The
numerical simulations show that the usual assumption of homogeneous
spherically symmetric perturbations collapsing to PBHs is not
appropriate.  Chapter \ref{chapterfive} provides a probabilistic
analysis of  the radial profiles of spherical cosmological
inhomogeneities that collapse to form PBHs.  
Based on the methods used to construct non-Gaussian PDFs, we 
derive the probability distribution for the central amplitude of $\R$
and for the second  radial derivative $\textrm{d}^2 \R/ \textrm{d}r^2$
at the centre of the spherically symmetric inhomogeneity used to
describe the radial profiles explored in studies of gravitational
collapse. We then consider the joint probability of both parameters
to compute the correction to $\beta_{\rm PBH}$.  
The results show how much the probability of PBH formation can be
reduced if we do not include all possible configurations forming PBHs. 

Chapter \ref{chaptersix} is the summary and conclusion of this
thesis. We also describe  future research which may follow. The key
achievement of this thesis is to combine for the first time the study
of two crucial probes of the early universe. The effects of nonlinear
non-Gaussian inhomogeneities and primordial black hole formation.

\clearemptydoublepage
\chapter{Non-Gaussian curvature perturbations}
\label{chaptertwo}


\section{Outline}

Observations of cosmological structure and CMB parameters are
best interpreted in the context of cosmological perturbation
theory. This is a useful tool to connect observations with models of
inflation derived or motivated by high energy  physics theories for
which there is no other available test. Surprisingly enough, the
simplest inflationary model, consisting of a single scalar field slowly
rolling down a quadratic potential, motivated
mainly by its simplicity, has passed all observational
tests. The future of cosmology relies on the extension of experimental
tests and predictions for new cosmological parameters, mostly beyond
linear order. This is crucial if we want to achieve a better
understanding of the physics dominating the early universe. 

This is enough motivation to study the nonlinear regime of
cosmological inhomogeneities. Among the observable effects, the
non-Gaussianity of perturbations has been widely studied in
inflationary models. Non-Gaussianity is an important observational
test as it might eliminate models of inflation even for a null
detection. Our goal in this chapter is to compute the nonlinear
correlations of a general isocurvature field which is valid for all
models.

We first introduce the theory of perturbations and then focus on  the
situation in which the curvature perturbation is generated by the
quantum fluctuations of an isocurvature scalar field. The isocurvature
or entropy perturbations are transformed into curvature
inhomogeneities at the end of a period of inflation or shortly after
it. We will show that only the presence of entropy fluctuations can
affect the evolution of curvature fluctuations   on superhorizon
scales.  

At linear order, we will show under which conditions the observed
power spectrum of curvature fluctuations can be attributed to the
action of the isocurvature field. Subsequently we present a method of
deriving such correlations from the solutions to the Klein-Gordon
equation of the isocurvature field. For specific cases we are able to
derive an explicit expression for the nonlinear parameter $\fnl$. The
prospects of observationally testing the predictions for the  models
of structure formation  presented here are also briefly discussed.

The introductory sections of this chapter present a review of the
elements of the standard inflationary scenario, including the linear
perturbation theory. We present the relevant definitions and
conventions to be used, with particular attention to those results of
linear perturbation theory which will be used in this and subsequent
chapters. From Section \ref{non-gaussianity} onwards, we focus on the
description of the non-Gaussian correlators of an auxiliary
isocurvature  field $\chi$. The expressions for the curvature
perturbation three-point correlators and the $\fnl$ values are
presented in the last section of this chapter.  

\section{Linear perturbations}

 In cosmological perturbation theory,  the universe is described to a
 lowest order by a homogeneous, isotropic background
 spacetime. The large-scale inhomogeneities and anisotropies observed
 in the real universe result from the growth of density fluctuations,
 the amplitudes of which are small in the early stages of the
 universe. (See \cite{peebles80} for a textbook description of the
 development of perturbation theory.)   
 
 In the framework of perturbation theory, the homogeneous background
 spacetime is accounted for by an ansatz metric. The most useful
 ansatz in this case is the Friedmann-Robertson-Walker (FRW) metric: 
\begin{equation}
  g_{\mu\nu} = a^2(\eta) \left( \begin{array}{cc}
    -1 & 0\\
    0 & \gamma_{ij}
  \end{array}
  \right),
\label{FRW:metric}
\end{equation}

\noindent where the conformal time $\eta$ is given in differential
form by
\beq
   d\eta = \frac{dt}{a(t)},
\eeq

\noindent $a$ is the scale factor and $\gamma_{ij}$ is the metric of
the three-dimensional space. In our notation Greek indices have values
$0,1,2,3$, while Latin ones have values $1,2,3$. We assume throughout
a  flat space, relying on the observational limit
$|\Omega_\kappa|<10^{-2}$ \citep{wmap2008-params}. The FRW metric
describes the isotropic space-time expanding at a uniform rate. The
expansion rate is conventionally characterised by the Hubble parameter  
\begin{align}
  {H} = \dtotal{ \ln{a}}{ t} =
  \frac{1}{a}\dtotal{\ln{a}}{\eta} = \frac{1}{a} \hcon,
\end{align}

\noindent where $H$ is defined with respect to coordinate time $t$ and
$\hcon$ with respect to conformal time $\eta$.  

\subsection{Metric perturbations}

\noindent In perturbation theory, observed anisotropies and
inhomogeneities are considered as departures from the metric
\eqref{FRW:metric}. For a perturbed metric, the metric tensor can be
split as 
\beq
g_{\mu\nu} =g_{\mu \nu}^{(0)} + \delta g_{\mu\nu},
\eeq

\noindent where $g_{\mu\nu}^{(0)}$ is the homogeneous FRW background and
$\delta g_{\mu\nu}$ encodes the perturbed quantities. First order
scalar perturbations of the metric are expressed in terms of the
functions $\varphi$, $B$, $\psi$ and $E$, which are defined by 
\begin{align*}
\delta g_{00}^{\rm (s)} =& - 2a^2\varphi(\eta,\bx),\\
\delta g_{0i}^{\rm(s)} =& a^2 \,B(\eta,\bx)_{,i},\\
\delta g_{ij}^{\rm (s)} =&  2a^2 \big(\psi(\eta,\bx)\, \gamma_{ij} +
E_{,ij}(\eta,\bx)\big), 
\end{align*}

\noindent where the index $({\rm{s}})$ denotes scalar modes. The vector constructed from the scalar $B$ is necessarily curl-free, i.e. $B_{,[ij]} = 0$. The pure vector contributions to the metric perturbations are
\begin{align}
  \delta g_{0i}^{\rm(v)} = &\,-a^2 S_i, \qquad
  \delta g_{ij}^{\rm(v)} = \,2a^2 F_{(i,j)},\notag
\end{align}

\noindent where we demand $S_{[i,j]} \neq 0$. The symmetric derivative 
of the function $F_{i}$ is the vector contribution to $g_{ij}$. To
distinguish scalar and vector contributions, the
vector part is forced to be divergence-free, i.e., $\gamma^{ij}S_{i,j} = 0$.
(The decomposition of a vector field into curl- and divergence-free
parts is formally known as Helmholtz's theorem.) The tensor
contribution to the perturbation quantities is $\delta~g_{ij}^{\rm
  (t)}=a^2 h_{ij}$. This is constructed as a transverse, traceless
tensor, which guarantees that it cannot be constructed from scalar or
vector perturbations.

The perturbation functions  $\varphi$, $B $, $\psi$ and $E$, represent
four degrees of freedom. The
divergenceless vectors $S_i$ and $F_j$ each have two degrees of
freedom and the transverse traceless tensor $h_{ij}$ has two
more.  We therefore have $10$ degrees of freedom in total.
The contravariant metric tensor of the perturbed metric is 
constructed, to first order, from the condition,
$g_{\mu\alpha}g^{\mu\beta}=\delta_{\alpha}^{\beta}$. Finally, the line
element of the metric is   
\begin{eqnarray}
\nonumber
ds^2 = a^2(\eta)\big\{-(1 + 2\varphi) d\eta^2 + 2(B_{,i} - S_{i})d\eta\,
dx^i \qquad \qquad  \qquad \\ 
\qquad \qquad  \qquad +[(1 + 2\psi)\gamma_{ij} + 2E_{ij} + 2F_{i,j} +
  h_{ij}]\,   
dx^i\,dx^j \big\}.
\label{perturbed:element}
\end{eqnarray}

In the present work we will study the nonlinear perturbations as the
quantum fluctuations of scalar matter fields. We will establish the
correspondence between scalar matter fluctuations and scalar
perturbations in the metric at first and second order in perturbation
hierarchy. We will then derive statistical parameters of
nonlinearity. 

In contrast to the scalar metric fluctuations, the vector and tensor
perturbations in the metric are not sourced by scalar matter
perturbations at first order. In the standard picture, they are only
related at second or higher order in perturbation theory (see 
e.g. \cite{ananda-vectors}), therefore their contribution to the
statistical parameters of nonlinearity are sub-dominant and henceforth
we neglect their contributions to the perturbations in the metric.

\subsection{Gauge {freedom}}

 In general relativity, the mathematical relations between physical
 quantities are manifestly independent of the coordinate
 choice. However, there is no covariant way of splitting background
 and perturbed variables. There is always an unphysical coordinate or
 gauge dependence associated with perturbed spacetimes. This issue of
 gauge ambiguity was disregarded in the initial works of
 perturbation theory \citep{lifshitz46,lifshitz-khalatnikov}. This
 could lead to erroneous results which were eventually resolved in a
 systematic way by \cite{bardeen80}. The importance of determining the
 gauge changes  that equations and perturbations undergo leads us to
 look at this problem in detail. In the following we adopt a
 `{passive}' approach to gauge transformations (For a recent review of
 these results, see \cite{malik-wands08}). Let us consider the general
 coordinate transformation, 
 \bea
 \tilde{\eta} = \eta + \xi^0, \qquad \tilde{x}^i = x^i+
 \xi_{,}^{~i}+ \bar{\xi}^i,
 \label{gauge:transf}
 \eea

 \noindent where $\xi^0 = \xi^0(\eta,x^i)$ is a scalar that determines
 the choice of constant-$\tilde{\eta}$ hypersurfaces. The scalar $\xi$
 and the divergence-free vector $\bar{\xi}^i$ are also functions of
 the original coordinates within these hypersurfaces. 

 The principle of relativity states that any physically meaningful
 measurement must be invariant for all observers, in particular, for
 observers with different coordinate systems. One
 of these invariants is the line element $ds^2$, where coordinates
 enter via differentials. Such differentials and the scale factors in
 both coordinate systems are related in the following way: 
 \begin{align}
 d\eta  =\,& d \tilde{\eta} - \xi^{0\prime} \, d \tilde{\eta} - \xi^0_{~,i}
 \, d\tilde{x}^i,\notag \\
 dx^i =\,&  d\tilde{x}^i -  \left(\xi^{\prime~i}_{~,} +
 \bar{\xi}^{\prime i}\right) \,d\tilde{\eta} - \left(\xi_{,~j}^{~i} +
 \bar{\xi}^i_{~,j} \right),\label{coord:transform}\\
 a(\eta) =\,& a(\tilde{\eta}) - \xi^0 a^{\prime}(\tilde{\eta}).\notag
 \end{align}

\noindent Where a $^\prime$ is the derivative with respect to conformal
time. To first order in the metric perturbations and coordinate
transformations, the perturbed line element,
Eq.~\eqref{perturbed:element} is written in the `{shifted}' coordinate
system as 
\begin{eqnarray}
  \nonumber
  ds^2 = a^2(\tilde{\eta})\big\{-\big(1 + 2 (\varphi - \hcon\xi^0 -
  \xi^{0\prime})\big)\,d\tilde{\eta}^2 + 2\big[(B + \xi^0 -
  \xi^{\prime})_{,i} - S_i + \bar{\xi}^{\prime}_{i}\big]
  \,d\tilde{\eta}\, d\tilde{x}^i  
  \\
  + \big[\left( 1 + 2(\psi -
   \hcon\xi^0)\right)\gamma_{ij} + 2\left(E - \xi\right)_{,ij} +2
   F_{i,j} - 2 \bar{\xi}_{i,j} + h_{ij}\big]  d\tilde{x}^j
 d\tilde{x}^i \big\},\qquad 
\end{eqnarray}

 \noindent where, as before, $\hcon \equiv a^{\prime}/a$ is the
 Hubble parameter in terms of conformal time. This metric can also be written
 using the initial definitions in terms of the  `{shifted}' coordinates: 
 \begin{eqnarray}
 ds^2 = a^2(\tilde{\eta})\big\{-(1 + 2\tilde{\varphi}) d\tilde{\eta}^2 +
 2(\tilde{B}_{,i} - \tilde{S}_{i})d\tilde{\eta}\,d\tilde{x}^i 
 \\ 
 + [(1 + 2\tilde{\psi}) \tilde{\gamma}_{ij} + 2\tilde{E}_{ij} +
 2\tilde{F}_{i,j} + \tilde{h}_{ij}]\, d\tilde{x}^i\,d\tilde{x}^j\big\}.\qquad
 \label{shifted:ds}
 \end{eqnarray}
 
\noindent This shows that the coordinate transformation
Eq.~\eqref{coord:transform} induces a transformation of the metric
perturbations. Comparing Eqs. \eqref{perturbed:element} and
\eqref{shifted:ds}, the change is given to first order by 
 \begin{align}
   \tilde{\psi} =& \,\psi - \hcon \xi^0, \label{transf:psi}\\
   \tilde{\varphi} =&\, \varphi - \hcon \xi^0
   -\xi^{0\prime},\label{transf:varfi}\\ 
   \tilde{B}=&\, B + \xi^0 - \xi^{\prime},\label{transf:b}\\
   \tilde{E} = &\, E -\xi.  
   \label{transf:e}
 \end{align}
 
 \noindent It must be stressed that the gauge transformations are, in effect, a change of the correspondence between the perturbed spacetime and the unperturbed background spacetime. 

A first exercise concerning gauge transformations is to find those quantities which remain invariant after a gauge transformation. To first order in perturbation variables, 
gauge-invariant quantities are linear combinations of the gauge-dependent quantities presented above. For scalar perturbations \cite{bardeen80} shows that only two independent gauge-invariant quantities can be configured purely from the metric perturbations:
 \begin{align}
 	\Phi_B = &\varphi + \hcon (B - E^{\prime}) + (B - E^{\prime})^{\prime},
	\label{bardeen:Phi} \\
	\Psi_B = &-\psi - \hcon (B - E^{\prime}). 
	\label{bardeen:Psi}
 \end{align}
 
\noindent Any other gauge invariants in the metric are linear combinations of these two quantities because the gauge freedom allows only two arbitrary scalar functions $\xi^0$ and $\xi$ \citep{malik00}. The Bardeen invariants will be useful in relating curvature perturbations in different gauges, as we will show below.
 
\subsection{Perturbations of the matter sector}
 
  Before displaying conservation equations for the curvature
  perturbations, we will discuss the perturbations of the matter
  sector. For a perfect fluid, that is, a fluid with no heat conduction or
  viscosity, the stress-energy tensor is 
  \beq
  T^\mu_{~\nu} = (p + \rho)u^\mu u_{\nu} + p \delta^{\mu}_{\nu} \,,
  \label{perfect:fluid}
  \eeq
  
  \noindent where the 4-velocity is defined with respect to proper time $\tau$ as 
  \beq
  	u^{\mu} =\, \dtotal{x^{\mu}}{\tau}
  \eeq

\noindent and is  subject to the normalisation $u^{\mu}u_{\mu} = -1$. 
Anisotropic stresses would be encoded in a stress tensor $\Pi_{\mu\nu}$, but are absent for perfect fluids and for scalar fields minimally coupled to gravity. These are precisely the kinds of matter considered here, so we ignore the tensor  $\Pi_{\mu\nu}$  in the subsequent analyses. 
  
 Using the normalisation $u^\mu u_{\mu} = -1$, the perturbed velocity
 has components 
\begin{align}
	u^{0} = \frac{1}{a}(1 - \varphi), \quad u^{i} = \frac{1}{a}( v^{,i} + v^{i}) , \\
	u_{0} = -a (1 + \varphi), \quad u_{i} =-a( v_{i} +  v_{,i} + B_{,i} - S_i),
\end{align}
  
\noindent  where the spatial parts are written in terms of the gradient of a scalar $v_{,i}$ and a (solenoidal) vector $v_{i}$.
  The perturbed energy-momentum tensor is: 
  \begin{align}
    T^0_0 =&\, - \left(\rho_0 + \delta\rho\right), \\
    T^0_i =&\, (\rho_0 + p_0) (B_{,i} + v_{,i} + v_{i} - S_{i}), \quad
     T^i_0 =\, -(\rho_0 + p_0) (v^{,i} + v^{i}),\\
     T^i_{~j} =&\,(p_0 + \delta)\delta^i_{~j} ,
  \end{align}

  \noindent where $p_0$ and $\rho_0$ represent the uniform
  pressure and matter density. In general our scalar stress-energy components
  can be written as $f(\eta,x^i) = f_{0}(\eta) +  \delta
  f(\eta,x^i)$, with the subscript $0$ denoting the background
  homogeneous part. As in the case of metric perturbations,
  coordinate transformations will affect the matter
  perturbations. This means that the matter density, velocity and
  pressure perturbations are gauge dependent. Under the
  transformation Eq.~\eqref{gauge:transf}, perturbed scalar functions
  of the form are thus transformed as 
  \beq
  \widetilde{\delta f} = \delta f - f^{\prime}_{0}\xi^0.
  \eeq

  \noindent The vector perturbations are derived either from a potential, which
  will transform with the shift $\xi^{\prime}$, or from
  a pure divergence-free vector, whose transformation depends on
  $\xi^{i}$. In particular, the velocity potential $v$ 
  transforms as:
  \beq
  \tilde{v} =  v + \xi^{\prime}, 
  \label{transf:vel} 
  \eeq
 
  \noindent and the vector function $v^i$ is transformed as
  \beq
  \tilde{v}^{i} =  v^{i} + \bar{\xi}^{i \prime}.
  \label{transf:vectvel} 
  \eeq

\subsection{Physical quantities and scales}

Before addressing the characteristics and governing equations of the
perturbed spacetime, let us define the physical scales and the
quantities that determine of the size and age of the universe. 

The time-like 4-vector field
\beq
N_{\mu} = -a(1 + \varphi)\delta^0_{\mu}, \label{covariant:n}
\eeq 

\noindent defines the direction perpendicular to the hypersurfaces of
constant time. In consequence, this vector field defines a coordinate
system. This vector is unitary ($N^{\nu}N_{\nu}=-1$) and the
contravariant vector $N^{\nu}= g^{\nu\mu}N_{\mu}$ has components  
\beq
N^{0} = \frac{1}{a} \left(1 - \varphi \right),   \qquad	
N^{i} = \frac{1}{a} \left( S^{i}  - B_{,}^{~i}\right).
\eeq

\noindent The expansion rate of the spatial hypersurfaces with respect
to the proper time of observers with 4-velocity $N^{\mu}$ is $\theta =
N^{\mu}_{~~;\mu}$.  Considering only scalar perturbations, this is given
by 
\beq
	\theta  = 3\frac{a^{\prime}}{a^2}(1 - \varphi) + 3\frac{1}{a}
	\psi^{\prime} - \frac{1}{a} 
\nabla^2\left( B - E^{\prime}\right), \label{confrormal:theta}
\eeq

\noindent where the operator $\nabla$ denotes the usual
three-dimensional gradient.  

By looking at the relation between proper and coordinate
time, $d\tau = (1 + \varphi)\,dt $, we extract the expansion with
respect to coordinate time from the above expression by writing    
\beq
	{\theta}_{(\rmt)} = (1 + \varphi)\theta = 3H + 3\dot{\psi} +
	\nabla^2 {\sigma}_{(\rmt)},
	\label{theta:coord}
\eeq

\noindent where the Hubble parameter, $H=\dot{a}/a$ is the
background uniform expansion rate with respect to the coordinate
time. The shear scalar in coordinate time is
${\sigma}_{(\rmt)}=(\dot{E}-B/a)$. 

For the sake of completeness, we include here the definition of some
useful scales. A comoving observer is one moving with the expansion of
the universe, i.e., one who measures zero net momentum density. The
distance of a comoving point  from our location (taken to be at the
origin of coordinates), is given by $r(t) = a(t)x$, where $x$ is the
comoving distance.  
 
The Hubble radius $ r_{H} = H^{-1}$ provides a good estimate for the distance
light has travelled since the Big Bang. Formally, the integral 
\beq
	\eta = \int_{0}^{t} \, \frac{ds}{a(s)},
\eeq 

\noindent defines the comoving distance travelled by a free photon
since $t=0$ and until time $t$. This is important because no
information could have travelled further than $\eta$. This define the
`comoving particle horizon'.  In the above integral $\eta$ can be
taken also as the conformal time. In a matter dominated universe $\eta
\propto a^{1/2}$, while in radiation domination $\eta \propto a$. In a
de Sitter inflationary universe 
\beq
\eta = \int \frac{ds}{a(s)} = \int \frac{da}{H a^2}  = - \frac{H^{-1}}{a(t)}.
\label{eta:desitter}
\eeq

\noindent This shows that in an inflationary phase $\eta \to -\infty$ as the
universe approaches the initial singularity $a = 0$, and increases
monotonically towards $0$. This leads us to consider the magnitude
$|\eta|$ when we use the conformal time in our calculations for
inflation. The maximum distance light travels from time $t = 0$ to us
is simply the comoving horizon times the scale factor,   
\beq
	d_{H} = a(t_{\rm now})\int_{0}^{t_{\rm now}}\,\frac{ds}{a(s)}.
	\label{horizon:def}
\eeq 

\noindent This is called the particle horizon, i.e., the radius of the
region which is in causal contact with us. This equation can be
applied to find the horizon radius at times different from $t_{\rm
  now}$.

\noindent Considering the current dark energy domination and a cold dark matter component (with density parameter $\Omega_{m}$), the horizon size does not coincide exactly with the Hubble scale. However, an approximate solution to the  integral Eq.~\eqref{horizon:def} shows that,
\beq
	d_{H} (t_{\rm now}) \approx \, 2H_{\rm now}^{-1} \frac{1 +
          0.084 \ln{\Omega_{m}}}{\sqrt{\Omega_{m}}} 
        \simeq 3.5 H_{\rm now}^{-1},
	\label{today:hor}
\eeq

\noindent where the last expression assumes  $\Omega_{m} =0.25$  \citep{hu97}. 

\subsection{Particular gauges}

Let us now focus on the expressions for the curvature perturbation on
three useful choices of time slicing and threading. These are the
uniform curvature gauge, the uniform density gauge and the comoving
gauge. 

For the first case the spatial hypersurfaces present an unperturbed
3-metric, which means $\tilde{\psi} = \tilde{E} = 0$, in other words,
curvature perturbations of the three-metric are set to zero. We
distinguish the quantities written in this gauge with a subscript
$\kappa$ indicating a constant curvature. So for a general coordinate
system we require the following transformations: 
\beq
\xi_{\kappa= {\rm const}}^0 =  \frac{\psi}{\hcon},\qquad \quad \xi_{\kappa}=
E.  
\eeq

\noindent In this case, the scalar perturbation becomes 
\beq
{\delta f}_\kappa~=~\delta f - f^{\prime}_0\frac{\psi}{\hcon}.
\eeq

\noindent In particular, for scalar fields, this is the gauge-invariant
Sasaki-Mukhanov variable \citep{sasaki86,mukhanov88}, explicitly,
\beq
\delta\phi_{\kappa} = \delta\phi - \phi_0^{\prime}\frac{\psi}{\hcon}.
\label{sasaki:mukhanov}
\eeq

In the uniform density gauge, the requirement $\tilde{\delta \rho} = 0$
for constant-time hypersurfaces implies
\beq
 \xi_{\delta\rho}^0 = \frac{\delta \rho}{\rho^{\prime}}.
\eeq

\noindent The gauge-invariant curvature perturbation on these hypersurfaces
is denoted by $\zeta$ and defined as
\beq
\zeta \equiv  \tilde{\psi}_{\delta\rho} = \psi - \hcon
\frac{\delta\rho}{\rho}. 
\label{curv:isodensity}
\eeq

\noindent In this case, there is another degree of freedom and one can
pick either $\tilde{B}$, $\tilde{E}$ or $\tilde{v}$ to be zero. The
gauge invariance is made explicit when the curvature perturbation is
written in terms of the Bardeen variables. We will return to this
point once we have defined the curvature in the comoving gauge.  

The comoving gauge is subject to the condition that the spatial
coordinates comove with the fluid, that is, for a constant-time slice
the 3-velocity of the fluid vanishes, $ v_{,}^{~i} = 0$. The
threading is chosen so that the constant-$\eta$ hypersurfaces are
orthogonal to the 4-velocity $u^{\mu}$, which demands $\tilde{v} +
\tilde{B} = 0$. An immediate consequence of this choice of gauge is
that the total 3-momentum vanishes on constant-time hypersurfaces. For
this reason several authors call this gauge the zero-momentum
gauge. Using Eqs. \eqref{transf:vel} and \eqref{transf:b}, the chosen
conditions imply that  
\beq
\xi^0_\rmm = \, - (v + B), \qquad
\xi_{\rmm} = \, - \int \, v\, d\eta + \tilde{\xi} (x^i),
\eeq

\noindent with $\tilde{\xi}(x^i)$ the residual coordinate gauge
freedom. This quantity is not specified at this stage because it is
not required for the determination of the scalar quantities like
curvature, expansion and shear. For arbitrary coordinates, the scalar
perturbations in the comoving orthogonal gauge are given by 
\begin{align}
	\tilde{\varphi}_{\rmm} = \varphi + \frac{1}{a} \left[(v+ B)a
	\right]^{\prime}, \quad \tilde{\psi}_{\rmm} = \psi + \hcon (v + B),\quad
	\tilde{E}_{\rmm} = E + \int\,v\,d\eta - \tilde{\xi}. 
\end{align}

\noindent The scalars $\tilde{\varphi}_{\rmm}$ and
$\tilde{\psi}_{\rmm}$ defined in this way are gauge-invariant.
The density perturbation in the comoving
gauge is also given in gauge-invariant form by
\beq
	\delta\tilde{\rho}_{\rmm} = \delta\rho + \rho^{\prime} (v + B).
\eeq

\noindent Some authors use the gauge-invariant density perturbation in the
comoving gauge by defining the combination $\Delta  \equiv
\widetilde{\delta \rho}_{\rmm,i}^{~~~~~i} / \rho_{0}$
\citep{bardeen80,kodama-sasaki}.    

The curvature perturbation $\psi$ in the comoving gauge was first used
by \cite{lukash80}  and first denoted as $\R$ by
\cite{liddle-lyth-paper}. It is mathematically defined as   
\beq
	\R \equiv \tilde{\psi}_{\rmm} = \psi + \hcon (v + B). 
	\label{def:R}
\eeq

 \noindent In the next section we will find that, through the Einstein
 equations and gauge-invariant quantities $\Psi_B$ and $\Phi_B$,
 defined in Eqs.~\eqref{bardeen:Psi} and \eqref{bardeen:Phi}, one can
 establish an equivalence at large scales between the curvature
 perturbation to linear order in the uniform density gauge and the
 same perturbation defined in the comoving gauge. Taking this
 equivalence for granted, in the meantime, allows us to relate $\R$
 and $\delta\rho$ directly. Indeed, if we consider the gauge
 transformation \eqref{curv:isodensity} from an initial flat
 hypersurface, then 
\beq
	\R = \zeta  = - \hcon \frac{\delta\rho_{\kappa}}{\rho},
	\label{R:delrho}
\eeq

\noindent at scales beyond the cosmological horizon (as will be made
explicit below).  

This last transformation shows the way of avoiding the gauge
anomaly. One can always change the gauge (or frame of reference) and
establish the equivalence between the perturbations of any two
gauges as long as a particular gauge is chosen at the start and all
quantities are initially defined in this gauge.  

\section{Evolution of perturbations and conserved quantities}
\label{gr:dynamics}

Just as the perturbed scalars in the metric are gauge dependent, so are the
evolution equations for these quantities and they must be treated
carefully in order to avoid spurious gauge modes.   

The equations governing the dynamics of space-time are found by varying
the action $\mathcal{I}$ with respect to the metric and matter
components. The action is defined as  
\beq
\mathcal{I} = \int_{-\infty}^{\infty} \mathcal{L}\, \sqrt{-g}\, d^4x,
\eeq

\noindent where $\mathcal{L}$ is the Lagrangian density of the matter
and the gravitational field. The gravitational Lagrangian of general
relativity is, 
\beq
	\mathcal{L} = -R / 2, 
\eeq

\noindent where $R$ is the Ricci scalar.  The Lagrangian
density for a classical matter field minimally coupled to gravity is 
\beq
\mathcal{L} = K - V - R/2,   \label{general:lagrangian}
\eeq

\noindent with $K$ the kinetic energy and $V$ the potential energy. For the matter sector of the Lagrangian we can define the energy-momentum tensor as  
\beq
T_{\mu\nu} = - 2 \dpartial{\mathcal{L}}{g^{\mu\nu}}+ g_{\mu\nu} \mathcal{L}.
\label{lagrangian:em}
\eeq

\noindent In particular, the energy-momentum tensor for a perfect
fluid with density $\rho$ and isotropic pressure $p$ and 4-velocity
$u^\mu$ is given by Eq.~\eqref{perfect:fluid}. 

Let us now look at the Lagrangian
density of a single scalar field $\phi$, minimally coupled to
gravity. Its kinetic energy is $ K = - 1/2
g^{\mu\nu}\partial_{\mu}\phi \partial_{\nu}\phi$, so from
Eq.~\eqref{general:lagrangian} we find  a canonic action 
\beq
\mathcal{L}_M =  -\frac{1}{2} \left[g^{\mu \nu} \partial_{\mu} \phi
\partial_{\nu} \phi + 2V(\phi) \right].
\label{lagrangian:inflaton}
\eeq

\noindent Using the definition Eq.~\eqref{lagrangian:em}, it is
easy to show that the scalar field energy-momentum tensor is
\beq
T_{\mu}^{~\nu} = g^{\alpha\nu}\phi_{,\mu} \phi_{,\alpha} - \delta^{~\nu}_{\mu}
\left(V(\phi) + \frac{1}{2}
g^{\alpha\beta}\phi_{,\alpha}\phi_{,\beta}\right).
\eeq

\noindent The comparison between the last expression with
Eq.~\eqref{perfect:fluid}, $T_{\mu\nu}$ for a perfect fluid, shows
that we can define the density, isotropic pressure and velocity as
\citep{tabensky-taub}  
\begin{align}
  u_\mu = \frac{\phi_{,\mu}}{|g^{\mu\nu}\phi_{,\mu} \phi_{,\nu}|}, \quad\rho =
  - g^{\mu\nu}\phi_{,\mu} \phi_{,\nu} + V, \quad
p = -g^{\mu\nu}\phi_{,\mu} \phi_{,\nu} - V.
\label{scalar:matter}
\end{align}

\noindent This identification provides an easy way to quantify the
energy density of a scalar field and its perturbations.
The Lagrangian density for two real fields is
\beq
\mathcal{L}_M =  -\frac{1}{2}(g^{\mu\nu}\phi_{,\mu} \phi_{,\nu}) -
\frac{1}{2}(g^{\mu\nu}\chi_{,\mu} \chi_{,\nu}) - U(\phi,\chi), 
\label{lagrangian:twofield}
\eeq

\noindent which features  the joint potential $U(\phi,\chi)$ of the
participating fields, each minimally coupled to gravity.  This case is
what concerns us in the rest of the chapter and we shall focus on its
dynamical equations.

\subsection{Background equations}

The Einstein equations  are  found by varying the action
\eqref{general:lagrangian} with respect to the metric. Under regular
conditions, with no variations of the fields at the boundaries, the
equations are found by applying the operator 
\beq
	\left[\frac{\delta}{\delta g} -
          \partial_{\mu}\frac{\delta}{\delta( \partial_{\mu} g)}
          \right] 
\eeq

\noindent to the Lagrangian. 
The Einstein equations dictate the dynamics relating the local
spacetime curvature to the local energy-momentum. In the adopted
natural units,   
\beq
G_{\mu \nu} =  T_{\mu \nu},
\label{def:einstein}
\eeq

\noindent where the left-hand side is the Einstein tensor, defined as
\beq
	G_{\mu\nu} = R_{\mu\nu} - \frac{1}{2} g_{\mu\nu} R.
\eeq
 
\noindent The Einstein equations can be split into components that are
parallel or orthogonal to the time-like  field $N^{\mu}$ at any order
in perturbation expansion.  
The two independent equations obtained at the background level are the
Friedmann and acceleration equations:  
\begin{align}
  \hcon^2 = &\frac{1}{3} a^2 \rho_0,
  \label{first:friedmann}\\
  \hcon^{\prime} = & - \frac{1}{6} a^2 \left(3 p_0 + \rho_0\right).
  \label{second:friedmann}
\end{align}

\noindent Additionally, the Bianchi identities $G^{\mu}_{~\nu;\mu} =
0$ imply the local conservation of energy and momentum, 
\beq
T^{\mu}_{~\nu;\mu} = 0,
\eeq

\noindent where $;$ denotes a covariant derivative with respect to the metric $g_{\mu\nu}$. 

For the background quantities,  the energy-momentum
conservation equations provide an expression for the expansion in terms of the matter fields. For the case of a single fluid in the
background FRW universe, the equation $T^{\nu}_{~~0;\nu} = 0$ gives
\begin{align}
	\rho^{\prime}_0 = - 3\hcon (p_0 + \rho_0). \label{rho:cons}
\end{align}

\noindent Note that the isotropy assumption means there is no net
background momentum and thus no other conservation equation at zeroth
order. Moreover, Eq.~\eqref{rho:cons} can also be obtained as a
combination of the Einstein equations \eqref{first:friedmann} and
\eqref{second:friedmann}.  

The homogeneous Einstein equations can be solved for the variables $a(t),\rho
(t),p(t) $ when an equation of state for the matter components is provided. This is dictated by the microphysics of the matter. In particular, the equation of state
\beq
 p = w \rho,
 \label{eqof:state}
\eeq 

\noindent describes most of the relevant cases of the
post-inflationary cosmology. For the case of pure radiation $w = 1/3
$, while for pressureless dust $w = 0$. For fluids with such an
equation of state, the solutions to the Einstein equations are 
\beq
 \frac{\rho}{\rho_\mathrm{i}} = 
 \left(\frac{a}{a_\mathrm{i}}\right)^{-3(w + 1)},\qquad
\qquad \frac{a}{a_\mathrm{i}} =
\left(\frac{\eta}{\eta_\mathrm{i}}\right)^{2 / (3w + 5)}, 
\eeq

\noindent with initial conditions $\rho = \rho_\mathrm{i}$ and $a = a_\mathrm{i}$ at $ \eta = \eta_\mathrm{i}$.

When more than one fluid is present, we account for the contribution of each component $\rho^{(j)}$ to the total matter density by defining a dimensionless density parameter
\beq
	\Omega_{(j)} = \frac{a^2 \rho^{(j)}}{3\hcon^2},
	\label{def:omega}
\eeq

\noindent where  the factor $3 \hcon^2 / a^2$ is the critical density. In the flat universe that concerns us, the curvature contribution $\Omega_{\kappa}$ is zero and the sum of all matter contributions is unity, i.e., 
\beq
	\sum_j \Omega_{(j)}  = \Omega_{0} = 1.
\label{omega:tot}
\eeq

\noindent  (note that the dark energy that dominates the expansion in
the present stage of our universe should also be included in this
sum. This component is customarily denoted by $\Omega_{\Lambda}$.) With
these definitions Eq.~\eqref{first:friedmann} can be written in the
form 
\beq
	\left(\frac{\mathcal{H}}{\mathcal{H}_\mathrm{i}}\right)^2  = a^2
        \left\{\sum_j \Omega_{(j)}  \left(\frac{a}{a_\mathrm{i}}\right)^{-3(1 +
          w^{(j)})}\right\}. 
\eeq 

\noindent For the case in which the matter is dominated by a single
scalar field $\phi$, energy density conservation leads to the
Klein-Gordon equation, 
\beq
 \phi_{;\mu}^{~~\mu} = \dtotal{V}{\phi},
\label{gen:kg}
\eeq

\noindent which can also be derived from the variation of the action
with respect to $\phi$. The potential $V$ is assumed to be an explicit
function of the field alone. In the case where there is more than one
field, a  Klein-Gordon equation is obtained for every scalar field,
with  an interaction potential  $U$. 
Note that the Klein-Gordon equation is valid at all orders in the
perturbation expansion. We will rely on this important fact to derive
the contribution of nonlinear perturbations to the non-Gaussianity of
the primordial fluctuations, the ultimate objective of this chapter.  

The Klein-Gordon equation for a homogeneous scalar field $\phi_0$ for
a FRW background metric is  
\beq
\ddot{\phi}_0 + 3H\dot{\phi}_0 + \dtotal{V}{\phi} = 0.
\label{hom:kg}
\eeq

\noindent The solutions of this equation will be explored in the
context of inflation in Section~\ref{inflation}.  
 The inflationary behaviour is guaranteed when the dominating scalar
 field meets the so called slow-roll conditions. The dynamics of
 inflation will be discussed in more depth in the following
 sections. In the meantime we note that, as
 mentioned in Section~\ref{intro-inflation}, the perturbations
 produced during a period of inflation exit the  cosmological horizon
 due to the shrinking of the latter scale. In a super-horizon regime,
 under suitable conditions, the curvature inhomogeneities are
 time-invariant. 
 
 \subsection{Dynamics of perturbations}

At linear order, the scalar metric perturbations are
related to matter perturbations via the Einstein equations. The
density and momentum constraints are
\begin{align}
  3\hcon( \hcon \varphi -\psi^{\prime}) + \nabla^2 \left[\psi - \hcon
    \sigma \right] =& - \frac{1}{2} a^2\delta \rho,
  \label{first:friedmann1}\\
  \left[\psi^{\prime} - \hcon \varphi \right]
  =&  \frac{1}{2} a^2\left(\rho_0 + p_0\right)\left[v + B\right],
  \label{second:friedmann1}
\end{align}

\noindent and two evolution equations for the scalar metric perturbations
\begin{align}
	\psi^{\prime\prime} +2\hcon \psi^{\prime} - \hcon \phi^{\prime} - 
	\left(2\hcon^{\prime} + \hcon^2\right)\varphi
	 =& -\frac{1}{2}a^2\delta p ,
	\label{third:friedmann1}\\
	\sigma^{\prime}+ 2\hcon \sigma + \psi - \phi =& 0.
	\label{fourth:friedmann1}
\end{align}

\noindent The energy-momentum conservation equations for the perturbed spacetime
are related to the ones above via the Bianchi identities.
Specifically, the evolution for the energy density perturbation is 
\beq
 \delta\rho^{\prime} + 3\hcon(\delta p + \delta \rho) = -(p_0
 +\rho_0)\left[3\psi^{\prime} + \nabla^2 (v + E')\right],
 \label{delrho:cons} 
\eeq 

\noindent and the momentum conservation equation is
\beq
	\left[(p_0 + \rho_0)(v + B)\right]^{\prime} + \delta p  = -
        (p_0 + \rho_0)[\varphi + 
	  4\hcon(v + B)]. \label{mom:cons} 
\eeq

\noindent Instead of solving the equations at first order, let us show
how the dynamical linear equations encode two important implications
for cosmological perturbations. A convenient way to find conserved
quantities is to work in Fourier space (the Fourier transformation is here
denoted by $\mathcal{F}$.) In this case, a generic
coordinate-dependent perturbation $f(t, \bx)$ is decomposed into harmonic
functions of time: 
\beq
	f(t, \bx) = \frac{1}{(2\pi)^3} \int_0^{\infty} \,
        \exp\left(-\imag \bk \cdot \bx\right)f_{k}(t),
        \quad \text{i.e.,} \quad
        \mathcal{F}\left[f(t,\bx)\right] = f_k(t).    
\eeq

\noindent Each function $f_k(t)$ is referred as a
perturbation mode and labelled by its comoving wavenumber $k$ and has
an associated scale $\lambda_k = a(t) / k$. The only
characteristic scale of the unperturbed universe is the Hubble scale
or cosmological horizon as defined in Eq.~\eqref{horizon:def}.
When perturbation modes lie well outside the cosmological horizon, the
ratio 
\beq
\varepsilon \equiv d_{H} / \lambda_{k} = k / a(t) H(t)  
\label{small:param}
\eeq

\noindent is much smaller than one. Since the spatial derivatives
$\nabla$ are transformed to $\bk / a $, this shows that we
can neglect gradients terms in the equations compared to the time
derivative which scales as  $H$, i.e., for a given function $f(t,x)$
with Fourier transform $f_k(t)$, 
\beq
\left| \mathcal{F}\left[\nabla f(t,x)\right] \right| = 
\left| \frac{\bk}{a} f_k(t) \right|  
  \, \ll \,  H.
\label{grad:expansion}
\eeq

\noindent This simplifies the equations considerably  and shows under
which conditions the curvature perturbations are conserved in the
superhorizon regime.  

 Let us use this approximation to look at the equivalence of the
 curvature perturbation on large scales in different gauges. From the
 Einstein equations we can rewrite the gauge invariant  curvature
 perturbation $\R$ in terms of the curvature perturbation
 variables. Making use of the momentum constraint
 Eq.~\eqref{second:friedmann1}, we note that 
\beq
v + B =  \frac{\hcon\varphi - \psi^{\prime}}{\hcon^{\prime} - \hcon^2}.
\eeq

\noindent We can insert this in the definition of $\R$ in
Eq.~\eqref{def:R} and  write the latter in terms of the Bardeen
gauge-invariant quantities. That is,    
\beq
	\R = - \Psi_B + 
        \frac{\hcon(\hcon\Phi_B + \Psi_B^{\prime})}{\hcon^{\prime} -
          \hcon^2}, 
	\label{R:invariant}
\eeq

\noindent which represents an alternative form of Eq.~\eqref{second:friedmann1}.

We can follow a similar procedure for Eq.~\eqref{first:friedmann1} and
write the uniform density curvature perturbation $\zeta$ in terms of
Bardeen invariants: 
\beq
	\zeta  \left(\frac{\hcon^{\prime} - \hcon^2}{\hcon} \right)=
	\Psi_B^{\prime}+ \hcon\Phi_{B} -
	 \left(\frac{\hcon^{\prime} - \hcon^2}{\hcon}\right)\Psi_{B}
	  - \frac{1}{3\hcon} \nabla^2{\Psi_B}. 
	\label{zeta:invariant}
\eeq

\noindent Note that, because the curvature perturbations  in the last two equations are written in terms of gauge-invariant quantities, $\R$ and $\zeta$ are manifestly gauge-invariant themselves. Moreover, the combination of these equations leads to the gauge-invariant generalisation of the Poisson equation,
\beq
	\nabla^2 \Psi_B  = 3\left(\hcon^{\prime} - \hcon^2\right) 
( \R - \zeta) = 
	\frac{a^2}{2} \delta{\rho}_{\rm m}.
\eeq

\noindent As before, if gradients are discarded, both $\R$ and $\zeta$
coincide. This result is important in view of the consequent
correspondence \eqref{R:delrho}, which is used extensively throughout
this thesis.

A second important feature is the evolution of $\zeta$ on superhorizon scales. 
The energy conservation Eq.~\eqref{delrho:cons} can be written in coordinate time as 
\beq
	\dot{\delta\rho	} + 3H \left(\delta \rho + \delta p\right) = 
   - \left(p_0 + \rho_0\right) \left[3\dot\psi + \nabla^2(\frac{v}{a} +
   \dot{E})\right].
   \label{delrho:const}
\eeq

\noindent This leads to an evolution equation for the perturbed
energy density when we include Eq.~\eqref{rho:cons}:
\beq
 \left[\rho_0 +\delta\rho\right]^{\cdot} + 3\left(H +
 \dot{\psi}\right)\left[p_0 + \delta p + \rho_0 + \delta \rho \right] =
 -\nabla\left(\frac{v}{a} + \dot{E}\right) + \mathcal{O}(\delta^2),
 \label{rho_delrho:const}
\eeq

\noindent which in view of the definition of the expansion $\theta_{(\rmt)}$, 
Eq.~\eqref{theta:coord}, gives to first order
\beq
\dot{{\rho}}  + {\theta}_{(\rmt)} \left[{p}+
  {\rho}\right] = \nabla^2\left[\frac{v}{a} + \dot{E}\right] +
\nabla^2\left(\sigma_{(\rmt)}\right)\left[{p_0} +
  {\rho_0}\right] + \mathcal{O}(\delta^2).
\label{rhotot:const}
\eeq
 
\noindent We emphasise that, in our notation, the density and
pressure with no subscript represent the sum of the background
function and its perturbation, i.e., $\rho =  \rho_0 + \delta \rho$. The
$\mathcal{O}(\delta^2)$ term indicates that the above equation is valid to
first order in perturbation expansion.

Since we are discarding
spatial gradients in the equations of motion on super-horizon scales, Eq.~\eqref{rho_delrho:const} provides an evolution equation for the curvature perturbation:  
\beq	
\dot{\psi}\left[p_0 + \rho_0 \right] = - \frac{\dot{\rho}}{3}
- H \left(p + \rho\right),
\eeq

\noindent or
\beq
\dot{\psi} = \,- \frac{1}{3}\frac{\dot{\delta \rho}}{p_0 + \rho_0}  + 
\frac{\dot{\rho}_0}{3}\left(\frac{\delta p  + \delta\rho}{(p_0
  + \rho_0)^2} \right). 
  \label{psi:evol}
\eeq
 
\noindent If we work in a gauge where the time slices are
uniform-density hypersurfaces, i.e.,  the uniform-density gauge, we may set  
\begin{align*}
\delta\rho  \rightarrow \, 0,\quad
\delta p   \rightarrow \, \delta p_{\delta{\rho}},\quad
\psi \rightarrow \, \psi_{\delta\rho} \,\equiv\, \zeta.
\end{align*}

\noindent All perturbations, including the pressure perturbation in
this gauge are independent of the density perturbation. From
thermodynamics we know that in fluids where entropy is constant the
pressure is a function of the density. In this category fall the 
barotropic fluids, defined as those fluids in which the pressure is
only a function of the density $\rho$ and vice-versa. In general the
pressure of a thermodynamic system (in our case the universe) is a
function of both the density and the entropy,      
\beq
\delta p = c_s^2 \delta \rho|_{s} + \delta p|_{\rho}, 
\eeq

\noindent where $c_s^2 = \partial p/ \partial{\rho}|_{s}$ is the
adiabatic sound speed in the system (see e.g. \cite{balloon} for a
careful treatment of thermodynamics of fluids in cosmology). If one defines
the entropy perturbation $\delta s$ from the identity $\delta
p|_{\rho} = \dot{p}\delta s$, then one has
\beq
	\delta s = \frac{\delta p}{\dot{p}} - \frac{\delta \rho}{\dot{\rho}}.
\eeq

\noindent In view of this, Eq.~\eqref{psi:evol} for the curvature
 perturbation reduces to 
 \beq
 \dot{\zeta}  = \,\frac{\dot{\rho}_0}{3}\left(\frac{\delta p_{\delta\rho}}{(p_0 + \rho_0)^2} 
\right) = \,-\frac{H \dot{p}_0}{p_0 + \rho_0}\delta s.  
 \label{zeta:evol}
\eeq

\noindent This result is important  in the description of the
evolution of perturbations after inflation. It indicates that for
inhomogeneities with characteristic scales much larger than the size 
of the cosmological horizon, the curvature can only be
modified if the matter content of the universe has  a
non-adiabatic or entropy component \citep{wands-malik}. It is also
remarkable that this argument requires the conservation of the
energy-momentum tensor, and not necessarily the Einstein equations.  This
means that the described property is valid for any theory of gravity
in which energy is conserved.   

This result has been extended beyond linear order in the perturbation
expansion. On superhorizon scales, the expression at second order is
\citep{malik-wands03}, 
\beq
	\dot{\zeta}_2 = \frac{\dot{\rho}_0}{3 (p_0 + \rho_0)^2} 
	\delta p_2|_{\rho} - 
	\left[\frac{2}{(p_0 + \rho_0)} \delta p_1|_\rho - 
	  2 \left(p_0 + \rho_0 \right)\zeta_1  
	\right] \dot{\zeta}_1, 
	\label{zeta2:evol}
\eeq

\noindent where numerical indices indicate the order of each
quantity in the perturbation expansion. This result shows that, as in
the case of the linear $\zeta$, the evolution at second order depends
only on the entropy perturbation and its derivatives.  This has also
been proved to all orders by \cite{lyth-zeta}. This result generalises
the special cases of a constant equation of state, i.e., $p / \rho  =
{\rm const.}$ and the single field inflationary case, for which the
conservation of $\zeta$ had been previously been verified
\citep{shibata-sasaki,salopek-bond}.   

 Eqs.~\eqref{zeta:evol} and~\eqref{zeta2:evol} have motivated several
 studies searching for significant growth of $\zeta$ on superhorizon
 scales during and after inflation. In particular, theories of
 multi-field inflation have been proposed to generate the the curvature
 perturbation and, at the same time, an observable signature of
 non-Gaussianity  \citep{mollerach,lyth-curvatonI,enqvist-curvaton}.    
 We will now study the effects of considering an auxiliary field, with
 special attention to those models where a large non-Gaussian
 contribution arises. We start with a short description of inflation
 in the next section.

\section{Inflation}
\label{inflation}

One can define cosmological inflation as the epoch
when the scale factor of the universe is accelerating\footnote[1]{This
  does not include the late time acceleration at the current epoch
  which is attributed to dark energy. Whereas inflation-like scalar
  fields may be responsible for such behaviour (see, e.g.,
  \cite{quintessence}), in this thesis we are not concerned with the
  dynamics of the late universe}:  
\beq
	\ddot{a} > 0.
\eeq

\noindent This condition can be written in terms of a more physical
quantity. The period of inflation can be considered as an epoch in
which the comoving Hubble horizon decreases with time: 
\beq
	\frac{d}{dt}\left(  \hcon^{-1}\right) < 0.
	\label{hubble:shrink}
\eeq

\noindent Within general relativity, the above conditions on the time
dependence of the scale factor give conditions on the
matter content through the Einstein equations. In
particular, Eq.~\eqref{second:friedmann} can be used to write the
condition \eqref{hubble:shrink} as
\beq
	3 p_0 + \rho_0 < 0. 
	\label{inflation:condition}	
\eeq

\noindent Demanding  a positive energy density $T^{0}_{0}= \rho_0$ is a sensible
physical condition. The condition $T^{00} > 0 $ is known as the weak
energy condition. In view of this, the above equation demands that the
dominant matter must have negative pressure during a period of
inflation. The simplest matter field with this property is a scalar
field, which is composed of spin-0 particles. The concept of a scalar
field is prevalent in particle physics where scalars such as the Higgs
scalar are essential in the construction of the standard
model. Although no scalar particle has so far been observed,  they 
play a fundamental role in cosmology, as they possess the unusual
feature of their potential energy dominating over their kinetic
energy.  

As indicated by Eq.~\eqref{scalar:matter}, the Lagrangian definition of
the energy-momentum tensor requires that the energy density and
pressure for a homogeneous scalar field be 
\beq
\rho_0 = \frac{1}{2} \dot{\phi}^2  + V(\phi), \qquad p_0 = \frac{1}{2}
\dot{\phi}^2 - V(\phi). 
\eeq  

\noindent This shows that in order to meet condition
\eqref{inflation:condition} we require the potential $V(\phi)$ to
dominate over the `{kinetic}' term. This can be dynamically achieved
with a sufficiently flat potential provided the field is displaced
away from its minimum. Such physical conditions are controlled by two
parameters: 
\beq
 \epsilonsr \equiv \,\frac{\planck^2}{2} \left(\frac{{V}^{\prime}}{V}\right)^2,
	\qquad 
	\etasr \equiv\, \planck \frac{{V}^{\prime\prime}}{V}.
	\label{sr:def}
\eeq

\noindent These are called the slow-roll parameters and, during
inflation, they are subject to the slow-roll and friction-dominated
conditions 
\beq
	\epsilonsr \ll \,1,\qquad \etasr \ll\,1.
	\label{sr:cond} 
\eeq

\noindent The first condition, the `slow-roll' condition, ensures the
          {slow rolling} of the field down its potential. The second, a
          `friction-domination' condition constrains the potential to
          be very flat one for the period of inflation. This condition
          is imposed to allow for an extended period of inflation
          (which should last for over 60 e-folds of expansion),
          required to recover a sufficiently flat and homogeneous
          universe in the observed scales.  
          When both slow-roll parameters are much smaller than one, the
          dynamics of the single field $\phi$ guarantees an
          accelerated expansion with a shrinking comoving Hubble horizon.  
 
 The scalar field satisfying these properties is called the inflaton. The
 equations dictating  the background dynamics of a universe dominated
 by the inflaton are 
\begin{align}
	H^2 = &\, \frac{V(\phi)}{3}, 
	\label{sr:fried1}\\
	3H\dot{\phi} = &\, - V^{\prime}(\phi).
	\label{sr:fried2}
\end{align}

\noindent The homogeneous Klein-Gordon equation \eqref{hom:kg},
reduces to Eq.~\eqref{sr:fried2} in the slow-roll regime.
The exact solution to these equations requires the specification of
the potential as a function of the homogeneous scalar field
$\phi(t)$. However, the first equation already shows that, if we
assume a constant $V$, 
\beq
	{a(t)}\propto\, \exp(\sqrt{V/3}t). 
\eeq

\noindent  This illustrates explicitly the exponential growth of the
scale factor in the inflationary regime. Additionally, the field
depends only linearly on time to lowest order, as expected for a slow
rolling field. An expansion in powers of the slow-roll parameters
shows that any deviation from this behaviour should be orders of
magnitude smaller than the form expressed here \citep{stewart-lyth}.
In the following we focus on the study of the tiny inhomogeneities
produced by quantum fluctuations of the inflationary field. This
aspect is crucial in  understanding the origin of the observed
structure in the universe. 

\subsection{Inflationary field power spectrum}

The mean amplitude of matter or curvature perturbations is inferred
from their power spectrum. This is constructed through the
quantization of real perturbation fields, as prescribed by quantum
field theory \citep{birrell-davies}. Following a perturbative
expansion, we split the scalar field as
\beq
	\phi(t,\bx) = \phi_0(t) + \delta \phi(t,\bx).
\eeq
 
 \noindent Such an expansion separates the general Klein-Gordon
 equation \eqref{gen:kg} into its homogeneous part \eqref{hom:kg} and
 the perturbation equation 
 \beq
\ddot{\delta \phi}+ 3 H \dot{\delta \phi}  - \frac{\nabla^2 \delta
  \phi}{a^2} + m_{\delta \phi}^2 \delta \phi  =0,
\label{kg:deltaphi}
\eeq

\noindent where the effective mass of the field fluctuation is defined as
\beq
m_{\delta \phi}^2 = m_\phi^2 + {V}_{NL}^{\prime\prime},
\eeq

\noindent and  ${V}_{NL} = V - m_{\phi}^2 \phi^2/2$  represents the
nonlinear part of the potential. Note that here we have neglected the
perturbations of the metric which enter the Klein-Gordon equation
through the operator $_{;\nu}^{~~~\nu}$. In this case, 
Eq.~\eqref{kg:deltaphi} is a valid approximation because we consider
$\phi$ to be subject to the slow-roll conditions.     

In order to compute the power spectrum of the field perturbations,
we need to consider the free field $\delta \phi_I$, or the field
in the so-called {\em interaction picture} of quantum field theory 
\citep{peskin}. This field is the solution to the Klein-Gordon
equation \eqref{kg:deltaphi} in the absence of the nonlinear term, i.e. 
\beq
\ddot{\delta \phi}_I+ 3 H \dot{\delta \phi}_I  - \frac{\nabla^2
\delta  \phi_I}{a^2}
+ m_{\phi}^2 \delta \phi_I =0. 
\label{kg:free}
\eeq

\noindent In the quantum framework we Fourier decompose this field as
\beq
\delta \phi_I (\bx,t) = \int \frac{d k^3 }{(2 \pi)^3} \left(
e^{\imag\bk \cdot \bx} a_k \delta \phi_{I \bk} (t)+
e^{- \imag \bk \cdot \bx} a^\dagger_k \delta \phi^*_{I \bk} (t) \right),
\label{fourier:quantum}
\eeq

\noindent where $*$ denotes complex conjugation and $\dagger$ the
hermitian adjoint operator. $a_k$ and $a^\dagger_k$ are operators
satisfying the usual canonical commutation relations,
\beq
[a_k,a_{p}^\dagger] = \delta^{(3)} (\bk - \bp), \qquad [a_k,a_{p}] =
\ [a_k^\dagger,a_{p}^\dagger]= 0,
\eeq

\noindent and $\delta^{(3)}(\bk)$ is the three-dimensional Dirac delta
function. Thus the field fluctuations in Fourier space are solutions
of the linear equation
\beq
\ddot{\delta \phi}_{I \bk}+ 3 H \dot{\delta \phi}_{I \bk}  +
\left(  \frac{k^2 }{a^2} + m_{ \phi}^2 \right) \delta \phi_{I \bk} =0.
\label{kg:fourier}
\eeq

\noindent For simplicity, we only consider de Sitter inflation, where
$H$ is constant, and the scale factor is  $a=-1/(H\eta)$. As mentioned
above, this is a good approximation in slow-roll inflation. In terms
of conformal time, the previous equation can be written as 
\beq
{\delta \phi}_{I \bk}^{\prime\prime}+ 3 \eta^{-1}{\delta \phi}^{\prime}_{I \bk}  +
\left[ {k^2 } + \eta^{-2} \left(\frac{m_{ \phi}}{H}\right)^2 \right]
\delta \phi_{I \bk} =0. 
\label{conformalkg:fourier}
\eeq

\noindent The solution to this equation involves the set of Bessel
complex functions. After proper normalisation, i.e., taking the
Bunch-Davies vacuum for the de Sitter spacetime \citep{bunch-davies}, 
one finds
\beq
\delta \phi_{I \bk}(\eta)= \frac{\sqrt{\pi}}{2a} \sqrt{-\eta}
\mathrm{H}_\nu^{(1)} (|k \eta|), \label{fourier:phi_i}
\eeq

\noindent where $\mathrm{H}^{(1)}_\nu$ is the Hankel function of the
first kind and of order 
\beq
\nu = \sqrt{\frac{9}{4}-\frac{m_{ \phi}^2}{H_*^2}}\,.
\eeq

\noindent The star indicates that we are evaluating $H$ just after
Hubble horizon exit\footnote[2]{In the perturbed KG equation the mass
  term is negligible because of the slow-roll condition
  $V^{\prime\prime} / V \ll V$ , which is equivalent to $ m_{\phi} \ll
  H$. Well before horizon exit the harmonic flat spacetime equation is
  recovered in the solution Eq.~\eqref{fourier:phi_i}. On the other
  hand,  there is no need to compute the correlation at times well
  after horizon exit. The relation \eqref{R:delrho} and the fact that
  $\R$ is conserved well outside the horizon indicate that the
  required field power spectrum can be evaluated a few Hubble times
  after horizon crossing.}, i.e., when $k = k_* = |1/\eta_* | \lsim aH
$. 

The two-point function of the field fluctuations reads
\beq
\langle \delta \phi_{I \bk_1} (\eta) \delta \phi_{I \bk_2} (\eta)
\rangle =  (2 \pi)^3 \delta^{(3)}\left( \bk_1 + \bk_2 \right)
\frac{\pi H_*^2 }{4 k_1^3} (|k_1 \eta|)^3 \left|\mathrm{H}_\nu^{(1)}
(|k_1 \eta|)\right|^2. 
\label{two-point:phi} 
\eeq

\noindent where the angled brackets indicate the expectation value, in
this case, of two modes of the perturbation field, evaluated at time
$\eta$. In terms of  the two-point function, the power spectrum
$P_\phi$ is defined as  
\beq
\langle \delta \phi_{I \bk_1} (\eta) \delta \phi_{I \bk_2} (\eta)
\rangle =  (2 \pi)^3 \delta^{(3)} \left( \bk_1 + \bk_2 \right) P_\phi(\eta, k_1)
\label{two-point:power},
\eeq

\noindent and the dimensionless power spectrum in this case is
\beq
\P_\phi(\eta,k)=\frac{k^3}{2 \pi^2} P_\phi (\eta,k)  = \frac{H^2}{8\pi} \left(| k\eta|\right)^3
\left| {\rm H}_{\nu}^{(1)}(|k\eta|)\right|^2.
\label{ps:def}
\eeq

By expanding the Hankel function in Eq.~\eqref{two-point:phi} on large scales, i.e. for  $|k \eta| \ll 1$, one obtains 
\beq \P_\phi
( \eta,k) = 2^{-2\omega} \frac{\Gamma\left(\frac{3}{2} -\omega
\right)^2}{\pi^3} {H_*^2 } \left( |k \eta| \right)^{2 \omega}
\simeq \left(
 1+ {\cal O} (\omega) \right) \left( \frac{H_*}{2
\pi} \right)^2 \left( |k \eta| \right)^{2 \omega} ,
\label{phi:ps}
\eeq
where
\beq
\omega \equiv \frac{3}{2}-\nu \sim \frac{m_{\phi}^2}{3H_*^2}~.
\label{alpha}
\eeq

\noindent  This is negligible in the massless case which corresponds
to a de Sitter inflationary phase.  Note that, in this case, the
approximation of a linearised potential is guaranteed by the slow-roll
approximation. Higher order terms in the Klein-Gordon equation are
suppressed by powers of the  slow-roll parameters. In this way,
interaction between Fourier modes are absent, i.e., the vacuum
fluctuations of different Fourier modes are decoupled and the field
fluctuations are Gaussian.   

\subsection{Observables}

In the standard picture of inflation,  the perturbations of the
inflaton field are stretched out of the horizon and subsequently
transferred into curvature perturbations which survive after the
universe has reheated.  Observationally, the temperature
inhomogeneities in the CMB are related to the mean amplitude of the
curvature perturbations through the Sachs-Wolfe effect
\citep{sachs-wolfe}. Thus, the power spectrum of
curvature perturbations is observed to be \citep{wmap2008-params} 
\beq
\P_{\R} = \,(2.45_{-0.093}^{+0.092})\times 10^{-9}\quad (\mathrm{\, 95\%\, CL}).
\label{obs:ps}
\eeq 
 
 \noindent at a pivot scale $k_0 = 7.5 a_0H_0  \approx 0.001\, {h}_0\,
 \Mpc^{-1}$. The power spectrum is also probed at other scales in the
 CMB with various filter functions and also with the power spectrum of
 galaxies and clusters. The rate of change of the observed value of
 $\P$ with $k$ is parametrised by the spectral index $n_s$. This is
 defined by   
 \beq
 	n_s -1 \equiv\, \dtotal{\log{\P_\R}}{\log{k}},
	\label{ns:def}
 \eeq
 
 \noindent where $1$ is subtracted by convention due to the fact that
 the matter density power spectrum has the form $P_{\rho} \propto
 k^{n_s}$. Observationally, the output from WMAP
 \citep{wmap2008-params}, the distance measurements from type 1a
 supernovae \citep{riess06,astier05} and the baryon acoustic
 oscillations \citep{percival07} constrain the spectral index to be 
 \beq
 	-0.256\, < \,1 - n_s\, < \,0.025 \quad (\mathrm{ 95\%\, CL.}).
 \eeq
 
 \noindent for  $0.001 \Mpc^{-1} < k  < 0.1 \Mpc^{-1}$.
 
In single-field inflation, the power spectrum of curvature perturbations can be calculated from $\P_{\phi}$ and Eq.~\eqref{R:delrho}, which for a single scalar field can be written as
\beq
	\R = - \frac{H}{\dot{\phi}}\delta \phi.
\label{R:delphi}
\eeq

\noindent This relation, used at linear order in Eqs.~\eqref{two-point:power} and \eqref{ps:def}, shows that 

\beq
\P_{\R} =\left(\frac{H}{\dot{\phi}}\right)^2 \P_\phi \simeq \left(
 1+ {\cal O} (\omega) \right) \left( \frac{H_*^{2}}{2
\pi \dot{\phi}} \right)^2 \left| k \eta \right|^{2 \omega}.
\label{linear:ps}
\eeq

\noindent In terms of the slow-roll parameters, using Eqs. \eqref{sr:fried1} and \eqref{sr:fried2} and evaluating the previous expression at horizon crossing, we have
\beq
\P_{\R} = \,\frac{2^8}{3}\frac{V_*^3}{V_*^{\prime2}} = \frac{8}{3}\frac{V_*}{\epsilon_*}~.
\label{ps:sr}
\eeq

\noindent Note that the evaluation of the power spectrum at horizon exit is justified by the fact that $\R$ is constant on super-horizon scales. In this regime we can define the 
root-mean-square (${\rm RMS}$) value of $\R$ (or $\zeta$) as
\beq
	\R_{\rm RMS} = \zeta_{\rm RMS} = \frac{H}{\dot{\phi}} \delta\phi_{\rm RMS} \approx 
	\frac{H^2}{2\pi \dot{\phi}}~.
\label{r:rms}	
\eeq

\noindent This quantity will play an important role throughout this thesis. 

We can  also write the spectral index as  \citep{stewart-lyth}
\beq
 n_s  = 1 + 2\etasr - 6 \epsilonsr,
 \label{ns:sr}
\eeq

\noindent to lowest order in slow-roll expansion. When $n_s  = 1$,
the power spectrum is called scale-invariant or Harrison-Zeldovich
\citep{harrison,zeldovich72}. When $n_s \neq 1$, the spectrum is
described as tilted  and, for $n_s > 1$, it is called  `blue' because
the power is enhanced at small wavelengths
\citep{mollerach-blue}. With the precision reached by the latest
probes of cosmological structure, it has been possible to constrain
the field parameters and discard some models of inflation alternative
to the simplest picture presented above
\citep{alabidi06a,alabidi08}. Current observations of the CMB and
large-scale structure, however, are compatible with the 
predictions of many other models of inflation. It is therefore crucial
to study additional observables which provide further insight into the
characteristics of the early universe.  At the level of scalar
perturbations, the most convenient observables for discriminating
between these models are the tensorial perturbations mean amplitude and
spectral index, the running of the spectral index and the
non-Gaussianity of perturbations. In this thesis we focus on the
effects of the latter.

\subsection{The $\delta N$ formalism}

We now present a formalism to account for the contributions of multiple fields to the curvature perturbation at all orders. 
An important feature of the perturbed curvature on superhorizon scales is that we can calculate its magnitude by considering the change in the number of e-folds of expansion of the relevant patch of universe with respect to a uniform background expansion. This in turn allows us to compute the amplitude of the curvature fluctuations from the matter fluctuations. The idea behind this technique is to consider $\zeta$ as a perturbation in the local expansion \citep{starobinsky-multifield,salopek-bond,sasaki-linear,sasaki-multifield,lyth-zeta}, i.e. 
\beq
 \zeta = \delta N,
 \label{deltaN:def}
\eeq
  
 \noindent where $\delta N$ is the perturbed expansion of the uniform-density hypersurfaces with respect to spatially flat hypersurfaces. 
 
We now describe the elements of the above formalism. The number of e-folds of expansion between two moments in proper time $\tau_1$ and
$\tau_2$ is given in the homogeneous background by 
\beq
 N = \int_{\tau_1}^{\tau_2}\frac{1}{3}\theta_0 \, d\tau =
 \int_{t_1}^{t_2} \frac{1}{3} \theta_{(\rmt)}\, dt = 
 \int_{t_1}^{t_2} H \,dt,  
\label{N:FRW}
\eeq

\noindent where $\theta_0$ refers to the homogeneous expansion, that
is,  $\theta$ considered to lowest order in
Eq.~\eqref{confrormal:theta}. In the perturbed metric, the number of
e-foldings  along an integral curve of the 4-velocity, i.e., along a
comoving worldline between $\tau_1$ and $\tau_2$, is
\begin{align}
\mathcal{N} = &\int_{\tau_1}^{\tau_2} \frac{1}{3} \theta \,d\tau =
\int_{t_1}^{t_2} \frac{1}{3}\theta_{(\rmt)} (1 + \varphi)(1 - \varphi)\,dt 
= \int_{t_1}^{t_2} (H + \dot{\psi}) \, dt.
\label{N:comov}
\end{align}

\noindent The last equality holds on superhorizon scales. It is
clear  that the difference $ \mathcal{N} - N$ will 
provide the change in the curvature perturbation from an initial
hypersurface at time $t_1$ and a final one at time $t_2$, i.e.,
\beq 
\delta N = \mathcal{N} - N =  \Delta \psi.
\label{delta:N}
\eeq

\noindent In particular, we can choose to
integrate the expansion starting from an initial uniform-curvature hypersurface at time $t_1$, that is, $\psi(t_1) = 0$. Then we can find the
amplitude of the curvature perturbation at the later time $t_2$ by
choosing a trajectory with an endpoint $t_2$ fixed on a comoving or uniform-density hypersurface. Denoting the difference between the background
and the perturbed expansion as $\delta \theta$, we can write
\beq
\delta{N} = \int_{t_1}^{t_2}\delta \theta_{(\rmt)} \, d\eta = 
 \psi_2 . 
\label{N:psi}
\eeq

\noindent In particular, when we consider the endpoint embedded in a uniform-density hypersurface, then Eq.~\eqref{deltaN:def} is recovered.  

 On large scales, where spatial gradients can be
 neglected, the local physical quantities like density and expansion
 rate obey the same evolution equations as in a homogeneous FRW
 universe \citep{wands-malik,sasaki-linear}. By using  homogeneous FRW
 universes to describe the evolution of local patches, we can evaluate
 the perturbed expansion in different parts of our universe with
 particular initial values for the fields during inflation. This is
 known as the `separate universe' approach and means that, when we
 neglect the decaying mode for the field perturbations on superhorizon
 scales, we can consider the local integrated expansion as a function
 of the local field values on the initial hypersurface. In particular,
 one can expand Eq.~\eqref{deltaN:def} as  
 \beq
 	\zeta = \delta{N}(\phi_i(t_1)) = \sum_i
        \dpartial{N(t_2)}{\phi_i} \delta \phi_i(t_1)  
	\label{linear:zeta}
 \eeq

\noindent where the initial time $t_1$ again corresponds to some
initial spatially-flat hypersurface.  We can use this formula to
construct the curvature power spectrum in multi-field inflation: 

\beq
	\P_{\zeta} = \sum_{i}
        \left(\dpartial{N(t_2)}{\phi_i}\right)^2 \P_{\delta\phi}\,. 
\eeq
   
 \noindent  This formalism can be extended to establish the
 equivalence between nonlinear matter and metric perturbations. This
 is done  by first assuming Eq.~\eqref{deltaN:def} as the definition of
 the curvature perturbation and then using Eq.~\eqref{R:delrho}
 to integrate $\zeta$. This gives \citep{lyth-zeta} 
 \beq
 	\zeta = \frac{1}{3(w+1)}
        \ln\left(\frac{\rho_{\kappa}}{\rho_0}\right). 
 \eeq
 
\noindent This nonlinear curvature perturbation can be written as a function of
 the initial field fluctuations, evaluated again at an initial flat
 hypersurface labelled by the time $t_1$. We use a Taylor expansion 
 \beq
 	\zeta = \sum_{i} \dpartial{N}{\phi_i} \delta \phi_i(t_1) + 
        \frac{1}{2}\frac{\partial^2 N}{\partial\phi_j\partial\phi_i}
          \delta \phi_j(t_1)\delta \phi_i(t_1) + \dots 
	\label{non-lin:zeta}
 \eeq
 
 \noindent where the leading order term coincides with the expansion
 \eqref{linear:zeta}. This last expansion greatly simplifies the
 derivation of the higher-order curvature correlations from the scalar
 field bispectrum.  
 
\section{Non-Gaussianity from isocurvature fields}
\label{non-gaussianity}

The perturbed energy conservation equations show the conditions under
which the curvature perturbation $\zeta$ may vary over time in a
regime in which the perturbation modes lie well outside the
horizon. Specifically, Eq.~\eqref{zeta:evol} shows that the
evolution of $\zeta$ is directly related to the presence of an entropy
perturbation $\delta s$. This quantity can be the intrinsic
non-adiabatic pressure of a single field, or the difference in the
density perturbations of any two fields which contribute to the
curvature perturbation. This effect motivates the study of models of
inflation in which the observed inhomogeneities are the result of
non-adiabatic field fluctuations  
\citep{mollerach,linde-curvaton,enqvist-curvatonI,lyth-curvatonI,moroi}.
Here we are interested specifically in  models in which a highly
nonlinear $\zeta$ can be 
generated from the aforementioned entropy perturbation. Out of the
multiple stages in which this field may have  influenced $\zeta$, we
focus on the case of an auxiliary scalar field during inflation
generally referred as the isocurvature field $\chi$ 
\citep{zaldarrion,enqvist-curvaton,enqvist-jokinen}.  

A first approximation to nonlinear $\zeta$ involves the first and
second order perturbations in real space: 
\beq
	\zeta(t,\vect{x}) = \zeta_1(t,\vect{x}) +
        \frac{1}{2}\zeta_2(t,\vect{x}). 
\eeq

\noindent The second order perturbation is conventionally written in
terms of the first order perturbation and the parameter $\fnl$
\citep{komatsu-spergel}. This gives 
\beq
	\zeta(t,\bx)  = \zeta_1(t,\bx) - \frac{3}{5} \fnl
        \left(\zeta_1^2(t,\bx) - \langle\zeta^2_{1}(t,x)\rangle\right),
	\label{fnl:real}
\eeq

\noindent which in Fourier space is written as a convolution 
\beq
	\zeta(\bk) = \zeta_1(\bk)- \frac{3}{5} \fnl
        \left([\zeta_1\star\zeta_1](\bk)  -  
        \langle\zeta^2_{1}(k)\rangle\right).
	\label{fnl:fourier}
\eeq

\noindent Note that Eq.~\eqref{fnl:fourier} shows explicitly the
superposition of modes that characterise non-Gaussian statistics. 
  
 Statistically, non-Gaussianity refers to the non-vanishing higher
 order moments of the quantity in question. In quantum mechanics this
 corresponds to the n-point correlation functions with $n \geq 3$. To
 lowest order, the bispectrum $B_\zeta(k_1,k_2,k_3)$ is defined by the
 expectation value of the product of three copies of the curvature
 field: 
 \beq
 	\langle  \zeta(\bk_1) \zeta(\bk_2) \zeta(\bk_3)\rangle = (2\pi)^3 B_{\zeta}(k_1,k_2,k_3) \delta^{(3)} (\bk_1,\bk_2,\bk_3).
\eeq
             	
\noindent The amplitude of $\fnl$ is given in terms of the bispectrum by substituting 
Eq.~\eqref{fnl:fourier} in this:
 \beq
\frac{6}{5} f_{\rm NL} = \frac{\Pi_i k_i^3}{\sum_i k_i^3}
\frac{B_\zeta}{4 \pi^4 \P_\zeta}. 
\label{fnl:def}
\eeq

\noindent In this section we present a method to compute the nonlinear
$\zeta$ by determining nonlinear solutions to the Klein-Gordon
equation of the field fluctuation. Then with the aid of the $\delta N$
formalism we will construct the three-point correlator of
$\zeta$. Special attention will be paid to the cases in which a large
$\fnl$  can be obtained. 

\subsection{Two-field inflation}

Here we consider a spacetime inflating by the action of a potential which depends on two minimally coupled fields. In this case, in addition to the canonical inflationary field, we consider a second field $\chi$ described by the action
\beq
{\cal S}_{\chi} = \int dx^4 \sqrt{-g} \left[ - \frac{1}{2} g^{\mu\nu}\partial_{\mu}
  \chi \partial_{\nu}\chi
- \frac{1}{2} m_\chi^2 \chi^2 - W(\chi) \right] ,
\label{chi_action}
\eeq

\noindent where $W(\chi)$ plays the role of a nonlinear potential. The
joint Lagrangian density of the two scalar fields in this model is
given by Eq.~\eqref{lagrangian:twofield}.  
The model in question demands that there is no contribution of $\chi$
to the background matter content. This is guaranteed by the following
condition on the potential of the two-field Lagrangian defined in
Eq.~\eqref{lagrangian:twofield}:  
\beq
	U(\phi,\chi) \approx V(\phi)  \quad\Rightarrow 
	\quad W(\chi) + \frac{1}{2} m_{\chi}^2 \dot{\chi}^2 \ll V(\phi).
 	\label{potential:cond}
\eeq

\noindent To analyse the dynamics of this auxiliary field, let us consider a
 background homogeneous part and a coordinate-dependent field fluctuation,
\beq
\chi(t,\bx) = \chi_0(t) + \delta \chi(t,\bx).
\eeq

\noindent The Klein-Gordon equation derived by varying the action
\eqref{chi_action} with respect to $\chi$ yields for the background field
\beq
\ddot{ \chi}_0+ 3 H \dot{ \chi}_0 + m_\chi^2 \chi_0 +
W'( \chi_0) =0, \label{background_KG}
\eeq

\noindent and for the field fluctuation
\beq
\ddot{\delta \chi}+ 3 H \dot{\delta \chi}  - \frac{\nabla^2 \delta
  \chi}{a^2}
+ m_{\delta \chi}^2 \delta \chi + \sum_{n=3}
\frac{W^{(n)}}{(n-1)!} \delta \chi^{n-1} =0.
\label{deltachi_equation}
\eeq

\noindent Here a superscript $(n)$ denotes the $n$-th derivative and
$m_{\delta \chi}^2$ is the effective mass for the field fluctuation defined as 
\beq
m_{\delta \chi}^2 = m_\chi^2 + W^{\prime\prime}.
\eeq

\noindent Following the same steps as with the inflaton power
spectrum, we recover a solution similar to
Eq.~\eqref{fourier:phi_i}. Specifically, 
\beq
\P_\chi  ( \eta,k) =  
\frac{H^{2}}{8\pi} (|k\eta|)^3 \left| {\rm H}^{(1)}_\mu(|k \eta|)\right|
= 2^{-2\alpha} \frac{\Gamma\left(\frac{3}{2} -\alpha 
\right)^2}{\pi^3} {H_*^2 } \left( |k \eta| \right)^{2 \alpha},
\label{chi:ps}
\eeq

\noindent with $\alpha$ and $\mu$ defined through the equation
\beq
\alpha \equiv \sqrt{\mu^2 - \frac{9}{4}} \approx \frac{m_{\delta
    \chi}^2}{3 H^2} > 0 .  
\label{alpha:chi}
\eeq

\noindent Because of this, the spectrum of $\chi$ is blue. The power
spectrum $\P_{\chi}$ vanishes on large scales for large values of
$\alpha$, i.e. when $m_{\delta \chi} \gsim H_*$, so we only consider
the case  
\beq
m_{\delta \chi} \lsim H_*.  \label{mass:condition}
\eeq

\noindent Indeed, we assume $m_\chi \ll H_*$. As shown below, the
interesting values of $W''$ are those which are large enough to
generate a large nonlinear coupling, but sufficiently small to satisfy
condition \eqref{mass:condition}, which also implies 
\beq
W'' \lsim H_*^2.
\eeq

\noindent For example, for
\beq
W=\frac{M}{3!} \chi^3+ \frac{\lambda}{4!} \chi^4 ,
\label{power:potential}
\eeq
this condition turns into conditions on $M$ and $\lambda$:
\beq
M \lsim \frac{H_*^2}{\chi_0}, \qquad \lambda^{1/2} \lsim
\frac{H_*}{\chi_0}.
\eeq

\noindent Typically we have a nonzero expectation value
$\chi_0$. Working in a perturbative expansion then requires that the
quantum fluctuations $\delta\chi$ do not exceed this expectation
value, i.e., we require  $\chi_0 \gsim H_*$. This imposes
additional constraints on the parameters $M$ and $\lambda$ in the
specific model of Eq.~\eqref{power:potential}, as  described below.

\subsection{Non-Gaussianity and nonlinear evolution}

The expectation value of the product of three fields was first
computed by \cite{maldacena} for the case of single-field inflation,
including slow-roll scalar field and gravitational interactions. In
our case, since the field gives a negligible contribution to the
energy density of the Universe, we will neglect its coupling to
gravity. 

The three-point correlation function of the field perturbations
$\delta \chi$ can
be computed using the expression given by Maldacena:
\beq
\langle \delta \chi (t,\bxf) \delta \chi (t,\bxs) \delta \chi
(t,\bxt) \rangle = i \int_{-\infty}^t dt' \langle \left[ H_I(t'),
\delta \chi_I (t,\bxf) \delta \chi_I (t,\bxs) \delta \chi_I
(t,\bxt) \right]  \rangle,
\label{maldacena:correlator}
\eeq
where $H_I$ is the interaction Hamiltonian written in terms of the
field perturbation in the interaction picture, i.e. in our case
\beq
H_I(t') = - \int d x^3 \sqrt{-g} {\cal L}_{I} = \sum_{n=3} \int d
x^3 a(t')^3 \frac{W^{(n)}}{n!} \delta \chi_I^{n}.
\label{hamilton:chi}
\eeq
Note that on the left-hand side of Eq.~\eqref{maldacena:correlator} the
expectation value is taken with respect to the vacuum of the
interacting theory, while on the right-hand side it is taken with
respect to the vacuum of the free theory. A generalisation of this
expression to higher order correlators, including loop
corrections, has been provided by \cite{weinberg-qcI}.

Recently,  \cite{musso} showed that the
general expression for the correlators of $m$ field fluctuations
can also be derived by solving perturbatively the field equation
of motion and using the expression for the two-point function of
the free field fluctuation. Here we make use of this formalism to
show that Eq.~\eqref{maldacena:correlator} can be derived by solving 
Eq.~\eqref{deltachi_equation} perturbatively.

As mentioned before, the evolution equation for $\delta \chi_I$ is given in identical form to the inflaton case, providing  $\phi \to \chi$ in Eq.~\eqref{conformalkg:fourier}.
On the other hand, the evolution equation for the full nonlinear
$\delta \chi_{}$ can be given perturbatively providing $|(\delta \chi-\delta \chi_I)/\delta\chi| \ll 1$. Rewriting Eq.~\eqref{deltachi_equation} and re-expressing
the nonlinear term in terms of $\delta \chi_I$, we obtain
\beq
\ddot{\delta \chi}_{} + 3 H \dot{\delta \chi}_{}  - \frac{\nabla^2
\delta
  \chi_{}}{a^2}
+ m_{\delta \chi}^2 \delta \chi_{} =- \sum_{n=3}
\frac{W^{(n)}}{(n-1)!} \delta \chi_I^{n-1}  . \label{chi_1_evol}
\eeq
This equation can be then solved by Green's method and its solution is
\beq
\delta \chi_{}(t,\bx) = \delta \chi_I (t,\bx) - \sum_{n=3}^{\infty}
\frac{1}{(n-1)!} \int d^3y \int dt' a(t')^3 \mathrm{G}_R (t,\bx;t',\by)
W^{(n)} \delta \chi_I^{n-1}(t',\by), \label{dchi1}
\eeq
where $\mathrm{G}_R(t,\bx;t',\by')$ is the  retarded Green's function
\beq
\mathrm{G}_R(t,\bx;t,'\by) = \imag \Theta(t-t') \left[ \delta \chi_I (t,\bx) ,
\delta \chi_I (t',\by)\right].
\eeq

\noindent where 
\beq
\Theta(x) = \int_{-\infty}^x \diracd(z) \, \d z
\label{heaviside:def}
\eeq
 is the Heaviside step function. By using
the fact that the free-field perturbation is Gaussian, i.e.  
\beq
\langle \delta \chi_I (\bxf ,t) \delta \chi_I (\bxs ,t) \delta
\chi_I (\bxt ,t) \rangle=0,
\eeq
 
 \noindent one can rewrite the three-point
correlation function of $\delta \chi$ making use of Eq.~\eqref{dchi1}:
\bea
\langle \delta \chi (t,\bxf ) \delta \chi (t,\bxs ) \delta \chi
(t,\bxt ) \rangle = - \sum_{n=3}^{\infty} \frac{\imag}{(n-1)!} \int dy^3 \int
dt' a(t')^3 W^{(n)} \times  \nonumber \\
\left[ \delta \chi_I(t,\bxf),\delta \chi_I (t',\by)\right] \langle
\delta \chi_I^{n-1} (t',\by) \delta \chi_I(t,\bxs)\delta
\chi_I(t,\bxt) \rangle  + {\{\rm Perms.\}} 
\label{semifinal}
\eea

\noindent Using this equation and the result \citep{musso},
\begin{align}
n \left[ \delta \chi_I(t,\bxf),\delta \chi_I (t',\by)\right] &
\langle \delta \chi_I^{n-1} (t',\by) \delta \chi_I(t,\bxs) \delta
\chi_I(t,\bxt) \rangle + \perms =
\notag \\
- \langle  & \left[ \delta \chi_I^n (t',\by), \delta \chi_I(t,\bxf)
\delta \chi_I(t,\bxs) \delta \chi_I(t,\bxt) \right] \rangle,
\label{musso:eq}
\end{align}

\noindent one can derive Maldacena's formula
\eqref{maldacena:correlator} for the $\chi$-Hamiltonian \eqref{hamilton:chi} after some manipulation. This shows that one can obtain Maldacena's formula either from considering perturbations in the action or from the perturbative expansion of the equations of motion.
In the case of slow-roll inflation, this equivalence was shown by \cite{seery-kg}. Equation \eqref{maldacena:correlator} can be generalised to higher correlation functions:
\begin{align}
\langle \delta \chi (t,\bxf) \delta \chi (t,\bxs ) \ldots \delta
\chi (t,\bx_m ) \rangle =&  
\label{Maldacena_higher} \\
\imag \int_{-\infty}^t dt' \Big< \big[&
H_I(t'), \delta \chi_I (t,\bxf ) \delta \chi_I (t,\bxs) \ldots
\delta \chi_I (t,\bx_m) \big] \Big>.\notag
\end{align}

\subsection{Field bispectrum}

Here we are interested in the three-point function of the scalar
field, also called the field  bispectrum $F( k_1,k_2,k_3)$, defined as 
\beq
\langle \delta \chi_{\bk_1}(\eta) \delta \chi_{\bk_2} (\eta)
\delta \chi_{\bk_3} (\eta) \rangle = (2 \pi)^3 \delta^{(3)} (\sum_i
\bk_i ) F(\eta; k_1,k_2,k_3). 
\label{bi}
\eeq

\noindent For simplicity, we assume that the nonlinear potential of
the scalar field is dominated by the cubic interaction
$W^{\prime\prime\prime}$ and that this is approximately constant. This
reduces the expansion in Eq.~\eqref{hamilton:chi} to the first term
only. Substituting this in Eq.~\eqref{maldacena:correlator} and using
Eqs.~\eqref{dchi1} and \eqref{musso:eq}, the field bispectrum becomes 
\begin{align}
F(\eta; k_1,k_2,k_3) =  & \label{masslessF2}\\
{W^{\prime\prime\prime}} \eta^{9/2}& \frac{H_*^2}{4 } Re
\left\{ -\imag \frac{\pi^3}{2^3} \prod_i \mathrm{H}_\mu^{(1)} (|k_i \eta|)
\int_{-\infty}^{\eta} d \eta' \sqrt{|\eta'|} \prod_j   \mathrm{H}_\mu^{(2)}
(| k_j \eta'|) \right\}. \notag
\end{align}

\noindent During de Sitter inflation, the ratio $k/ aH $ can be
written as the product $|k \eta|$ and we will use this to parametrise
the various stages of evolution of the bispectrum. 

It is instructive to first evaluate the bispectrum (\ref{masslessF2})
during inflation ($\eta\le \eta_{\rm reh}$) for the case of a massless field
fluctuation (i.e.,  $m_{\delta \chi} =\alpha =0$,
$\mu=3/2$). In this case,
the integral in Eq.~(\ref{masslessF2}) can be evaluated
analytically \citep{bernardeau-ng,zaldarrion} using the
expressions for the Hankel functions,
\beq
\mathrm{H}_{3/2}^{(1)} (z) =-\imag \sqrt{\frac{2}{\pi}} (1-\imag z)
\frac{e^{iz}}{z^{3/2}},
\qquad \mathrm{H}_{3/2}^{(2)} (z) =\imag \sqrt{\frac{2}{\pi}} (1+\imag z)
\frac{e^{-iz}}{z^{3/2}}.
\eeq

\noindent One finds\footnote[3]{Our result coincides with the one found in
\cite{zaldarrion}, modulo the overall sign and the factor $1/3$ inside the parentheses.}
\beq
F(\eta; k_1,k_2,k_3)=  \frac{W^{\prime\prime\prime} H_*^2 }{4\prod_i k_i^3} \left[
-\frac{4}{9}k_t^3 + k_t \sum_{i<j} k_i k_j + \frac{1}{3}
\left(\frac{1}{3}+ \gamma+\ln (\left|k_t \eta\right|) \right) \sum_i k_i^3
\right], \label{I3/2} 
\eeq

\noindent where $k_t = \sum_i k_i$. The integral in
Eq.~(\ref{masslessF2}) has been evaluated for three different stages:
when the modes are well inside the Hubble radius, around the Hubble
radius and outside the Hubble radius. The first stage does not give
any contribution to the integral because the Hankel functions
oscillate rapidly for $ \left| k_i \eta\right| \gg 1$ and the fields
can be taken as free in the asymptotic past. At late time, the
integral is dominated by the modes that are around the Hubble scale or
larger, $\left| k \eta \right| \lsim 1$. In particular, we will show
below that the time-dependent term in Eq.~\eqref{I3/2}, with the
typical local momentum dependence, essentially comes from the
nonlinear and classical super-Hubble evolution of the field
perturbation. The finite part, with the non-trivial momentum
dependence, comes from integrating over times corresponding to
Hubble-crossing.

Let us now consider the case with $m_{\delta \chi} \neq 0 $, and  $1/2 < \mu
< 3/2$. The integral
inside Eq.~\eqref{masslessF2} cannot be integrated
analytically in this case. One can only evaluate it over the period
when all modes are well outside the Hubble radius because, for small arguments, Hankel functions can be written in terms of Bessel functions (which for real arguments are real). Decomposing Eq.~\eqref{masslessF2} in this fashion, using the Bessel function expansion for small arguments and retaining the dominant real part, then yields
\begin{align}
F(\eta, k_1, k_2,k_3) = &F(\eta_*, k_1, k_2,k_3) + \frac{W^{\prime\prime\prime}
H_*^2}{12} \frac{2^{3-4\alpha} \Gamma(\mu)^4  }{\mu \pi^2  \alpha}
\times \label{massive}
\\ &\left\{ \frac{2\mu}{3(\mu-\frac{1}{2})}\left[1 - \frac{\alpha
    \left({\eta}/{\eta_*} \right)^{3-3 \alpha} }{2 \mu}\right] -
(\eta/ \eta_*)^{-\alpha} \right\}\frac{(- \eta)^{4 \alpha} \sum_i
  (k_i)^{2 \mu}  }{ \prod_i (k_i)^{2 \mu}}. 
\notag
\end{align}
The first term on the right-hand side is the non-Gaussianity at
Hubble exit and it can only be computed numerically.
The second term represents
the non-Gaussianity generated at late times
during inflation, when the large-scale local
term dominates over the finite Hubble-crossing term.

Now we will show that the large-scale local contribution to the
non-Gaussianity of Eqs. \eqref{I3/2} and \eqref{massive}
can be derived by solving the equation of motion of the field
perturbation on large scales.
The nonlinear evolution of the field fluctuation is derived
by taking the large-scale limit $ \left| k\eta \right| \to 0$ in Eq.~\eqref{conformalkg:fourier} for the free-field  $\delta\chi_I$ and Eq.~\eqref{chi_1_evol} for the nonlinear $\delta \chi$. This leads to the equations
\begin{align}
\ddot{\delta \chi}_{I \bk} + 3 H \dot{\delta \chi}_{I\bk}
+ m_{\delta \chi}^2 \delta \chi_{I\bk} =0 , 
\label{homo} \\
\ddot{\delta \chi}_{ \bk} + 3 H \dot{\delta \chi}_{ \bk} +
m_{\delta \chi}^2 \delta \chi_{ \bk} ={\cal S}_\bk  ,
\label{chi_1_evol_ls}
\end{align}

\noindent where the source on the left-hand side is given in terms of the linear
solution,
\beq
{\cal S}_\bk=- \frac{W^{\prime\prime\prime}}{2} (\delta \chi_I \star \delta
\chi_I)_\bk, 
\label{source}
\eeq

\noindent and where $\star$ denotes the convolution operation.
The growing solution of the homogeneous Eq.~(\ref{homo}) is
\beq
{\delta \chi}_{I \bk}(\eta) = {\delta \chi}_{I \bk} (\eta_*) (| k
\eta|)^{\frac{3}{2} - \mu}. 
\label{homo_sol_0}
\eeq

\noindent Only in the massless limit  is this constant. One can find the
solution of the inhomogeneous equation by using
the method of variation of parameters, which yields
\begin{align}
\delta \chi_{ \bk} (\eta) = \delta \chi_{ \bk}  (\eta_*)&(|k
\eta|)^{\frac{3}{2} - \mu} + \frac{1}{2 \mu H_* } \left[ (|k
\eta|)^{-\frac{3}{2}+\mu} \int_{\eta_*}^\eta d\eta' a(\eta') (|k
\eta|)^{-\frac{3}{2}-\mu}  {\cal S}_\bk(\eta') \right. \notag
\\ &\left.
 -
(|k \eta|)^{-\frac{3}{2}-\mu}  \int_{\eta_*}^\eta d\eta' a(\eta')
(|k \eta|)^{-\frac{3}{2}+\mu}
{\cal S}_\bk(\eta') \right].
\label{ls_solution}
\end{align}

Using the source \eqref{source} with Eq.~\eqref{homo_sol_0} and integrating over conformal time, one obtains
\beq
\delta \chi_\bk (\eta) = \delta \chi_\bk(\eta_*) (|k
\eta|)^{\alpha} + \frac{W^{\prime\prime\prime}}{6H_*^2} f(\eta/\eta_*) [\delta
\chi_I (\eta) \star \delta \chi_I(\eta)]_\bk~,
\label{nonlinear_relation}
\eeq
where
\beq
 f(x) = \left\{ \begin{array}{ll}
\frac{1}{3} \left(1 - x^3 \right) + \ln(x)
& \mbox{for }~ \mu = 3/2
, \\
\frac{1}{\alpha(\mu -1/2)}\left(1 - \frac{\alpha}{2
\mu}x^{3-3\alpha} \right) - \frac{3}{2 \mu \alpha} x^{-\alpha}  &
\mbox{for }~ 1/2 < \mu < 3/2 .\end{array} \right. \label{f_def}
\eeq

\noindent We can now use Eq.~\eqref{nonlinear_relation} to compute the
bispectrum from its definition (\ref{bi}).
In terms of the  power spectrum of the field perturbation,
this is
\beq
F(\eta; k_1, k_2,k_3) = F(\eta_*; k_1, k_2,k_3) + 
\frac{W^{\prime \prime \prime }}{3 H_*^2} f(\eta/\eta_*) 
\sum_{i<j} P(\eta,k_i) P(\eta,k_j) ,  
 \label{ls_bi}
\eeq

\noindent which correctly reproduces the large-scale contribution to the
non-Gaussianity in Eqs.~\eqref{I3/2} and \eqref{massive}. Note
that, in the limit $\mu \to 3/2$ and $\alpha \to 0$, the large-scale expression for a massive field converges to the massless
case, as can be checked by taking this limit in the lower expression on the right-hand side of Eq.~\eqref{f_def} and using
\bea
\frac{2\mu}{3(\mu-\frac{1}{2})} &\to& 1 + \frac{\alpha}{3}+ {\cal O}(\alpha^2), \\
x^{- \alpha} = \exp \left[{- \alpha} \ln (x)
\right] &\to& 1 - \alpha \ln (x) + {\cal O}(\alpha^2).
\eea

\noindent In summary, at late times the non-Gaussianity of the field is
dominated by the nonlinear evolution on large scales and thus the
bispectrum of the field perturbation is of the local form, i.e.
proportional to the product of two power spectra (See
Sec.~\ref{intro-ng} for the definition of triangulation of the
bispectrum). In the massless case, in which $\mu =3/2$, the power
spectrum of the field fluctuation is constant while the bispectrum
grows as $\ln (a)$. In the massive case, however, the power spectrum
decays as $a^{-2\alpha}$ and the bispectrum decays as $a^{ -3\alpha}$,
thus `{growing}' as $a^\alpha$ with respect to the product of two
power spectra. This relative growth is important for the
non-Gaussianity in the curvaton mechanism, as can be seen from
Eq.~\eqref{fnl:def}. The curvaton case will be discussed below in more
detail. 

Before making contact with curvature perturbations, let us extend the solution \eqref{nonlinear_relation} and consider the evolution of the non-adiabatic perturbations in a radiation-dominated era. This applies once inflation has ceased and  the inflaton field has been thermalised, but before the isocurvature perturbation is converted into an adiabatic one. 

During the radiation dominated era, $H = (2t)^{-1}$ and the field perturbation
evolves on large scales according to
\bea
\ddot{\delta \chi}_{I \bk} + \frac{3}{2t} \dot{\delta \chi}_{I
\bk} + m_{\delta \chi}^2 \delta \chi_{I \bk} = 0
, \label{chi_0_evol_ls_rad} \\
\ddot{\delta \chi}_{ \bk} + \frac{3}{2t} \dot{\delta \chi}_{ \bk}
+ m_{\delta \chi}^2 \delta \chi_{ \bk} ={\cal S}_\bk .
\label{chi_1_evol_ls_rad}
\eea

\noindent The growing mode of the homogeneous equation is a 
Bessel function of the first kind $J$ \citep{langlois-curvaton},
\beq
\delta \chi_{I \bk} (t) =  \delta \chi_{I \bk}(t_{\rm reh})
\frac{\pi}{2^{5/4} \Gamma(3/4)} \frac{J_{1/4}(m_{\delta
\chi}t)}{(m_{\delta \chi}t)^{1/4}}.
\eeq

\noindent For $m_{\delta \chi} t\ll1 $, well before decay, the growing mode is constant.
Indeed, by solving Eq.~\eqref{chi_1_evol_ls_rad} at lowest
order in $m_{\delta \chi}t$, we find\footnote[4]{This result is in
agreement with \cite{enqvist-curvaton} where the computation considered a general nonlinear potential up to order ${\cal
O}\left((m_{\delta \chi}t/2)^8\right)$.}
\beq
\delta \chi_{\bk}(t) = \delta \chi_{\bk}(t_{\rm reh})
\frac{\pi}{2^{5/4} \Gamma(3/4)} \frac{J_{1/4}(m_{\delta
\chi}t)}{(m_{\delta \chi}t)^{1/4}} - \frac{W^{(3)}}{10 m_{\delta
\chi}^2} (m_{\delta \chi}t)^2 [\delta \chi_I (t) \star \delta
\chi_I(t)]_\bk. \label{nonlinear_after_infl}
\eeq

\noindent This result shows that the isocurvature fluctuation
continues its nonlinear evolution throughout the radiation era. We
will use this result in the context of a curvaton field to account for
the consequences of considering a nonlinear source in the bispectrum
of the field and then compute the corresponding non-Gaussian parameter
$\fnl$.

\subsection{The curvaton}

The curvaton is an alternative inflationary mechanism to generate the
matter density fluctuations
\citep{mollerach,linde-curvaton,enqvist-curvatonI,lyth-curvatonI,moroi}. This
is achieved without appealing to the perturbations in the original
inflaton field. Instead, an auxiliary 'curvaton' field, subdominant
during inflation, generates isocurvaure fluctuations which transform into
adiabatic ones after the inflationary phase, during the decaying
oscillations of the curvaton field.  During inflation, the
isocurvature field presents a negligible contribution to the energy
density. After inflation the field still plays no significant role in
the background evolution as long as its mass $m_{\chi}$ is negligible
compared to the Hubble parameter. However, once $m_{\chi}^2 \approx
H^2$, the curvaton field starts oscillating at the bottom of its
potential. At this stage the potential can be approximated as
quadratic. The energy  density of the field decays during the
oscillations like a non-relativistic component ($\rho_{\chi} \propto 1
/ a^{3}$). The curvaton then   contributes significantly to the energy
density of the universe and the curvaton fluctuations are transformed
into adiabatic matter fluctuations (for the simplest version of this
mechanism see \cite{bartolo-curvaton}). 

We can write the number of e-folds of expansion in terms of the value
of the field  $\chi$:   
\beq
{N}(t_{\rm dec},t_{\rm in}) = \frac{1}{3}\ln \left(\frac{
\rho_{\chi_{\rm in}}}{\rho_{\chi_{\rm dec}}}
\right),\label{N_curvaton}
\eeq

\noindent where
\beq
\rho_{\chi_{\rm in}}=\frac{1}{2}m_\chi^2 \chi_{\rm in}^2, \qquad
\rho_{\chi_{\rm dec}}=\frac{1}{2}m_\chi^2 \chi_{\rm dec}^2,
\label{energy_in_dec}
\eeq
are the energy densities of the curvaton at the onset of the
oscillations (on a flat slicing) and at the moment of decay (on a
uniform density slicing), respectively.

The non-Gaussianities generated by the curvaton mechanism have
been studied in several papers. In the following  we consider and
combine all the possible effects, including  the intrinsic
non-Gaussianity of the curvaton field fluctuation and the nonlinear
relation between the curvature perturbation $\zeta$ and the curvaton
fluctuation. The intrinsic non-Gaussianity of the curvaton can be
generated inside and outside the Hubble radius due to its nonlinear 
potential. In particular, as discussed later in this section, we
take into account  the nonlinear evolution during both inflation
\citep{bernardeau-ng,zaldarrion} and the
radiation epoch \citep{enqvist-curvaton,lyth-curvatonII}.

In order to compute $\zeta$ and its n-point functions, one
can follow two equivalent procedures: 
\begin{enumerate}
\item Expand the perturbed number
of e-folds $\delta N$ on an initial flat slice at the onset
of the oscillations  ($t=t_{\rm in}$), in terms of the field
fluctuations $\delta \chi(t_{\rm in})$, and then use
Eqs.~\eqref{ls_bi} and \eqref{nonlinear_after_infl} to introduce the
three-point correlators of the field fluctuations. 

\item Expand the perturbed
number of e-folds $\delta N$ on an initial flat slice at
Hubble crossing $(t=t_*)$ in terms of the field fluctuations
$\delta \chi(t_{\rm in})$ and then take into account the nonlinear
relation between the field fluctuation at $t=t_{\rm in}$ and the
one at $t=t_*$ using Eqs.~(\ref{nonlinear_relation}) and
(\ref{nonlinear_after_infl}). 
\end{enumerate}

\noindent We will follow the latter procedure,
which has been used in the rest of the literature on the
curvaton. We write, as in Eq.~\eqref{non-lin:zeta}
\beq
\zeta = N_{,\chi_*} \delta \chi(t_*) + \frac{1}{2}
N_{,\chi_*\chi_*} \delta \chi^2(t_*),
\eeq
where $N$ is given by Eq.~\eqref{N_curvaton}.

In general, as we have seen in the previous section, $\chi_{\rm
in}$ is a nonlinear function of the field value at Hubble
exit, and we parameterise this dependence by the function
$g(\chi(t_*))$ and its derivatives:
\begin{align}
g = &\chi_0 (t_{\rm in}), \\
\delta \chi (t_{\rm in}) = &\sum_{n=1} \frac{g^{(n)}}{n!} \delta
\chi^n(t_*).
\label{def:g}
\end{align}

\noindent Using $\frac{\partial}{\partial \chi_*} = g'
\frac{\partial}{\partial g}$ and Eq.~\eqref{energy_in_dec},
one can differentiate $N$ in Eq.~\eqref{N_curvaton} to
obtain
\beq
N_{,\chi_*} = \frac{2}{3} \frac{g'}{g} \mathcal{C},
\eeq

\noindent where the prime denotes here a derivative with respect to
$\chi_*$ and
\beq
\mathcal{C}=1-\frac{\partial \ln \bar\rho_{\chi_{\rm dec}}}{\partial \ln
\bar  \rho_{\chi_{\rm in}}} \approx \left. \frac{3 \bar \rho_{\chi} }{4
\bar \rho - \bar \rho_{\chi}} \right|_{\rm dec},
\label{def:r}
\eeq

\noindent where $\bar \rho$ is the unperturbed total energy
density. Here we have taken a uniform $\bar \rho_{\rm
  dec}$ assuming that the radiation is unperturbed. Also, to arrive at
the last equality we used $(\bar \rho_{\chi_{\rm dec}} /\bar \rho_{\chi_{\rm
    in}})^{1/3} = [(\bar \rho_{\rm dec} - \bar \rho_{\chi_{\rm
      dec}})/(\bar \rho_{\rm in}  - \bar \rho_{\chi_{\rm in}})]^{1/4}$.

To compute the power spectrum we neglect the evolution during the
radiation dominated era, and from Eq.~\eqref{nonlinear_relation} we
obtain ${g'}/{g} =(|k\eta|)^\alpha$, which yields 
\beq
\P_{\zeta}(t,k) = \frac{4}{9} \mathcal{C}^2 \P_\chi (t,k).
\label{ps:pschi}
\eeq

\noindent The spectral index is given by \cite{lyth-curvatonI} as
\beq
n_s-1 = 2\frac{\dot H}{H^2} = 2\alpha ,
\eeq

\noindent where  Eq.~\eqref{alpha:chi} implies
\beq
\alpha \sim \frac{m_{\delta \chi}^2}{3 H_*^2} \ll 1.
\label{alpha_2}
\eeq

\noindent If $\alpha$ is not too small, the spectrum can be extremely
blue and this is ruled out by observations \citep{wmap2008-params}. 
Differentiating $N$  in Eq.~\eqref{N_curvaton} once more yields
\beq
N_{,\chi_*\chi_*} = N_{,\chi_*}^2 \left[\frac{3}{2\mathcal{C}} \left(
1+\frac{g g''}{g'^2}\right) -2 -\mathcal{C} \right] .
\eeq
If we neglect the non-Gaussianity of the field fluctuations at
Hubble crossing, which are subdominant with respect to the ones
accumulated during the super-Hubble evolution, and the
definition \eqref{fnl:def} gives

\beq
f_{\rm NL} = \frac{5}{4\mathcal{C}} \left(1+ \frac{ g g''}{g'^2} \right)
-\frac{5}{3} - \frac{5\mathcal{C}}{6}.
\eeq

\noindent We have arrived to a well known result obtained without assuming a
dominant contribution of $W^{\prime\prime\prime}$ in the perturbation
equations \citep{italians03,rodriguez-ngII,sasaki-curvaton}. 

By comparing the above equation with the nonlinear evolution given
by Eqs.~\eqref{nonlinear_relation} and
\eqref{nonlinear_after_infl}, and stopping the nonlinear evolution
of perturbations when the field starts oscillating at $t \simeq
1/m_{\delta \chi}$, the non-Gaussianity in the curvature
perturbation becomes
\beq
f_{\rm NL} = \frac{5}{4\mathcal{C}} \left[ 1+ \frac{ \chi_{0~\rm in}
W^{(3)}}{m_{\delta \chi}^2}\left( \frac{m_{\delta \chi}^2}{3
H^2_*} f(\eta_{\rm reh}/\eta_*) - \frac{1}{5} \right)\right]
-\frac{5}{3} - \frac{5\mathcal{C}}{6},
\label{curvaton:fnl}
\eeq
where the function $f$ has been defined in Eq.~\eqref{f_def}.
At late times, $f$ can be approximated by
\beq
 f(\eta_{\rm reh}/\eta_*) \simeq \left\{ \begin{array}{ll} -\Delta N & \mbox{for }~ \alpha 
\Delta N
 \lsim
1
, \\
 - \frac{1}{\alpha} e^{\alpha \Delta N } & \mbox{for }~ \alpha \Delta N \gsim 1 ,\end{array}
\right.  \label{f_def_2}
\eeq
where we have used Eq.~\eqref{alpha_2} and $\Delta N = N_{\rm reh}-N_* \simeq 60$ is the number of e-folds between Hubble crossing and the end of
inflation.

Note that our result, Eq.~\eqref{curvaton:fnl}, can also be obtained
by computing the  field bispectrum from the isocurvature field
evolution during radiation domination
Eq.~\eqref{nonlinear_after_infl}. In this context, the field
bispectrum is given by 
\beq
F(t; k_1, k_2,k_3) = F(t_{\rm reh} ; k_1, k_2,k_3) -
\frac{W^{(3)}}{5 m_{\delta \chi}^2}  \sum_{i<j} P(\eta,k_i)
P(\eta,k_j) ,
 \label{ls_rde}
\eeq

\noindent which clearly evolves with time before the decay of the curvaton. This non-Gaussianity is transferred to the curvature correlation by expanding the perturbed number of e-folds $\delta N$ on an initial flat slice taken at the onset
of the oscillations $(t=t_{\rm in})$. Then we can write $\zeta =
\delta N$ in terms of the field fluctuations $\delta \chi(t_{\rm in})$
as in Eq.~\eqref{non-lin:zeta}. With the aid of Eqs.~\eqref{ls_bi} and
\eqref{ls_rde}, we then replace the three-point correlators of the
field fluctuations in the curvature perturbation bispectrum. Our
result  in Eq.~\eqref{curvaton:fnl} is thus recovered.

\section{Model discrimination through observations}

In its simplest version, the curvaton model proposes an isocurvature field whose quantum fluctuations reproduce the spectrum of curvature fluctuations we observe in the CMB. Fixing  the required perturbation amplitude and spectral index imposes  important restrictions on the possible values of $\chi$. Additionally, the non-Gaussianity of the curvaton could constrain the parameter space further.

As an example, let us consider a curvaton with bare mass $m_{\chi} \ll
H_*$ and nonlinear potential $W=\frac{M}{3!}\chi^3$, such that
$m_{\delta \chi}^2 \simeq W''= \chi_{0~*} M$ and $W^{\prime\prime\prime} = M$. Note that the negligible mass in this case allows for a scale-invariant power spectrum in the field fluctuations and consequently the curvature perturbations.
For simplicity we take $\chi_{0~\rm osc} \simeq \chi_{0~*}$,
which is consistent with neglecting the nonlinear potential in Eq.~\eqref{background_KG}), since this would only contribute a term of order ${\cal O}(\alpha^2)$. When the nonlinear coupling
is very small, Eq.~\eqref{curvaton:fnl} reduces to
\beq
f_{\rm NL} = \frac{1}{\mathcal{C}} \left(1- \frac{5}{4}\alpha \Delta N \right) -\frac{5}{3} -
\frac{5\mathcal{C}}{6}, \qquad  \alpha \lsim \Delta N^{-1},
\eeq
and the intrinsic non-Gaussianity of the curvaton field gives a
negligible contribution to the total non-Gaussianity in the curvature
perturbation. However, when the nonlinear coupling is important, the
nonlinear parameter is
\beq
f_{\rm NL} = \frac{1}{\mathcal{C}} \left(1- \frac{5}{4} \exp
\left(\alpha \Delta N \right) \right) -\frac{5}{3} -
\frac{5\mathcal{C}}{6}, \qquad \Delta N^{-1} \lsim \alpha.
\label{large:ng}
\eeq
In this case the intrinsic non-Gaussianity of the curvaton can be the
main source of non-Gaussianity in the curvature perturbation. For
example, with $\frac{\chi_{0~*} M}{3H^2} \simeq 0.07$ and $\mathcal{C}=1$, one
finds $f_{\rm NL} \simeq - 100$, which is within reach of current
and future experiments. However, in this case, if
the curvaton is the only field responsible for the curvature perturbations,
the spectral index of scalar fluctuations will be largely blue. This
is in disagreement with current observations and is therefore excluded.

Let us finally consider the special case in which the curvaton is not
responsible for the linear fluctuations observed by CMB and
large-scale structure probes. Using the same potential as in the
example above, this case constrains the fraction $r$ in
Eq.~\eqref{def:r} to be small, as the contribution to the curvature
power spectrum is controlled by this parameter (see
Eq.~\eqref{ps:pschi}). On the other hand, $\alpha$ can take large
values without violating constraints on the power spectrum, which is
dominated by the inflaton perturbations. As for second order
perturbations, $\fnl$ is still given by the formula \eqref{large:ng}
and can be large due to the freedom in the mass and the small value of
$\mathcal{C}$. This happens at all scales and the non-Gaussianity is induced
through the evolution of fluctuations on superhorizon scales.   

It is important to note that, in this case, the blue spectrum of the
curvaton perturbations may dominate $\mathcal{P}_{\zeta}$ at small
scales. If this happens, and if the perturbations have enough power on
small scales, a significant amount of PBHs would be produced.  
In this special case, as in many other versions of inflation, an
important constraint on the model  comes from the probability of PBH
formation as we will study in the following chapters.

\clearemptydoublepage
\chapter{Statistics of non-Gaussian fluctuations}
\label{chapterthree}

\section{Introduction}
\label{sec:intro}

In the inflationary paradigm, the prediction that the spectrum of
fluctuations should exhibit Gaussian statistics has recently been
challenged. This prediction follows from the fact that the curvature
perturbation, which commonly  refers to the comoving
curvature perturbation defined in Eq.~\eqref{def:R},  is treated as a
free field during inflation, 
\begin{equation}
  \label{intro:freer}
  \R(t,\vect{x}) = \int \frac{\d^3 k}{(2\pi)^3} \; \R(t,\vect{k}) \e{\imag
  \vect{k}\cdot\vect{x}} ,
\end{equation}
where there is no coupling between the $\R(t,\vect{k})$ for different
$\vect{k}$.
With this understanding, Eq.~\eqref{intro:freer}
means that $\R$ does not interact with either  itself or any other particle
species in the universe.
The real-space field
$\R(t,\vect{x})$ is obtained by summing an infinite number of independent,
identically distributed, uncorrelated oscillators.
Under these circumstances the
Gaussianity of $\R(t,\vect{x})$ follows from the central limit theorem
\citep{bardeen-bond},
given reasonable assumptions about the individual distributions of the
$\R(t,\vect{k})$. The exact form of the distributions
of the $\R(t,\vect{k})$ is
mostly irrelevant for the inflationary density
perturbations.

In conventional quantum field theory, all details
of $\R$ and its interactions are encoded in the $n$-point correlation functions
of $\R$, written as
$\outbra \R(t_1,\vect{x}_1)  \cdots  \R(t_n,\vect{x}_n) \inket$.
Working in the Heisenberg picture, where the operators carry time dependence but
the states \\$\{ \inket$, $\outket \}$ do not, 
these functions express the amplitude for the early-time
vacuum $\inket$ to evolve into the late-time vacuum $\outket$ in the
presence of the fields $\R(t_i,\vect{x}_i)$.
Given the $n$-point functions for all $n$ at arbitrary $\vect{x}$ and $t$,
one can determine $\R(t,\vect{x})$ \citep{streater-wightman}, at least in
scattering theory.
In the context of the inflationary density perturbations, these vacuum evolution
amplitudes are not directly relevant. Instead, one is interested in the
equal time expectation values
$\inbra \R(t,\vect{x}_1) \cdots \R(t,\vect{x}_n) \inket$, which can be
used to measure gravitational particle creation
out of the time-independent early vacuum $\inket$
during inflation. These expectation values are calculated using
the so-called `closed-time-path formalism',
which was introduced by  \cite{schwinger-ctp}; see also
\cite{calzetta-hu,jordan,dewitt} and \cite{hajicek}.
In this formalism there is a doubling of degrees of freedom, which is also
manifest in finite temperature calculations \citep{lebellac,rivers}.
This method has recently been used
\citep{weinberg-qcI,weinberg-qcII,sloth,seery-loops} 
to extend the computation of the correlation functions of $\R$ to beyond
tree-level.

Knowledge of the expectation values of $\R$ in the state $\inket$ is sufficient
to predict a large number of
cosmological observables, including the power spectrum of the
density perturbations generated during inflation
\citep{guth-pi,hawking-fluct}, and the two- and three-point
functions of the CMB temperature anisotropies
\citep{hu-sugiyama,hu-trispectrum,komatsu-spergel,kogo-komatsu,
okamoto-hu,babich-creminelli, babich-zaldarriaga,babich,liguori-hansen,
cabella-hansen,creminelli-nicolis}. Because they are
defined as expectation values in the quantum vacuum, these observables all
have the interpretation of ensemble averages, as will be discussed in more
detail below.

On the other hand, one sometimes needs to know the probability that
fluctuations of some given magnitude occur in the curvature perturbation $\R$
\citep{press-schechter,bardeen-bond,peacock-heavens}.
This is not a question about ensemble averages, but about the probability measure on the ensemble itself.
As a result, such information cannot easily be
obtained from inspection or simple manipulation of the $n$-point functions.

For example, if we know by some a priori means that $\R$ is free,
then the argument given above, based on the
central limit theorem, implies that at any position $\vect{x}$, the probability density
of fluctuations in $\R$ of amplitude $\fluct$ must be
\begin{equation}
  \label{intro:gaussian}
  \Prob({\fluct})
  \simeq \frac{1}{\sqrt{2\pi} \sigma}
  \exp \left( - \frac{\fluct^2}{2\sigma^2} \right) ,
\end{equation}
where the variance in $\R$ is
\begin{equation}
  \label{intro:variance}
  \sigma^2 = \langle \R(t,\vect{x})^2 \rangle =
  \int \d \ln k \; \ps_{\R}(k) .
\end{equation}
The quantity $\ps_{\R}(k)$ is the dimensionless power spectrum,
which is defined in terms of the two-point function of $\R$,
calculated from the quantum field theory in-vacuum:
\begin{equation}
  \label{intro:twopt}
  \inbra \R(t,\vect{k}_1) \R(t,\vect{k}_2) \inket
  = (2\pi)^3\diracd(\vect{k}_1 + \vect{k}_2)\frac{2\pi^2}{k_1^3} \ps_{\R}(k_1) .
\end{equation}
This is the only relevant observable, because
it is a standard property of free fields that
all other non-vanishing correlation functions can be
expressed in terms of the two-point function \eqref{intro:twopt}, and hence
the power spectrum. Those expressions can be achieved through a
generalisation of Wick's theorem using an equal-time normal-ordering
\citep{luo93}. 
In practice, in order to give a precise meaning to \eqref{intro:gaussian},
it would be necessary to specify what it means for $\R$ to develop
fluctuations of amplitude $\fluct$, and whether it is the fluctuations
in the microphysical field $\R$ or some smoothed field $\bar{\R}$
which are measured. These details affect the exact expression
\eqref{intro:variance} for the variance of $\fluct$.

The average in Eq.~\eqref{intro:twopt},
denoted by $\inbra \cdots \inket$, is the expectation value in the
quantum in-vacuum. To relate this abstract expectation value
to real-world measurement probabilities, one introduces a notional ensemble of
possible universes, of which the present universe and the density fluctuations
that we observe are only one possible realisation 
(e.g.,\cite{lyth-curvaton-ng}). 
However, for ergodic processes, we may freely trade ensemble averages
for volume averages. The ergodicity of a system refers to that
property of processes by which the average value of a process characteristic
measured over time is the same as the average value measured over the
ensemble.   

If we make the common supposition that the inflationary density
perturbation is indeed ergodic, then we expect the volume average of
the density fluctuation to behave like the ensemble average: the
universe may contain regions where the fluctuation is atypical, but
with high probability most regions contain fluctuations with
root-mean-square amplitude close to $\sigma$. Therefore the
probability distribution on the ensemble, which is encoded in
Eq.~\eqref{intro:twopt}, translates to a probability distribution on
smoothed regions of a determined size within our own universe.  

In order to apply the above analysis, it is necessary to know in advance that
$\R$ is a free field. This knowledge allows us to use the
central limit theorem to connect the correlation functions of
$\R$ with the probability distribution \eqref{intro:gaussian}.
The situation in the real universe is not so simple. In particular,
the assumption that during inflation $\R$ behaves as a free field, and
therefore that the oscillators $\R(\vect{k})$ are uncorrelated and
independently distributed, is only approximately correct.
In fact, $\R$ is subject to
self-interactions and interactions with the other constituents of the universe,
which mix $\vect{k}$-modes. Consequently, the oscillators
$\R(\vect{k})$ acquire some phase correlation and are no longer independently
distributed.
In this situation the central limit theorem gives
only approximate information concerning the
probability distribution of $\R(\vect{x})$, and it is necessary to
use a different method to connect the correlation functions of $\R$ with
its probability distribution function (PDF).

In this chapter we give a new derivation of the PDF
of the amplitude of fluctuations in $\R$ which directly connects
$\Prob(\fluct)$ and the correlation
functions $\langle \R(\vect{k}_1) \cdots \R(\vect{k}_n) \rangle$, without
intermediate steps which invoke the central limit theorem or other
statistical results.
When the inflaton is treated as a free field, our method
reproduces the familiar prediction \eqref{intro:gaussian} of Gaussian statistics.
When the inflaton is `{not}' treated as a free field,
the very significant advantage of our
technique is that it is possible to directly calculate the corrections to
$\Prob(\fluct)$. Specifically,
the interactions of $\R$ can be measured by the departure of the correlation
functions from the form they would take if $\R$ were free.
Therefore, the first corrections to the free-field approximation
are contained in the three-point function, which is exactly zero when there
are no interactions.

The three-point function for single-field, slow-roll inflation
has been calculated by  \cite{maldacena},
whose result can be expressed in the form \citep{seery-lidsey-a}
\begin{equation}
  \label{intro:threept}
  \langle \R(\vect{k}_1) \R(\vect{k}_2) \R(\vect{k}_3) \rangle =
  4 \pi^4 (2\pi)^3 \diracd\big(\sum_i \vect{k}_i\big) \frac{\bar{\ps_{\R}}^2}
  {\prod_j k_j^3} \A(k_1,k_2,k_3) ,
\end{equation}

\noindent where $\A$ is Maldacena's $\A$-function divided by two
\citep{maldacena} \footnote[5]{In Maldacena's normalisation, the
  numerical prefactor in Eq.~\eqref{intro:threept} is not consistent
  with the square of the two-point function,
  Eq.~\eqref{intro:twopt}. We choose $\A$ so that the prefactor
  becomes $4\pi^2(2\pi)^3$. This normalisation of
  Eq.~\eqref{intro:threept} was also employed by
  \cite{seery-lidsey,seery-lidsey-a}, although the distinction from
  Maldacena's $\A$ was not pointed out explicitly.}. 
$\bar{\ps_{\R}}^2$ measures the amplitude of the spectrum when the
$\vect{k}_i$ crossed the horizon. (For earlier work on the derivation
of the three-point-function, see \cite{falk-rangarajan,gangui-lucchin,
  pyne-carroll,acquaviva-bartolo}.)
This result has since been extended to cover the non-Gaussianity produced
during slow-roll inflation with an arbitrary number of fields
\citep{maldacena,seery-lidsey-a,
creminelli,rodriguez-ngI,rodriguez-ngII,lyth-zaballa,
zaballa-rodriguez,vernizzi-wands},
preheating \citep{enqvist-jokinen,enqvist-jokinen-a,jokinen-mazumdar},
models where the dominant non-Gaussianity is produced by a light scalar which
is a spectator during inflation \citep{boubekeur-lyth,alabidi-lyth,
lyth-curvaton-ng},
and alternative models involving a small speed of sound for the
inflaton perturbation
\citep{seery-lidsey,
alishahiha-silverstein,calcagni-nongaussian,arkani-hamed-creminelli,
creminelli}.

For single-field, slow-roll inflation,
the self-interactions of $\R$ are suppressed by
powers of the slow-roll parameters. This means that the correction to Gaussian
statistics is not large. In terms of the $\A$-parametrised three-point function
\eqref{intro:threept}, this is most commonly expressed by writing,
in an equivalent form to $\zeta$ in Eq.~\eqref{fnl:real}
\beq
   \R(t,\bx)  = \R_1(t,\bx) - \frac{3}{5} \fnl
   \left(\R_1^2(t,\bx) - \langle\R^2_{1}(t,x)\rangle\right),
\eeq

\noindent with
\begin{equation}
  \fnl = - \frac{5}{6} \frac{\A}{\sum_i k_i^3} = \mathcal{O}(\epsilonsr,\etasr)
\end{equation}

\noindent  giving the relative contribution of the non-Gaussian piece
in $\R$    and $\R_1$ being a Gaussian random field
\citep{komatsu-spergel,verde-wang}. (Note that there are differing
sign conventions for $\fnl$ \citep{malik-lyth06}, here we stick to
that used by the WMAP team.) In models with more
degrees of freedom, much larger non-Gaussianities are expected,
perhaps with $\fnl \sim 10$
\citep{rigopoulos-shellard,rigopoulos-shellard-vantent,rigopoulos-shellard-vantent-b,rigopoulos-shellard-vantent-c,boubekeur-lyth,rodriguez-ngII,vernizzi-wands}.    
If the inflationary perturbation has a speed of sound different from
unity, then large non-Gaussianities may also appear
(e.g. \cite{seery-lidsey,loverde-shandera}), although in this case it
is difficult to simultaneously achieve scale invariance. 
The current observational constraint, as we mentioned in Chapter
\ref{chaptertwo}, is of order $|\fnl|\lesssim 100 $. In the absence of
a detection, the forthcoming PLANCK mission may tighten this
constraint to $|\fnl| \lesssim 3$
\citep{komatsu-spergel,liguori-hansen}. 

Non-Gaussian PDFs have been studied previously by several authors.
The closest analysis to the method developed in this chapter comes
from \cite{verde-ngII}, who worked with a path integral expression 
for the density fluctuation smoothed on a scale $R$
(which they denoted by `$\delta_R$'). Also, the analysis of  
\cite{bernardeau-mf,bernardeau-ng} has some features in common with our own,
being based on the cumulant generating function. Moreover, the
expression for the probability density in those papers is expressed as
a Laplace transform. Our final expression, Eq.~\eqref{hubble:prob},
can be interpreted as a Fourier integral, viz \eqref{genfunc:prob},
which (loosely speaking) can be related to a Laplace integral via a
Wick transformation. 
Despite these similarities, the correspondence between the two
analyses is complicated because \cite{bernardeau-mf,bernardeau-ng}
work in a multiple-field picture and calculate a probability density
only for the isocurvature field `$\delta s$', which acquires its
non-Gaussianity via a mixing of isocurvature and adiabatic modes long
after horizon exit, of which particular cases were presented in the
Chapter \ref{chaptertwo}. This contrasts with the situation in the present
chapter, where we restrict ourselves to a single-field scenario and
compute the PDF for the adiabatic mode $\R$. This 
would be orthogonal to $\delta s$ in field space and its non-Gaussianity
is generated exactly at horizon exit.

In the older literature it is more common to deal with the density
fluctuation $\delta_\rho$ measured on comoving slices, rather than
the curvature perturbation $\R$.
For slowly varying fields, on scales larger than the horizon,
$\R$ and $\delta_\rho$ can be related in the comoving gauge via 
Eq.~(25) of \citep{lyth-classicality}:
\begin{equation}
  \left( \frac{aH}{k} \right)^2\delta_{\rho} =
  - \left( \frac{3}{2} + \frac{1}{1+ w} \right)^{-1} \R ,
\end{equation}

\noindent to first order in cosmological
perturbation theory for a barotropic fluid.
(One may use the $\delta N$ formalism to go beyond leading order as in
Chapter \ref{chaptertwo}, but to obtain results valid on sub-horizon
scales one must use the full Einstein equations; see, e.g.,
\cite{langlois-zeta,langlois-vernizzi-b}.)
For fluctuations on the Hubble scale $(k \simeq aH)$, this means $|\R|
\simeq \delta_\rho$, so $\R$ provides a useful measure of the density
fluctuation on such scales. By virtue of this relationship with the
density fluctuation, the probability distribution $\Prob(\fluct)$ is
an important theoretical tool, especially in studies of structure
formation. For example, it is the principal object in the
Press--Schechter formalism \citep{press-schechter}. As a result, there
are important reasons why knowledge of the detailed form of the
PDF of $\fluct$, and not merely the approximate
answer provided by the central limit theorem,  is important. 

Firstly, large amplitude collapsed objects,
such as primordial black
holes (PBHs) naturally form in the high-$\fluct$ tail of the distribution
\citep{carr,carr-hawking}.
Such large fluctuations are extremely rare. This means that a small change in
the probability density for $|\fluct| \gg 0$ can make a large difference in the
mass fraction of the universe which collapses into PBHs
\citep{bullock-primack,ivanov}. Thus
one may hope to probe the form of the PDF for $\fluct$ using well-known and
extremely stringent constraints on PBH formation in the early universe
\citep{carr-constraints,carr-lidsey-constraints,green-constraints,zaballa-green,josan09}. 
The corrections calculated in this chapter are
therefore not merely of theoretical interest, but relate directly to
observations, and have the potential to sharply discriminate between models
of inflation.

Secondly, as described above, although the non-Gaussianities produced
by single-field, slow-roll inflation are small, this is not mandatory.
In models where non-Gaussianities are large,
it will be very important to account for the effect of non-Gaussian fluctuations
on structure formation
\citep{verde-wang,verde-ngII,verde-jimenez,verde-trispectrum}.
The formalism presented in this chapter provides
a systematic way to obtain such predictions, extending the analysis given by
\cite{verde-ngII}.

The outline of this chapter is as follows.
In Section~\ref{sec:measure} we obtain the probability measure on the ensemble
of possible fluctuations. This step depends on the correlation functions
of $\R$. In Section~\ref{sec:harmoniccurv}, we discuss
the decomposition of $\R$ into harmonics. This is a technical step, which is
necessary in order to write down a path integral for $\Prob(\fluct)$.
First, we Fourier decompose $\R$. Then we write the path integral
measure, and
finally we give a precise specification of $\fluct$, which measures the size of fluctuations.
We distinguish two interesting cases:
a `total fluctuation' $\fluct$,
which corresponds to $\R$ (or approximately $\delta_\rho$) smoothed over
regions the size of the Hubble volume; and the `spectrum' $\spect(k)$,
which describes the contributions to $\fluct$ from regions of the
primordial power spectrum around the scale described by wavenumber $k$.
In Section~\ref{sec:probdelta} we
evaluate $\Prob(\fluct)$. We give the calculation for the Gaussian case first,
in order to clearly explain our method with
a minimum of technical detail. This is
followed by the same calculation
but including non-Gaussian corrections which follow from a non-zero
three-point function.
In Section~\ref{sec:probrho} we calculate $\Prob[\spect(k)]$.
Finally, we summarise our results  in Section~\ref{sec:conclude}.

\section{The probability measure on the ensemble of $\R$}
\label{sec:measure}
Our method is to compute the probability measure $\Prob_t[\Rsp]$ on the
ensemble of realisations of the curvature perturbation $\Rsp(\vect{x})$,
which we define to be the value of $\R(t,\vect{x})$ at some fixed time $t$.
This
probability measure is a natural object in the Schr\"{o}dinger approach to
quantum field theory, where the elementary quantity is the wavefunctional
$\Psi_t[\Rsp]$, which is
related to $\Prob_t[\Rsp]$ by the usual rule
of quantum mechanics, that $\Prob_t[\Rsp] \propto |\Psi_t[\Rsp]|^2$. Once
the measure $\Prob_t[\Rsp]$ is known, we can directly calculate
(for example) $\Prob_t(\fluct)$
by integrating over all $\Rsp$ that produce fluctuations of amplitude
$\fluct$. Although the concept of a probability
measure on $\Rsp$ may seem rather formal, the Schr\"{o}dinger
representation of quantum field theory is entirely equivalent to the more familiar formulation in terms of a Fock space.
This representation is briefly discussed, for example, by \cite{polchinski} and \cite{visser}. A brief introduction to infinite-dimensional probability measures is given by \cite{albeverio}.
 Indeed, a similar procedure has been discussed by
 \cite{ivanov}, who calculated the probability measure on a stochastic
metric variable $a_{\mathrm{ls}}(\vect{x})$ which can be related to our
$\R(\vect{x})$. Although the approaches are conceptually
similar, our method is substantially
different in detail. In particular, the present
calculation is exact in the sense that we make no reference to the stochastic
approach to inflation, and therefore are not obliged to introduce a
coarse-graining approximation. Moreover, Ivanov's analysis appeared before
the complete non-Gaussianity arising from $\R$-field interactions around
the time of horizon crossing had been calculated \citep{maldacena},
and therefore did not include this effect.

\subsection{The generating functional of correlation functions}

The expectation values $\langle \R(\vect{x}_1) \cdots \R(\vect{x}_2)
\rangle$ in the vacuum $\inket$
at some fixed time $t$ can be expressed in terms of a
Schwinger--Keldysh path integral,%
\footnote[6]{Henceforth, we use the notation $\langle \cdots \rangle$ to
mean expectation values in the in-vacuum, and no longer write $\inket$
explicitly where this is unambiguous.}
\begin{align}
  \langle \R(t,\vect{x}_1) \cdots \R(t,\vect{x}_n) \rangle &= \notag\\
  \int [\d\R_- \d\R_+]_{\inket}^{\R_{+}(t,\vect{x}) = \R_-(t,\vect{x})}&
  \R(t,\vect{x}_1) \cdots \R(t,\vect{x}_n)
  \exp\left( \imag~\hat{\action}^{(n)}[\R_+] - \imag~\hat{\action}^{(n)}[\R_-] \right) .
    \label{genfunc:sk}
\end{align}

\noindent Here $[d\R_+]$ is the integrand of the path integral
over $\R$. 
  In cosmology we are generally interested in $\R$ evaluated
at different spatial positions on the same $t$-slice, so we have set
all the $t$ equal in \eqref{genfunc:sk}.
The path integral is taken over all fields $\R$ which begin in a
configuration corresponding to the vacuum $\inket$ at past
infinity. The correlator is equal to the expectation value of three
copies of $\R$ at time $t$ so we require two path integrals: the first
integral $[d\R_+]$ 
evolves the vacuum state $\inket$ from past infinity to the state
$\R_+(t,\vect{x})$ at time $t$ where the $n$ copies of the field $\R$ are
averaged, and a second path integral $[d\R_-]$ which will project back the
average  to the vacuum state through a second functional integral
. $\hat{\action}^{(n)}[\R]$ is the action for the fluctuation $\R$, which is
computed perturbatively to order $n$ in $\R$ when we want to compute
correlations of the same order. For example, $\hat{\action}^{(3)}[\R]$ is given
to third order in $\R$ by \cite{maldacena} in the context of slow-roll
inflation and by
\cite{seery-lidsey-a,seery-lidsey} in the inflationary models where the kinetic
energy is not negligible. The action $\hat{\action}^{(n)}[R]$
is time ordered for the argument $\R_-$ and anti-time ordered for
$\R_+$. (For details of the Schwinger--Keldysh or `closed time path'
formalism, see
\cite{calzetta-hu,jordan,weinberg-qcI,lebellac,hajicek,rivers}.)   

An expression equivalent to Eq.~\eqref{genfunc:sk} can be given in terms of the
`equal time' generating functional
\begin{align}
  \label{genfunc:def}
  Z_t[q] = \int& [\d\Rsp] \int [\d\R_-
    \d\R_+]_{\inket}^{\R_{\pm}(t,\vect{x}) = \Rsp(\vect{x})} \\
  &\exp\left( \imag~\hat{\action}^{(n)}[\R_+] -
  \imag~\hat{\action}^{(n)}[\R_-] + 
  \imag \int_{\Sigma_t} \d^3 x \; \Rsp(\vect{x})q(\vect{x}) \right),
  \notag
\end{align}

\noindent where $q$ is some arbitrary source field, also known in the
theory of special functions as the
formal argument of the generating function. $\Sigma_t$ is a
spatial slice at coordinate time $t$. 
The equal-time correlation functions $\langle \R(t,\vect{x}_1) \cdots
\R(t,\vect{x}_n) \rangle$ are recovered from $Z_t[q]$ by functional
differentiation, 
\begin{equation}
  \label{genfunc:corrls}
  \langle \R(t,\vect{x}_1) \cdots \R(t,\vect{x}_n) \rangle =
  \left.\frac{1}{\imag^n} \frac{\delta}{\delta q(\vect{x}_1)} \cdots
  \frac{\delta}{\delta q(\vect{x}_n)} \ln Z_t[q] \right|_{q = 0} .
\end{equation}

\noindent Up to normalisation,
this is merely the rule for functional Taylor coefficients, so it
is straightforward to invert Eq.~\eqref{genfunc:corrls} for $Z_t[q]$.
We obtain
\begin{equation}
  \label{genfunc:reconstruct}
  Z_t[q] = \exp\Big\{ \sum_{n=0}^\infty \frac{\imag^n}{n!}
  \int \cdots \int \d^3 x_1 \cdots \d^3 x_n \; q(\vect{x}_1)
  \cdots q(\vect{x}_n) \langle \R(t,\vect{x}_1) \cdots \R(t,\vect{x}_n)
  \rangle\Big\}.
\end{equation}

Eq.~\eqref{genfunc:def} for the generating functional
can be rewritten in a suggestive way. We define
the wavefunctional at time $t$ as
\begin{equation}
  \Psi_t[\Rsp] = \int [\d\R]_{\inket}^{\R(t,\vect{x}) = \Rsp(\vect{x})}
  \exp\left( \imag~\hat{\action}^{(n)}[\R] \right) . 
  \label{wave-func:R}
\end{equation}

\noindent This definition is simply the functional generalisation of the
familiar quantum-mechanical
wavefunction. It expresses the amplitude for the field $\R(t,\vect{x})$ to have
the spatial configuration $\Rsp(\vect{x})$ at time $t$, given the
boundary condition that $\R$ started in the vacuum state in the far past.
In terms of $\Psi_t[\Rsp]$, the generating functional can be rewritten as
\begin{equation}
  \label{genfunc:genfunctional}
  Z_t[q] = \int [\d \Rsp] \;  \Psi_t[\Rsp]^\dag \Psi_t[\Rsp]
  \exp\left( \imag \int \d^3 x \; \Rsp(\vect{x}) q(\vect{x}) \right)
  = \widetilde{|\Psi_t[\Rsp]|^2} \propto \widetilde{\Prob[\Rsp]} ,
\end{equation}

\noindent where a tilde denotes a (functional) Fourier transform, and $\dag$
denotes Hermitian conjugation.
Eq.~\eqref{genfunc:genfunctional} implies that $Z_t[q]$ is the complementary
function for the probability
distribution $\Prob_t[\Rsp]$ \citep{albeverio}, which can formally be obtained by inversion of $Z_t[q]$. Hence, up to an overall normalisation,
\begin{equation}
  \label{genfunc:prob}
  \Prob_t[\Rsp] \propto \int [\d q] \; \exp\left( - \imag \int \d^3 x \;
  \Rsp(\vect{x})q(\vect{x}) \right) Z_t[q] .
\end{equation}

\noindent The normalisation is not determined by this procedure.
We will fix the $\Rsp$-independent prefactor, which correctly normalises
the PDF, by requiring $\int \d \fluct \, \Prob(\fluct) =1$ at the end
of the calculation. 
For this reason, we systematically drop all field-independent prefactors
in the calculation that follows.

\subsection{The probability density on the ensemble}
So far, all our considerations have been exact, and apply for any
quantum field $\R(t,\vect{x})$. For any such field,
Eq.~\eqref{genfunc:prob} gives the probability density for a spatial
configuration $\Rsp$ at time $t$, and implies that to obtain $\Prob_t[\Rsp]$
we need to know {all} such functions for all $n$-point
correlations and at all spatial positions $\vect{x}$.
In practice, some simplifications occur when $\R$ is identified as the
inflationary curvature perturbation.

The most important simplification is the possibility of a perturbative
evaluation.
The dominant mode of the CMB fluctuation is constrained to be Gaussian
to high accuracy, so the non-Gaussian corrections to the leading order
cannot be large. Moreover, the amplitude of its spectrum is constrained by CMB observations.
Specifically, as mentioned in Section \ref{intro-inflation}, in the range of wavenumbers probed by the CMB,  the spectrum has amplitude
$\ps_{\R}^{1/2} \sim 10^{-5}$. Therefore, 
each higher-order correlation function is suppressed by an
increasing number of copies of
the spectrum, $\ps_{\R}(k)$, as we have shown in Chapter \ref{chaptertwo}. 

Provided the amplitude of $\ps_{\R}$ is small, it might seem
reasonable to truncate the exponential in Eq.~\eqref{genfunc:prob} for a 
given $n$ and to work with a perturbation series in $\ps$.
However, this simple approach is too na\"{\i}ve, because the integrals over
$q$ eventually make any given term in the series large, and this invalidates
simple perturbative arguments based on power-counting in
$\ps_{\R}$. The perturbation series can only be justified \emph{a
  posteriori}, a point to which we will return in
Section~\ref{sec:nongaussian}. 

We work to first-order in the three-point correlation, that is, we
consider non-vanishing two and three-point correlators in
Eq.~\eqref{genfunc:reconstruct},
\begin{align}
  Z_t[q] = \exp\Big\{-\frac{1}{2}& 
  \int \int \d^3 x_1 \d^3 x_2 \; q(\vect{x}_1)
  q(\vect{x}_2) \langle \R(t,\vect{x}_1) \R(t,\vect{x}_2)
  \label{genfunc:short}\\ 
  - \frac{\imag}{6}&  \int \int \int \d^3 x_1 \d^3 x_2 \d^3 x_3
  \; q(\vect{x}_1)  q(\vect{x}_2)   q(\vect{x}_3) \langle
  \R(t,\vect{x}_1) \R(t,\vect{x}_2) \R(t,\vect{x}_3) \rangle \Big\}.
  \notag
\end{align}

\noindent This generating functional is introduced in the expression
derived for the probability, Eq.~\eqref{genfunc:prob}, for which we
expand the exponential third order term in $q(\vect{x})$ in power
series to lowest order. We
finally arrive to the product 
\begin{equation}
  \label{invert:a}
  \Prob_t[\Rsp] \propto \int [\d q] \;
  \correction_t[q] \pregauss_t[q;\Rsp] ,
\end{equation}
where $\correction[q]$ and $\pregauss[q;\Rsp]$ are defined by
\begin{equation}
  \correction_t[q] =
  \left(1 - \frac{\imag}{6}
  \int \frac{\d^3 k_1 \, \d^3 k_2 \, \d^3 k_3}{(2\pi)^9} \;
  q(\vect{k}_1) q(\vect{k}_2) q(\vect{k}_3)
  \langle \R(t,\vect{k}_1) \R(t,\vect{k}_2) \R(t,\vect{k}_3) \rangle \right) ,
\end{equation}

\noindent and
\begin{equation}
  \pregauss_t[q;\Rsp] =
  \exp\left(- \int \frac{\d^3 k_1 \, \d^3 k_2}{(2\pi)^6} \;
  \frac{q(\vect{k}_1) q(\vect{k}_2)}{2}
  \langle \R(t,\vect{k}_1) \R(t,\vect{k}_2) \rangle - \imag \int
  \frac{\d^3 k}{(2\pi)^3} \;
  q(\vect{k}) \Rsp(\vect{k}) \right) .
  \label{pregauss:exp}
\end{equation}

\noindent The expression for $\pregauss_t$  gives rise to the Gaussian
part of the PDF. $\correction$ is of the form $1$ plus a
correction which is small when the perturbative analysis is valid.
Higher-order perturbative corrections in $\ps_{\R}$ can be accommodated
if desired by retaining higher-order terms in the power series
expansion of the exponential in \eqref{genfunc:prob}. Therefore our
method is not restricted to corrections arising from non-Gaussianities
described by three-point correlations, but can account for
non-Gaussianities which enter at any order in the correlations of
$\R$, limited only by the computational complexity. However, 
in this chapter, we work only with the three-point non-Gaussianity.

We now complete the square for $\pregauss_t[q;\Rsp]$ in
\eqref{pregauss:exp} and make the finite field redefinition 
\begin{equation}
  q(\vect{k}) \mapsto \hat{q}(\vect{k}) = q(\vect{k}) + (2\pi)^3
  \imag \frac{\Rsp(\vect{k})}{\langle \R(t,\vect{k}) \R(t,-\vect{k})
    \rangle'}\, , 
\end{equation}

\noindent where the prime in  $\langle \R(t,\vect{k}) \R(t,-\vect{k})
\rangle'$ indicates that the momentum-conservation $\delta$-function is omitted.
The measure $[\d q]$ is formally invariant under this
shift, giving $\int [\d q] = \int [\d \hat{q}]$,
whereas $\pregauss_t[q;\Rsp]$ can be split into an $\Rsp$-dependent piece,
which we call $\Gauss_t[\Rsp]$,
and a piece that depends only on $\hat{q}$ but not $\Rsp$,
\begin{equation}
  \pregauss_t[q;\Rsp] \mapsto
  \Gauss_t[\Rsp] \exp \left( - \frac{1}{2} \int \frac{\d^3 k_1 \, \d^3 k_2}
  {(2\pi)^6} \; \hat{q}(\vect{k}_1) \hat{q}(\vect{k}_2)
  \langle \R(t,\vect{k}_1) \R(t,\vect{k}_2) \rangle \right) ,
\end{equation}

\noindent where $ \Gauss_t[\Rsp]$ is a Gaussian in $\Rsp$,
\begin{equation}
  \Gauss_t[\Rsp] = \exp \left( - \frac{1}{2} \int \d^3 k_1 \, \d^3 k_2 \;
  \langle \R(t,\vect{k}_1) \R(t,\vect{k}_2) \rangle
  \frac{\Rsp(\vect{k}_1) \Rsp(\vect{k}_2)}{\prod_i
  \langle \R(t,\vect{k}_i) \R(t,-\vect{k}_i) \rangle'} \right) .
\end{equation}

\noindent Eq.~\eqref{invert:a} for the probability density becomes
\begin{equation}
  \label{invert:b}
  \Prob_t[\Rsp] \propto \Gauss_t[\Rsp] \int [\d \hat{q}] \;
  \correction_t[\hat{q}]
  \exp \left( - \frac{1}{2} \int \frac{\d^3 k_1 \, \d^3 k_2}
  {(2\pi)^6} \; \hat{q}(\vect{k}_1) \hat{q}(\vect{k}_2)
  \langle \R(t,\vect{k}_1) \R(t,\vect{k}_2) \rangle \right) ,
\end{equation}
One can easily verify that this is the correct expression, since if
we ignore the three-point contribution (thus setting $\correction_t = 1$),
one recovers (after applying a correct normalisation)
\begin{equation}
  \int [\d\Rsp] \; \Rsp(\vect{k}_1) \Rsp(\vect{k}_2)
  \Gauss_t[\Rsp] = \langle \R(t,\vect{k}_1) \R(t,\vect{k}_2) \rangle.
\end{equation}
The remaining task is to carry out the $\hat{q}$ integrations
in $\correction_t$. The only terms which contribute are those
containing an even power of $\hat{q}$, since any odd function
integrated against $\e{-\hat{q}^2}$ vanishes identically.
In the expansion of $\prod_i q(\vect{k}_i)$ in terms of $\hat{q}$,
there are two such terms: one which is quadratic in $\hat{q}$,
and one which is independent of $\hat{q}$. These are
accompanied by linear and cubic terms which do not contribute to
$\Prob_t[\Rsp]$. For any symmetric kernel $\mathsf{K}$ and vectors
$\vect{p}$, $\vect{q} \in \reals^m$, one has the general results \citep{rivers}
\begin{align}
  &\int [\d f] \; \exp \left( - \frac{1}{2} \int \d^m x \, \d^m y \;
  f(\vect{x}) f(\vect{y}) \mathsf{K}(\vect{x},\vect{y}) \right) =
  \left( \det \mathsf{K} \right)^{-1/2} , \\
  \int [\d f] & \; f(\vect{p}) f(\vect{q}) \exp \left( - \frac{1}{2}
  \int \d^m x \, \d^m y \; f(\vect{x}) f(\vect{y}) \mathsf
  {K}(\vect{x},\vect{y}) \right) =
  \mathsf{K}^{-1}(\vect{p},\vect{q}) \left( \det \mathsf{K}
  \right)^{-1/2} .
\end{align}
These rules allow us to evaluate the $\hat{q}$ integrals in
Eq.~\eqref{invert:b}, giving
\begin{equation}
  \label{invert:c}
  \Prob_t[\Rsp] \propto \Gauss_t[\Rsp] \left(1 + \correction_t^{(0)}[\Rsp]
  + \correction_t^{(2)}[\Rsp] \right) ,
\end{equation}
where 
\begin{align}
  \label{invert:d}
  \correction_t^{(0)}[\Rsp]
  = - \frac{1}{6} \int \d^3 k_1 \, \d^3 k_2 \, \d^3 k_3 \;
  &\langle \R(t,\vect{k}_1) \R(t,\vect{k}_2) \R(t,\vect{k}_3) \rangle
  \frac{\Rsp(\vect{k}_1) \Rsp(\vect{k}_2) \Rsp(\vect{k}_3)}
  {\prod_i\langle \R(t,\vect{k}_i) \R(t,-\vect{k}_i) \rangle'}\,, \\
  \correction_t^{(2)}[\Rsp]
  = \frac{1}{6} \int \d^3 k_1 \, \d^3 k_2 \, \d^3 k_3 \;
  &\langle \R(t,\vect{k}_1) \R(t,\vect{k}_2) \R(t,\vect{k}_3) \rangle\times
  \label{invert:e} \\
  \times &\frac{\Rsp(\vect{k}_1) \diracd(\vect{k}_2 + \vect{k}_3)}
  {\prod_{i \neq 3}\langle \R(t,\vect{k}_i) \R(t,-\vect{k}_i) \rangle'}
  + \perms. 
  \notag
\end{align}

\noindent In the last expression we include the possible permutations of
the labels $\{ 1, 2, 3 \}$ since these give rise to distinct integrands.

In fact, $\correction_t^{(2)}$ is negligible.
This happens because the three-point function
contains a momentum-conservation $\diracd$-function, $\diracd(\vect{k}_1 +
\vect{k}_2 + \vect{k}_3)$, which requires that the vectors $\vect{k}_i$
sum to zero in momentum space. [For this reason, it
is often known as the ``triangle condition'', and
we will usually abbreviate it schematically as
$\diracd(\triangle)$.] In combination with the $\diracd$-%
function, $\diracd(\vect{k}_2 + \vect{k}_3)$, the effect is to constrain
two of the momenta (in this example $\vect{k}_2$ and $\vect{k}_3$) to be
equal and opposite, and the other momentum (in this example,
$\vect{k}_1$) to be zero.
This corresponds to the extreme local or `{squeezed}' limit \citep{maldacena,
creminelli-zaldarriaga,allen-gupta},
in which the bispectrum reduces to the power spectrum evaluated on a perturbed
background, which is sourced by the zero-momentum mode. Written explicitly,
$\correction_t^{(2)}$ behaves like
\begin{equation}
  \label{invert:esti}
  \correction_t^{(2)}[\Rsp]
  \simeq \frac{1}{6} \int \frac{\d^3 k_1 \, \d^3 k_2}{(2\pi)^3}
  \Asq \Rsp(\vect{k}_1)
  \diracd(\vect{k}_1) + \perms ,
\end{equation}

\noindent where we have written $\displaystyle\lim_{k_1 \rightarrow 0} \A = \Asq k_2^3$,  for some known constant $\Asq$.
In particular, Eq.~\eqref{invert:esti} vanishes,
provided $\Rsp(\vect{k})$ approaches zero as $k \rightarrow 0$.
This condition is typically satisfied, since by construction $\Rsp(\vect{k})$
should not contain a zero mode. Indeed, any zero mode, if present, would
constitute part of the zero-momentum background, and not a part of
the perturbation $\Rsp$.
Accordingly, Eqs.~\eqref{invert:c}--\eqref{invert:d} with
$\correction_t^{(2)}=0$ give $\Prob_t[\Rsp]$ explicitly in terms of
the two- and three-point correlation functions.

\subsection{The smoothed curvature perturbation}
The probability density $\Prob_t[\Rsp] \propto (1 + \correction_t^{(0)}
[\Rsp])\Gauss_t[\Rsp]$ expressed in Eq.~\eqref{invert:b}
relates to the microphysical field $\R(t,\vect{x})$ which appeared in
the quantum field theory Lagrangian. A given $\vect{k}$-mode
of this field begins in the vacuum state at $t \rightarrow -\infty$.
At early times, the mode is far inside the horizon ($k \gg aH$).
In this (`subhorizon') r{e}gime, the $\vect{k}$-mode
cannot explore the curvature
of spacetime and is immune to the fact that it is living in a de Sitter
universe. It behaves like a Minkowski space oscillator.
At late times, the mode is far outside the horizon
($k \ll aH$). In this (`superhorizon') r{e}gime, the $\vect{k}$-mode
asymptotes to a constant amplitude, provided that
only one field is dynamically relevant during inflation
\citep{lyth-zeta,wands-malik}.%
\footnote[7]{Where multiple fields are present, there will typically be
an isocurvature perturbation between them: hypersurfaces of constant
pressure and density will not coincide. Under these circumstances
$\R$ will evolve \citep{wands-malik}. We do not consider the evolving case
in this chapter, but rather restrict our attention to the single-field
case where the superhorizon behaviour of $\R$ is simple.}
If we restrict attention to tree-level diagrams, then under reasonable
conditions
the integrals which define the expectation values of $\R$
are typically  dominated
by the intermediate (`horizon crossing') r{e}gime,
where $\R(\vect{k})$ is exiting the
horizon ($k \sim aH$) \citep{weinberg-qcI,weinberg-qcII}.
As a result, the correlation functions
generally depend only on the Hubble and slow-roll parameters around the
time of horizon exit.

The simple superhorizon behaviour of $\R$ means that we can treat the power
spectrum as constant outside the horizon.
As has been described,
its value depends only on the Hubble parameter and the
slow-roll parameters around the
time that the mode corresponding to $k$ exited the horizon.
For this reason, the time $t$ at which we evaluate the wavefunctional
$\Psi_t[\Rsp]$, the generating functional $Z_t[q]$ and the
PDF $\Prob_t[\Rsp]$ is irrelevant, provided it is taken
to be late enough that the curvature perturbation on interesting cosmological
scales has already been generated and settled down to its final value.
Indeed, we have implicitly been assuming that $t$ is the time
evaluated in comoving slices, so that
observers on slices of constant $t$ see no net momentum flux. Because $\R$
is gauge-invariant and constant outside the horizon, our formalism is
independent of how we choose to label the spatial slices.
The evolution of $\R$ outside the horizon is the principal obstacle
 involved in extending our analysis to
the multiple-field scenario.

When calculating the statistics of density fluctuations on some given
length scale $2\pi/\kmax$, one should smooth the perturbation field over
wavenumbers larger than $\kmax$.
To take account of this, we introduce a smoothed field $\Rsm$ which is related
to $\Rsp$ via the rule $\Rsm(\vect{k}) = \window(k,\kmax)
\Rsp(\vect{k})$, where $\window$ is some window function. The probabilities
we wish to calculate and compare to the real universe relate to $\Rsm$
rather than $\Rsp$.
The exact choice of filter $\window$ is mostly arbitrary. For the
purpose of analytical calculations, it is simplest
to pick a sharp cutoff in $\vect{k}$-space,
which removes all modes with $k < \kmax$. Such window function is given by
\beq
\window(k,\kmax) = \Theta(k - \kmax),
\eeq

\noindent where, $\Theta(x)$ is the Heaviside step function defined in
Eq.~\eqref{heaviside:def}.  
This choice of window function has the disadvantage that it is
non-local and oscillatory in real space, which makes physical
interpretations difficult.
The most common alternative choices, which do not suffer from such drawbacks,
are (i) a Gaussian or (ii)
the so-called `top hat', which has a sharp cutoff in real space.
We allow for a completely general choice of $\cont{0}$ function $\window$,
subject to the restriction that $\window \neq 0$ except at $k = \infty$
and possibly at an isolated set of points elsewhere (we will work with
more specific forms of the window function in the following Chapters
\ref{chapterfour} and \ref{chapterfive}). This restriction is made
so that there is a one-to-one relationship between $\Rsm$ and $\Rsp$. If this
were not the case, it would be necessary to coarse-grain over classes of
microphysical fields $\Rsp$ which would give rise to the same smoothed
field $\Rsm$.

In addition to this smoothing procedure,
the path integral must be regulated before carrying
out the calculation in the next Section. This is achieved by artificially
compactifying momentum space, so that the range of available wavenumbers
is restricted to $k < \Lambda$, where $\Lambda$ is an auxiliary hard cutoff or `regulator'.  At the end of the calculation we take $\Lambda \rightarrow \infty$.
Some care is necessary in carrying out this compactification.
We set $\Rsm = 0$ for $k > \Lambda$. In order to maintain continuity
at $k = \Lambda$, we introduce a 1-parameter family of
functions $\window_\Lambda$. These functions are supposed to
satisfy the matching condition $\displaystyle\lim_{\Lambda \rightarrow \infty}
\window_\Lambda(k) = \window(k)$, and are subject to the restriction
$\window_\Lambda(\Lambda) = 0$. (These conditions could perhaps be relaxed.)
The relationship between $\Rsp$ and $\Rsm$ becomes
\begin{equation}
  \label{filter:filter}
  \Rsm(\vect{k})
  = \Theta(\Lambda-k) \window_\Lambda(k;\kmax) \Rsp(\vect{k})
\end{equation}

\noindent To minimise unnecessary clutter in equations, we frequently
suppress the $\Lambda$ and $\kmax$ dependences in $\window$, writing
only $\window(k)$ with the smoothing scale $\kmax$ and hard cutoff
$\Lambda$ left implicit. Both the Gaussian and the `top-hat' window functions
approach zero as $k \rightarrow \infty$, and are compatible
with \eqref{filter:filter} in the $\Lambda \rightarrow \infty$ limit.
In this limit, the final result is independent of the exact choice of
$\window_\Lambda(k,\kmax)$.

We are interested in the probability of observing
a given filtered field $\Rsm$. One can express this via the rule
\citep{verde-ngII,taylor-watts}
\begin{equation}
  \label{filter:prob}
  \Probsm_t[\Rsm]
   = \int [\d \Rsp] \; \Prob_t[\Rsp] \diracd[\Rsm = \theta(\Lambda-k)
   \window \Rsp] .
\end{equation}

\section{Harmonic decomposition of the curvature perturbation}
\label{sec:harmoniccurv}
In the previous Section, we obtained the probability density for
a given smoothed spatial configuration of the
curvature perturbations. Given this probability density, the probability
$\Prob$ that the configuration exhibits some characteristic of
$\R$, such as fluctuations of amplitude $\fluct$ or a `fluctuation
spectrum' of the form $\spect(k)$, is formally obtained by integrating
over all configurations of $\Rsm$ which exhibit the criteria which
define $\fluct$ \citep{verde-ngII}.
In this section, we give a precise specification of these
criteria. Before doing so, however, we exploit the compactification of
momentum space introduced in \eqref{filter:filter} to define a
complete set of partial waves. The smoothed field $\Rsm$ can be
written as a superposition of these partial waves with arbitrary
coefficients. Moreover, the path integral measure can
formally be written as a product of conventional integrals over these
coefficients \citep{hawking-zeta}.

In the following we assemble the necessary
formulae for the partial-wave decomposition.
In particular, we will obtain
expressions for the decomposition of $\Rsm$, 
for  the characteristics $\fluct$ and $\spect(k)$,
and a precise specification of the path integral measure.

\subsection{Harmonic expansion of $\Rsm$}
\label{sec:harmonic}
We expand $\Rsm(\vect{k})$ in harmonics on the unit sphere and along the
radial $k = |\vect{k}|$ coordinate:
\begin{equation}
  \label{harmonic:expand}
  \Rsm(\vect{k}) = \sum_{\ell = 0}^\infty \sum_{m = - \ell}^{\ell}
  \sum_{n = 1}^\infty \Rsm^m_{\ell|n} Y_{\ell m}(\theta,\phi)
  \psi_n(k) .
\end{equation}
The $Y_{\ell m}(\theta,\phi)$ are the standard spherical harmonics
on the unit 2-sphere, while
the $\psi_n(k)$ are any complete, orthogonal set of functions on the finite interval
$[0,\Lambda]$. These harmonics should satisfy the following conditions%
\footnote[8]{When expanding functions on $\reals^3$ in terms of polar
  coordinates, a more familiar expansion involves spherical waves
  $Z_{\ell m|k} \propto J_{\ell}(kr) Y_{\ell m}(\theta,\phi)$, where
  $J_{\ell}$ is a spherical Bessel function. These waves are
  eigenfunctions of the Laplacian in polar coordinates, viz, $\nabla^2
  Z_{\ell m|k} = - k^2 Z_{\ell m |k}$. An arbitrary function on
  $\reals^3$ can be written in terms of spherical waves, which is
  equivalent to a Fourier expansion. We do not choose  spherical waves
  as an appropriate complete, orthogonal set of basis functions here
  because we do not wish to expand '{arbitrary}' functions, but rather
  functions obeying particular boundary conditions at $k=0$. The
  spherical waves for low $\ell$ behave improperly at small $k$ for
  this purpose. Moreover, it is not possible to easily impose the
  boundary condition $\Rsm(k) \rightarrow 0$ as $k \rightarrow \Lambda$.}:
\begin{enumerate}
  \item $\psi_n(k) \rightarrow 0$ smoothly as $k \rightarrow 0$, so that
        power is cut off on very large scales, and the universe
        remains asymptotically FRW with the zero-mode $a(t)$, which
        was used when computing the expectation values $\langle \R
        \cdots \R \rangle$; 
         \label{harmonic:asymp}
  \item $\psi_n(k) \rightarrow 0$ smoothly as $k \rightarrow \Lambda$,
        so that the resulting $\Rsm$ is compatible with
        Eq.~\eqref{filter:filter}; \label{harmonic:compact}
  \item $\psi_n(k)$ should have dimension $\dimension{M^{-3}}$, in order that
        Eq.~\eqref{harmonic:expand} is dimensionally correct;
         \label{harmonic:dims}
  \item the $\psi_n(k)$ should be orthogonal in the measure
        $\int_0^\Lambda \d k \, k^5 \ps_{\R}^{-1}(k)\window^{-2}(k)$.
        \label{harmonic:technical}
\end{enumerate}

\noindent In addition, there is a constraint on the coefficients $\Rsm^m_{\ell|n}$,
because $\Rsm(\vect{k})$ should be real in configuration space and
therefore must obey the Fourier reality
condition $\Rsm(\vect{k})^\ast = \Rsm(-\vect{k})$, where an asterisk
 denotes complex conjugation. The $\Rsm^m_{\ell|n}$
are generically complex, so it is useful to separate the real and imaginary
parts by writing $\Rsm^m_{\ell|n} = a^m_{\ell|n} + \imag b^m_{\ell|n}$.
The condition that $\Rsm$ is real in configuration space implies
\begin{eqnarray}
  \label{harmonic:reality}
  a^{-m}_{\ell|n} = (-1)^{\ell+m}a^m_{\ell|n} \,,\\
  b^{-m}_{\ell|n} = (-1)^{\ell+m+1}b^m_{\ell|n} .
\end{eqnarray}
These conditions halve the number of independent coefficients,
since the $a$ and $b$ coefficients with strictly
negative $m$ are related to those with
strictly positive $m$, whereas for the $m=0$ modes,
the $b$ coefficients vanish if $\ell$ is even and the $a$ coefficients
vanish if $\ell$ is odd.

Condition~\ref{harmonic:asymp} is made because, in the absence of this constraint, $\Rsm$ could develop unbounded fluctuations on extremely large scales, which would renormalise $a(t)$. Therefore, condition~\ref{harmonic:asymp}
can be interpreted as a consistency requirement, since the inflationary
two- and three-point functions are calculated using perturbation theory
on a FRW background with some given $a(t)$, which must be recovered
asymptotically as $|\vect{x}| \rightarrow \infty$.
It will later be necessary to sharpen
this condition to include constraints on the behaviour of $\ps_{\R}(k)$ near
$k=0$ beyond the weak requirement that $\sigma^2 = \int \ps_{\R}(k)
\, \d \ln k$ is finite.
Condition~\ref{harmonic:technical} is a technical requirement made for
future convenience. Any other choice of 
normalisation would work just as well, but this choice is natural,
given the $k$-dependence in the Gaussian kernel $\G[\Rsm]$. Indeed,
with this condition, the Gaussian prefactor in $\Prob
(\fluct)$ will reduce to the exponential of the sum of the squares of the
$a^m_{\ell|n}$ and $b^m_{\ell|n}$.
 Condition~\ref{harmonic:dims} ensures that the inner product
of two $\psi_n(k)$ in the measure $\int_0^\Lambda \d k \, k^5 \ps_{\R}^{-1}(k)
\window^{-2}(k)$ is dimensionless.
Condition~\ref{harmonic:compact} has less fundamental significance.
It follows from the condition $\window_\Lambda(\Lambda) = 0$ and the
artificial compactification of momentum space. However, as in the usual
Sturm--Liouville theory \citep{morse-feshbach}, the precise
choice of boundary condition is immaterial when $\Lambda \rightarrow \infty$, so this
does not affect the final answer.

To demonstrate the existence of a suitable set of $\psi_n(k)$,
we can adopt the definition
\begin{equation}
  \label{harmonic:psi}
  \psi_n(k) = \frac{\sqrt{2}}{J_{\nu+1}(\alpha_\nu^n)} \frac{\ps_{\R}(k)
  \window(k)}
  {\Lambda k^2} J_\nu\left( \alpha_\nu^n \frac{k}{\Lambda} \right) ,
\end{equation}

\noindent where $J_\nu(z)$ is the Bessel function of the first kind
and of order $\nu$, which is regular at the origin, and
$\alpha_\nu^n$ is its $n$-th zero. The order $\nu$ is arbitrary, except that
in order to obey condition~\ref{harmonic:asymp} above, we must have
$k^{\nu-2} \ps_{\R}(k) \rightarrow 0$ as $k \rightarrow 0$. This assumes that
$\window(k) \rightarrow 1$ as $k \rightarrow 0$, as is usual for a
volume-normalised window function.
The $\psi_n(k)$ obey the orthonormality condition
\begin{equation}
  \label{harmonic:orthonormal}
  \int_0^\Lambda \d k \; \frac{k^5}{\ps_{\R}(k) \window^2(k)}
  \psi_n(k) \psi_m(k) = \delta_{mn} ,
\end{equation}

\noindent where $\delta_{mn}$ is the Kronecker delta. The completeness relation can be
written
\begin{equation}
  \label{harmonic:complete}
  \diracd(k-k_0)|_{k \in [0,\Lambda]}
  = \frac{k_0^5}{\ps_{\R}(k_0) \window^2(k_0)} \sum_n \psi_n(k) \psi_n(k_0) ,
\end{equation}
where the range of $k$ is restricted to the compact interval $[0,\Lambda]$.

Although we have given an explicit form for the $\psi_n$ in order to
demonstrate existence, the argument does not depend in detail
on Eq.~\eqref{harmonic:psi}. The only important properties are
Eqs.~\eqref{harmonic:orthonormal}--\eqref{harmonic:complete}, which
follow from condition~\ref{harmonic:technical}.

\subsection{The path integral measure}

Since any real $\cont{0}$ function $\Rsm$
obeying the boundary conditions $\Rsm(\vect{k})
\stackrel{}{\longrightarrow} 0$  as $k \rightarrow 0$ and
$\Rsm(\vect{k}) \stackrel{}{\longrightarrow} 0$  as $k \rightarrow \Lambda$ can
be expanded in the form \eqref{harmonic:expand}, one can formally integrate
over all such $\Rsm$ by integrating over the coefficients $\Rsm^m_{\ell|n}$.
This prescription has been widely used for obtaining explicit results from
path integral calculations. (For a textbook treatment, see
\cite{kleinert}.) In the present case, one
should include in the integral only those $\Rsm(\vect{x})$ which are real and so correspond to a physical curvature perturbation in the
universe. Since the $Y_{\ell m}$ are complex,
this means that instead of integrating unrestrictedly over the
$\Rsm^m_{\ell|n}$, the reality conditions \eqref{harmonic:reality} must
be respected. A simple way to achieve this is to integrate only over
those $a^m_{\ell|n}$ or $b^m_{\ell|n}$ with $m \geq 0$. The $m=0$ modes
must be treated separately since the $a$ and $b$ coefficients vanish for
odd and even $\ell$, respectively.

The integral over real $\Rsm$ can now be written as
\begin{align}
  \label{harmonic:measure}
  \int_{\reals} [\d \Rsm] =
  \Bigg[ \prod_{\ell =0}^{\infty} &\prod_{m=1}^\infty \prod_{n=1}^{\ell}
  \mu \int_{-\infty}^{\infty} \d a^m_{\ell|n} \int_{-\infty}^\infty
  \d b^m_{\ell|n} \Bigg] \Bigg[ \prod_{\substack{r=0 \cr \even{r}}}
  ^\infty \prod_{s = 1}^\infty
  \tilde{\mu} \int_{-\infty}^\infty \d a^0_{r|s} \int_{-\infty}^\infty
  \d b^0_{r+1|s} \Bigg] ,
\end{align}
where the subscript $\reals$ on the integral indicates schematically that
only real $\Rsm(\vect{x})$ are included.
The constants $\mu$ and $\tilde{\mu}$ account
for the Jacobian determinant which arises in writing $\int [\d\Rsm]$ in terms
of the harmonic coefficients $\Rsm^m_{\ell|n}$. Their precise form is of no
importance in the present calculation as they will be absorbed by the
final normalisation factor.

As  noted above, the detailed
form of the measure \eqref{harmonic:measure} is not absolutely necessary
for our argument. The important point is that each $a$ or $b$ integral
can be carried out independently for $m \geq 0$.
For this purpose, it is sufficient that the spectrum of partial waves be
discrete, which follows from the (artificial) compactness of momentum space.
However, although it is necessary to adopt some `regulator' $\Lambda$
in order to write the path integral measure in a concrete form such as 
\eqref{harmonic:measure}, we expect the answer to be independent of the
specific regulator which is chosen. In the present context, this means that
our final expressions should not depend on $\Lambda$, so that the passage
to the $\Lambda \rightarrow \infty$ limit becomes trivial.

\subsection{The total fluctuation $\fluct$ and the spectrum $\spect(k)$}
\label{sec:condition}
There are at least two useful ways in which one might attempt to measure
the amplitude of fluctuations in $\Rsm$.
The first is the `{total smoothed fluctuation}'
at a given point $\vect{x} = \vect{x}_0$. By
a suitable choice of coordinates, we can always arrange that $\vect{x}_0$
is the origin, so the parameter becomes $\fluct \equiv \Rsm(\vect{0}) $.
When $\Rsm$ is smoothed on scales of order the horizon size this gives
a measure of the fluctuation in each Hubble volume, since distances of
less than a horizon size no longer have any meaning.
For example, \cite{shibata-sasaki} have proposed that $\fluct$
defined in this way represents a useful criterion for the formation of
PBHs, with formation occurring whenever $\fluct$ exceeds a threshold
value $\fluct_{\mathrm{th}}$ of order unity \citep{green-liddle}. 
This measure of the fluctuation is non-local in momentum space.
Making use of the relation $\int \d\Omega(\theta,\phi) \, Y_{\ell m}(\theta,
\phi) = \sqrt{4\pi} \delta_{\ell,0} \delta_{m,0}$ for the homogeneous mode
of the spherical harmonics, one can characterise the amplitude as
\begin{equation}
  \label{strength:nonlocal}
\fluct\,\equiv\,\Rsm(\vect{0}) = \int \frac{\d^3 k}{(2\pi)^3} \Rsm(\vect{k}) \e{\imag \vect{k}\cdot\vect{x}}|
  _{\vect{x}=0} = \frac{\sqrt{4\pi}}{(2\pi)^3} \int \d k \; k^2
  \sum_{n=1}^{\infty} a^0_{0|n} \psi_n(k)  .
\end{equation}

On the other hand, one might be interested in contributions to the
total smoothed fluctuation
in each Hubble volume which arise from features in the spectrum near some
characteristic scale of wavenumber $k$. For this reason, we consider
a second measure of the fluctuation, which we call the
`{fluctuation spectrum}',  defined by 
\beq
\spect(k) = \frac{\d \Rsm(\vect{0})}{\d \ln{k}}.
\eeq
(Thus the total smoothed fluctuation can be obtained by integrating
its spectrum according to the usual rule, viz, $\fluct = \int \spect(k) \, \d
\ln k$.) This condition is local
in $\vect{k}$-space. Differentiating \eqref{strength:nonlocal},
one can characterise $\spect(k)$ by the  functional constraint
\begin{equation}
  \label{strength:local}
  \spect(k) = \frac{\sqrt{4\pi}}{(2\pi)^3}
  \sum_{n=1}^{\infty} a^0_{0|n} k^3 \psi_n(k) .
\end{equation}

\noindent We will calculate the statistics of both the total
fluctuation $\fluct$ and the spectrum $\spect(k)$. In each
case, the calculation is easily adapted to other observables which are
non-local or local in momentum space\footnote[9]{The local and
  non-local variables defined here should not be confused with the
  local and equilateral triangulations which are specifically defined
  for the bispectrum in Chapter \ref{chaptertwo}.}. Indeed, both the
non-local $\fluct$ and the local $\spect(k)$ are members of a large class
of observables, which we can collectively denote by $\vartheta$, and
which all share nearly-Gaussian statistics. Specifically, Eqs.~\eqref{strength:nonlocal} and \eqref{strength:local} can be written in a unified
manner in the form
\begin{equation}
  \label{strength:condition}
  \sum_{n=1}^\infty a^0_{0|n} \Sigma_n(k)
  = \frac{(2\pi)^3}{\sqrt{4\pi}} \vartheta(k) ,
\end{equation}
where 
\begin{equation}
  \Sigma_n = \left\{ \begin{array}{ll}
  \int_0^\Lambda \d k \; k^2 \psi_n(k)\qquad & \qquad
    (\vartheta = \fluct); \\
  k^3 \psi_n(k) \qquad & \qquad
    (\vartheta = \spect(k)) .
  \end{array} \right.
  \label{sigman:psin}
\end{equation}
Note that, in the first case, the $\Sigma_n$ are independent of $k$.
Any characteristic which can be put in this form, coupling only to the
real zero-modes $a^0_{0|n}$ of $\Rsm$, will necessarily develop
nearly-Gaussian ({i.e.}, weakly non-Gaussian) statistics.
More general choices of characteristic are possible,
which cannot be cast in the form \eqref{strength:condition}.
For example, one can consider characteristics which depend non-linearly on
the $a^0_{0|n}$. Such characteristics
will generally lead to strongly non-Gaussian probabilities.
The Gaussianity of the final PDF depends on the geometry
of the constraint surface in an analogous way to the decoupling of the
Fadeev-Popov ghost fields in gauge field theory \citep{weinberg-qcI}.
These non-Gaussian choices of characteristic can also
be handled by generalising our technique, but we do not consider them here.

\section{The probability density function for $\fluct$}
\label{sec:probdelta}

We first calculate the probability density for the non-local constraint
$\fluct$, given by Eq.~\eqref{strength:nonlocal}.
The expression is
\begin{equation}
  \label{prob:defn}
  \Prob(\fluct) \propto \int_{\reals} [\d \Rsm] \; \Probsm[\Rsm]
  \diracd\left[ \sum_{n=1}^\infty a^0_{0|n} \Sigma_n -
  \frac{(2\pi)^3}{\sqrt{4\pi}} \fluct \right] .
\end{equation}
To obtain this density, one treats $\fluct$ as a collective coordinate
parameterising part of $\Rsm$. The remaining degrees of freedom, which are
orthogonal to $\fluct$, are denoted by $\Rsm^\perp$. Therefore the
functional measure $[\d \Rsm] $ can be broken into
$ [\d \Rsm^\perp]$ and $ \d\fluct$. After integrating
the functional density $\Prob[\Rsm] \, [\d \Rsm]$ over
$\Rsm^\perp$, the quantity which is left is the probability density
$\Prob(\fluct)\,\d\fluct$.
In this case, the integration over the orthogonal degrees of freedom
$\Rsm^\perp$ is accomplished via the
$\diracd$-function, which filters out only those members of the ensemble which
satisfy Eq.~\eqref{strength:nonlocal}.
We emphasise that this is a conventional
$\diracd$-function, not a $\diracd$-functional. There is no need to
take account of a Fadeev--Popov type factor because the
Jacobian associated with
the constraint \eqref{strength:condition} is field-independent, in virtue
of the linearity of Eq.~\eqref{strength:nonlocal} in $a^0_{0|n}$.

\subsection{The Gaussian case}
\label{sec:gaussian}
We first give the calculation in the approximation that only the two-point
function is retained. In this approximation, the PDF of $\fluct$ will
turn out to be purely Gaussian, which allows us to develop our method
without the extra technical difficulties introduced by including
non-Gaussian effects.

If all correlation functions of order three and higher are set to
zero, then we are in a Gaussian regime and hence $\Prob[\Rsm] \propto
\G[\Rsm]$. Using \eqref{intro:twopt}, one can write 
\begin{eqnarray}
  \G[\Rsm] = \exp \Big( - \frac{1}{2} \int \d\Omega \int k^2 \, \d k \;
  \frac{k^3}{(2\pi)^3 2\pi^2} \frac{1}{\ps_{\R}(k)\window^2(k)}
  \qquad\qquad \qquad 
  \\ \qquad \times
  \sum_{\ell_1, m_1, n_1} \sum_{\ell_2, m_2, n_2} \Rsm^{m_1}_{\ell_1|n_1}
  \Rsm^{m_2\dag}_{\ell_2|n_2} Y_{\ell_1 m_1}(\theta,\phi)
  Y^\dag_{\ell_2,m_2}(\theta,\phi) \psi_{n_1}(k) \psi_{n_2}(k)
  \Big).
  \nonumber
\end{eqnarray}
The harmonics $Y_{\ell m}$ and $\psi_n$ integrate out of this expression
entirely, using the orthonormality relation \eqref{harmonic:orthonormal}
and the spherical harmonic completeness relation 
\beq
\int \d \Omega \,
Y_{\ell_1 m_1} Y^\dag_{\ell_2 m_2} = \delta_{\ell_1 \ell_2}
\delta_{m_1 m_2}.
\label{harmonics:deltas}
\eeq

\noindent Moreover, after rewriting the $a$ and $b$ coefficients with
$m<0$ in terms of the $m>0$ coefficients, we obtain
\begin{eqnarray}
  \G[\Rsm] = \exp \Big( - \frac{1}{2\pi^2 (2\pi)^3}
  \sum_{\ell =0}^\infty \sum_{m=1}^{\ell} \sum_{n=1}^\infty
  \left[|a^m_{\ell|n}|^2 + |b^m_{\ell|n}|^2 \right] -\qquad 
\label{gaussian:harmonics}\\
  \qquad\frac{1}{4\pi^2 (2\pi)^3} \sum_{\substack{\ell = 0 \cr \even{\ell}}}
  ^\infty \sum_{n=1}^\infty  \left[|a^0_{\ell|n}|^2 + |b^0_{\ell+1|n}|^2 \right]
  \Big) .
 \nonumber
\end{eqnarray}

The $\diracd$-function in \eqref{prob:defn} constrains one of the
$a^0_{0|n}$ (e.g. $a^0_{0|0}$) in terms of $\fluct$ and the other coefficients. It would then be possible to evaluate $\Prob(\fluct)$ by integrating out the $\diracd$-function
immediately. 
However, this does not turn out to be a
convenient procedure.
Instead, we introduce the Fourier representation of the
$\diracd$-function and rewrite \eqref{prob:defn} as
\begin{equation}
   \Prob(\fluct) \propto \int_{\reals} [\d \Rsm] \int_{-\infty}^\infty
  \d z \; \G[\Rsm] \exp \left( \imag z \left[ \sum_{n=1}^\infty a^0_{0|n}
  \Sigma_n - \frac{(2\pi)^3}{\sqrt{4\pi}} \fluct \right] \right) ,
  \label{delta:zet}
\end{equation}
where the functional measure is understood to be Eq.~\eqref{harmonic:measure}.
The final answer is obtained by integrating out $z$ together with all of
the $a$ and $b$
coefficients. In order to achieve this, it is necessary to separate
$a^0_{0|n}$, $z$ and $\fluct$ from each other by successively
completing the square in $a^0_{0|0}$ and $z$. Working with $a^0_{0|0}$ first,
we find
\begin{eqnarray}
  \exp\left( - \frac{1}{4\pi^2} \frac{1}{(2\pi)^3} \sum_{n=1}^\infty
  |a^0_{0|n}|^2 + \imag z \sum_{n=1}^\infty a^0_{0|n} \Sigma_n \right) 
  \qquad \qquad \qquad \qquad\qquad \qquad 
  \label{gauss:asquare}\\  \qquad
  {=} \exp \left( - \frac{1}{4\pi^2} \frac{1}{(2\pi)^3} \sum_{n=1}^\infty
  (a^0_{0|n} - 2 \pi^2 (2\pi)^3 \imag z \Sigma_n )^2 -
  (2\pi)^3 \pi^2 z^2 \Sigma^2 \right) ,
  \nonumber
\end{eqnarray}
where we have introduced a function  $\Sigma^2 \equiv \sum
_{n=1}^\infty \Sigma_n^2$. In the final PDF,
$\Sigma^2$ will turn out to be the variance
of $\fluct$. From Eq.~\eqref{gauss:asquare},
it is clear that making the transformation
$a^0_{0|n} \mapsto a^0_{0|n} + 2\pi^2 (2\pi)^3 \imag z \Sigma_n$ suffices
to separate $a^0_{0|n}$ from $z$. The measure, Eq.~\eqref{harmonic:measure},
is formally invariant under this transformation. Exactly the same procedure
can now be applied to $z$ and $\fluct$, giving
\begin{equation}
  \exp \left( - (2\pi)^3 \pi^2 z^2 \Sigma^2 - \frac{(2\pi)^3}{\sqrt{4\pi}}
  \imag \fluct z \right) = \exp \left[ - (2\pi)^3 \pi^2 \Sigma^2 \left(
  z + \frac{\imag \fluct}{2 \pi^2 \sqrt{4\pi} \Sigma^2} \right)^2
  - \frac{\fluct^2}{2\Sigma^2} \right] .
\end{equation}
As before, the finite shift $z \mapsto z - \imag \fluct / 2\pi^2 \sqrt{4\pi}
\Sigma^2$ leaves the measure intact and decouples $z$ and $\fluct$. The
$a$, $b$ and $z$ integrals can be done independently, but since they do not
involve $\fluct$, they contribute only an irrelevant normalisation to
$\Prob(\fluct)$. Thus, we obtain Gaussian statistics for $\fluct$:
\begin{equation}
  \label{gauss:gauss}
  \Prob(\fluct) \propto \exp \left( - \frac{\fluct^2}{2\Sigma^2} \right) .
\end{equation}

It remains to evaluate the variance $\Sigma^2$.
In the present case, we have $\Sigma_n = \int_0^\Lambda
\d k \, k^2 \psi_n(k)$. From the completeness relation
Eq.~\eqref{harmonic:complete}, it follows that
\begin{equation}
  \sum_n k_0^2 \psi_n(k_0) k^2 \psi_n(k) = \frac{k^2 \ps_{\R}(k_0)
  \window^2(k_0)} {k_0^3} \diracd(k-k_0) .
\end{equation}
$\Sigma^2$ is now obtained by integrating term-by-term under the summation.
The result coincides with the `{smoothed}' conventional variance
(cf. Eq.~\eqref{intro:variance}),
\begin{equation}
  \label{total:variance}
  \Sigma^2_{\Lambda}(\kmax)
  = \int_0^\Lambda \d \ln k \; \window^2(k;\kmax) \ps_{\R}(k) .
\end{equation}
Thus, as expected, Eq.~\eqref{gauss:gauss} reproduces the
Gaussian distribution \eqref{intro:gaussian}
which was derived on the basis of the central limit theorem, with the
proviso that parameters (such as $\Sigma^2$) describing the distribution
of $\fluct$ are associated with the smoothed field $\Rsm$ rather than the
microphysical field $\Rsp$. $\Sigma^2$ is therefore implicitly
a function of scale, with the scale-dependence entering through the window function.
Note that 	it was only necessary to use the completeness relation to obtain this
result, which follows from condition~\ref{harmonic:technical} in
Section~\ref{sec:harmonic}.

\subsection{The non-Gaussian case}
\label{sec:nongaussian}
The non-Gaussian case is a reasonably straightforward extension of the
calculation described in the preceding section, with the term
$\correction^{(0)}$ in Eq.~\eqref{invert:c} now being included. However,
some parts of the calculation become algebraically long, and there are subtleties
connected with the appearance of the bispectrum.

The inclusion of $\correction^{(0)}$ corrects the pure Gaussian statistics
by a quantity
proportional to the three-point function, $\langle \R \R \R \rangle$,
which is given in Eq.~\eqref{intro:threept}.
This correction is written in terms of the
representative spectrum $\bar{\ps_{\R}}^2$, which prescribes when the slow-roll
prefactor, given by the amplitude of the spectrum, should be evaluated
\citep{maldacena}.
For modes which cross the horizon
almost simultaneously,  with size $ k_1 \sim k_2  \sim k_3 $, this prefactor should be $\bar{\ps_{\R}}^2 = \ps_{\R}(k)^2$, where $k$ is the common magnitude of the $k_i$. In the alternative case, where one
$\vect{k}$-mode crosses appreciably before the other two, $\bar{\ps_{\R}}^2$
should be roughly given by
\begin{equation}
  \bar{\ps_{\R}}^2 = \ps_{\R}(\max k_i) \ps_{\R}(\min k_i) .
    \label{intro:prefactor}
\end{equation}

\noindent Since the difference between this expression and the expression when all the $k$ are of the same magnitude is very small, it is reasonable to adopt
Eq.~\eqref{intro:prefactor} as our definition of $\bar{\ps_{\R}}^2$. We stress
that this prescription relies on the conservation of $\R$ outside the
horizon \citep{allen-gupta}, and it would 
therefore become more complicated if extended
to a multiple-field scenario.

With this parametrization, the probability measure on the ensemble is
obtained by combining \eqref{intro:twopt}, \eqref{invert:c},
\eqref{invert:d} and \eqref{intro:threept}:
\begin{equation}
  \Probsm[\Rsm] \propto \G[\Rsm] \left( 1 - \frac{1}{6}
  \int \frac{\d^3 k_1 \, \d^3 k_2 \, \d^3 k_3}{(2\pi)^6 2\pi^2}
  \diracd(\triangle) \frac{\bar{\ps_{\R}}^2 \A}{\prod_i \ps{\R}(k_i)}
  \frac{\Rsm(\vect{k}_1) \Rsm(\vect{k}_2) \Rsm(\vect{k}_3)}
  {\window(k_1) \window(k_2) \window(k_3)} \right) .
\end{equation}
This expression should be integrated with the constraint
\eqref{strength:nonlocal} and measure \eqref{harmonic:measure}
to obtain the probability $\Prob(\fluct)$.
At first this appears to lead to an undesirable consequence, since the
integral of any odd function of $\Rsm$ multiplied by $\G[\Rsm]$ must be zero.
It may therefore seem that the non-Gaussian corrections we are trying
to obtain will cancel out. This would certainly be correct if the
integral were unconstrained. However, the presence of the
$\diracd$-function  constraint
means that the shifts of $a^0_{0|n}$ and $z$ which are
necessary to decouple the integration variables give rise to a non-vanishing
correction.

The finite shift necessary to decouple $a^0_{0|n}$ and $z$ is not changed
by the presence of non-Gaussian corrections, since it only depends on the
argument of the exponential term. This is the same in the Gaussian and
non-Gaussian cases. After making this shift, which again leaves the
measure invariant, the integration becomes
\begin{equation}
  \label{nongauss:a}
  \Prob(\fluct) \propto \int_{\reals} [\d \Rsm] \int_{-\infty}^\infty
  \d z \; \G[\Rsm] \exp\left( - (2\pi)^3 \pi^2 \Sigma^2 z^2 -
  \frac{(2\pi)^3}{\sqrt{4\pi}} \imag z \fluct \right)
  (1 - \mathcal{J}_0 - \mathcal{J}_2) ,
\end{equation}

\noindent where
\begin{eqnarray}
  \mathcal{J}_0 =\, \int \d^3 k_1 \, \d^3 k_2 \, \d^3 k_3
  \frac{2\pi^4(2\pi)^3}{3 (4\pi)^{3/2}}
  \diracd(\triangle) \frac{\bar{\ps_{\R}}^2 \A}{\prod_i \ps_{\R}(k_i)}
  \qquad \qquad \qquad \\
  \nonumber
  \sum_{n_1, n_2, n_3} \imag^3 z^3
  \Sigma_{n_1} \Sigma_{n_2} \Sigma_{n_3}
  \frac{\psi_{n_1}(k_1) \psi_{n_2}(k_2) \psi_{n_3}(k_3)}
  {\window(k_1) \window(k_2) \window(k_3)} ,
\end{eqnarray}

\noindent and 
\begin{align}
  \mathcal{J}_2 =\, \Big[
  \int \frac{\d^3 k_1 \, \d^3 k_2 \, \d^3 k_3}{6(2\pi)^3 \sqrt{4\pi}}
  \diracd(\triangle) \frac{\bar{\ps_{\R}}^2 \A}{\prod_i \ps_{\R}(k_i)}
  \sum_{n_1} \sum_{\ell_2, m_2, n_2} \sum_{\ell_3, m_3, n_2} &
  \\ \notag \times\,
  \imag z \Sigma_{n_1}
  \frac{\psi_{n_1}(k_1)}{\window(k_1)}
  \Rsm^{m_2}_{\ell_2|n_2} \Rsm^{m_3}_{\ell_3|n_3}
  Y_{\ell_2 m_2}(\theta_2,\phi_2)Y_{\ell_3 m_3}(\theta_3,\phi_3)&
  \frac{\psi_{n_2}(k_2) \psi_{n_3}(k_3)}{\window(k_2)\window(k_3)} \Big] \\
    &+ \swaplabel{1}{2}+ \swaplabel{1}{3} .
  \notag 
\end{align}

\noindent The symbol $\swaplabel{1}{2}$ represents the expression in square
brackets with the labels $1$ and $2$
exchanged, and similarly for $\swaplabel{1}{3}$.
The range of the $m_2$ and $m_3$ summations is from $-\ell_2$ to $\ell_2$
and $-\ell_3$ to $\ell_3$, respectively. In addition, the shift
of $a^0_{0|n}$ generates other terms linear and cubic in the
$\Rsm^{m}_{\ell|n}$,
but these terms do not contribute to $\Prob(\fluct)$ and we have omitted them
from \eqref{nongauss:a}.

After shifting $z$ to decouple $z$ and $\fluct$, the integrals $J_0$ and $J_2$
develop terms proportional to $z^0$, $z$, $z^2$ and $z^3$. Of these, only
the $z^0$ and $z^2$ survive the final $z$ integration. Consequently, we
suppress terms linear and cubic in $z$ from the following expressions.
The integral $\mathcal{J}_0$ becomes
\begin{eqnarray}
  \mathcal{J}_0 = \int \d^3 k_1 \, \d^3 k_2 \, \d^3 k_3 \;
  \frac{\pi^2(2\pi)^3}{3(4\pi)^2} \left( \frac{1}{16\pi^5}
  \frac{\fluct^3}{\Sigma^6} - 3 \frac{z^2 \fluct}{\Sigma^2} \right)
  \diracd(\triangle) \frac{\bar{\ps_{\R}}^2 \A}{\prod_i \ps_{\R}(k_i)}
  \\ \mbox{} \times
  \sum_{n_1, n_2, n_3} \Sigma_{n_1} \Sigma_{n_2} \Sigma_{n_3}
  \frac{\psi_{n_1}(k_1) \psi_{n_2}(k_2) \psi_{n_3}(k_3)}
  {\window(k_1) \window(k_2) \window(k_3)} ,
  \nonumber
\end{eqnarray}

\noindent while $\mathcal{J}_2$ simplifies to
\begin{align}
  \mathcal{J}_2 = \Big[ \int \frac{\d^3 k_1 \, \d^3 k_2 \, \d^3
      k_3}{48\pi^3(2\pi)^3} \frac{\fluct}{\Sigma^2} \sum_{n_1}
    \sum_{\ell_2, m_2, n_2} \sum_{\ell_3, m_3, n_3} & \\ 
    \notag \mbox{} \times
  \Sigma_{n_1} \frac{\psi_{n_1}(k_1)}{\window(k_1)} \Rsp^{m_2}_{\ell_2|n_2}
  \Rsp^{m_3}_{\ell_3|n_3} Y_{\ell_2 m_2}(\theta_2,\phi_2)
  &Y_{\ell_3 m_3}(\theta_3,\phi_3) \frac{\psi_{n_2}(k_2) \psi_{n_3}(k_3)}
  {\window(k_2)\window(k_3)} \Big]
  \\
 &\qquad\qquad + \swaplabel{1}{2} + \swaplabel{1}{3} ,
  \notag
\end{align}

\noindent the $m$ summations being over the same range as before. Thus
$\mathcal{J}_0$ contains corrections proportional to $\fluct$ and
$\fluct^3$, whereas $\mathcal{J}_2$ 
only contains corrections proportional to $\fluct$.

The $a$, $b$ and $z$ integrations can now be performed, with the integrand written entirely in terms of the $a^m_{\ell|n}$ and $b^m_{\ell|n}$
with $m \geq 0$. There are no $a$ or $b$ integrations in $J_0$.
 There are no $z$ integrations in $\mathcal{J}_2$ but
the $a$ and $b$ integrations involved in the product
$\Rsm^{m_2}_{\ell_2|n_2} \Rsm^{m_3}_{\ell_3|n_3}$ fix $\ell_2 = \ell_3$,
$m_2 = m_3$ and $n_2 = n_3$. One then uses the spherical harmonic completeness
relation,
\begin{equation}
  \sum_{\ell = 0}^\infty \sum_{m = -\ell}^\ell
  Y_{\ell m}(\theta_1,\phi_1) Y_{\ell m}^\dag(\theta_2,\phi_2) =
  \diracd(\phi_1 - \phi_2) \diracd(\cos \theta_1 - \cos\theta_2)
\end{equation}
and the equivalent relationship for the $\psi$-harmonics,
Eq.~\eqref{harmonic:complete}, to obtain 
\begin{align}0
  \mathcal{J}_2 = \Bigg[ \int \frac{\d^3 k_1 \, \d^3 k_2 \d^3
      k_3}{24\pi} \frac{\fluct} {\Sigma^2} \diracd(\triangle)
    &\frac{\bar{\ps_{\R}}^2 \A}{\prod_i \ps_{\R}(k_i) \window(k_i)}
  \frac{\ps_{\R}(k_2) \window^2(k_2)}{k_2^3}\\
  \sum_n \Sigma_n \psi_n(k_1)
  \diracd(\vect{k}_2 + \vect{k}_3) \Bigg]  &  +
  \swaplabel{1}{2} + \swaplabel{1}{3} .
  \notag
\end{align}

\noindent The terms with 1 exchanged with 2 and 3 generate the same
integral as the first term and can be absorbed into an overall factor of 3.

$\mathcal{J}_0$ involves only $z$ integrations. It can be written as 
\begin{align}
  \mathcal{J}_0 = \int \frac{\d^3 k_1 \, \d^3 k_2 \, \d^3
    k_3}{96\pi^2} 
  & \left(
  \frac{\fluct^3}{\Sigma^6} - 3 \frac{\fluct}{\Sigma^4} \right)
  \diracd(\triangle) \frac{\bar{\ps_{\R}}^2 \A}{\prod_i
    \ps_{\R}(k_i)\window(k_i)} 
  \\ &\qquad \times
  \sum_{n_1, n_2, n_3} \Sigma_{n_1} \Sigma_{n_2} \Sigma_{n_3}
  \psi_{n_1}(k_1) \psi_{n_2}(k_2) \psi_{n_3}(k_3) .
  \notag
\end{align}

\noindent To simplify these expressions further, it is necessary to obtain
the value of the sum $\sum_{n=1}^\infty \Sigma_n \psi_n(k)$.
Reasoning as before from the completeness relation
Eq.~\eqref{harmonic:complete}, it follows that
\begin{equation}
  \sum_{n=1}^\infty \Sigma_n \psi_n(k) = \frac{\ps_{\R}(k)\window^2(k)}{k^3} .
\end{equation}

\noindent From this, it is straightforward to show that 
\begin{equation}
  \mathcal{J}_0 = \int \frac{\d^3 k_1 \, \d^3 k_2 \, \d^3 k_3}
  {96\pi^2\prod_i k_i^3 \window^{-1}(k_i)}
  \diracd(\triangle) \bar{\ps_{\R}}^2 \A \left(
  \frac{\fluct^3}{\Sigma^6} - 3 \frac{\fluct}{\Sigma^4} \right) ,
\end{equation}

\noindent where $\Sigma^2$ is the smoothed variance, Eq.~\eqref{total:variance}.
On the other hand $J_2$ becomes
\begin{equation}
  \mathcal{J}_2 = \int \frac{\d^3 k_1 \, \d^3 k_2 \, \d^3 k_3}{24\pi/3}
  \frac{\fluct}{\Sigma^2} \diracd(\triangle) \frac{\bar{\ps_{\R}}^2}
  {\ps_{\R}(k_2)}
  \window(k_1) \A \frac{\diracd(\vect{k}_2 + \vect{k}_3)}{k_1^3 k_2^3} .
\end{equation}

\noindent After integrating out $\vect{k}_3$ and the angular part of
$\vect{k}_1$ and $\vect{k}_2$, this gives
\begin{equation}
  \mathcal{J}_2 = 2\pi \int \d k_2 \; k_2^2 \int \d k_1 \; \diracd(k_1)
  \frac{\fluct}{\Sigma^2} \window(k_1) \frac{1}{k_2^3 \ps_{\R}(k_2)}
  \lim_{k_1 \rightarrow 0} \A \frac{\ps_{\R}(k_1)}{k_1^3} ,
\end{equation}

\noindent where we have used the fact that $k_1$ is constrained to zero by the
$\diracd$-function to evaluate the bispectrum $\A$ in the `squeezed'
limit where one of the momenta goes to zero \citep{maldacena,
allen-gupta,creminelli-zaldarriaga}. In this limit, $\min (k_i) = k_1$
and $\max (k_i) = k_2 = k_3$, so it is possible to expand $\bar{\ps_{\R}}^2$
unambiguously. Moreover, $\displaystyle\lim_{k_1 \rightarrow 0} \A = \Asq k_2^3$, so $\mathcal{J}_2 = 0$ if $\ps_{\R}(k)/k^3 \rightarrow 0$
as $k \rightarrow 0$.
This more stringent condition on how strongly large-scale power is
suppressed was anticipated in Section~\ref{sec:harmonic}. It requires
that $\ps_{\R}(k)$ falls at small $k$ faster than $k^3$.
If this does not occur, then the integral diverges. (There is a marginal
case when $\ps_{\R}(k)/k^3$ tends to a finite limit as $k$ approaches
zero. We assume that this is not physically relevant.)

The $\mathcal{J}_2$ integral contains a $\diracd$-function
$\diracd(\vect{k}_2+\vect{k}_3)$. It can therefore be interpreted as
counting contributions to the bispectrum which come from
a correlation between the modes $\vect{k}_2$ and $\vect{k}_3$, in a background
created by $\vect{k}_1$, which exited the horizon in the asymptotic past.
As we have already argued, modes of this sort are included in the FRW
background around which we perturb to obtain
the correlation functions of $\R$, so we can anticipate that its contribution
should be zero, as the above analysis shows explicitly. In this interpretation,
the condition $\ps_{\R}(k)/k^3 \rightarrow 0$ as $k \rightarrow 0$ ensures
that the perturbation does not destroy the FRW background.
Indeed, fluctuations on very large scales in effect describe
transitions from one FRW world to another via a shift in the zero-momentum
modes of the background metric. In this case, there is only one such mode,
which is the scale factor $a(t)$.
These transitions are rather like changing
the vacuum state in a quantum field theory. As a result, fluctuations of
a large volume of the universe between one FRW state and another
are strongly suppressed.

For fluctuations on the Hubble scale, therefore,
the PDF should be
\begin{equation}
  \label{hubble:prob}
  \Prob(\fluct) = \frac{1}{\sqrt{2\pi} \Sigma} \left[ 1 -
  \left( \frac{\fluct^3}{\Sigma^6} - 3 \frac{\fluct}{\Sigma^4} \right)
  \mathcal{J} \right]
  \exp\left( - \frac{\fluct^2}{2\Sigma^2} \right) ,
\end{equation}
where we have used the fact that the corrections are odd in $\fluct$
and therefore do not contribute to the overall normalisation of
$\Prob(\fluct)$. The (dimensionless) coefficient $J$ is
\begin{equation}
  \label{hubble:j}
  \mathcal{J} = \int \frac{\d^3 k_1 \, \d^3 k_2 \, \d^3 k_3}{96\pi^2
    \prod_i k_i^3 \window^{-1}(k_i)}
  \diracd(\triangle) \bar{\ps_{\R}}^2 \A .
\end{equation}
This explicit expression is remarkably simple. Although it is preferable for calculation, it can be recast directly as the integrated bispectrum with respect to $\window$:
\begin{equation}
  \mathcal{J} = \frac{1}{48(2\pi)^3(2\pi^2)^3} \int \d^3 k_1 \, \d^3
  k_2 \, \d^3 k_3 \; \langle \R(\vect{k}_1) \R(\vect{k}_2)
  \R(\vect{k}_3) \rangle 
  \window(k_1) \window(k_2) \window(k_3).
\end{equation}
  
As a consistency check, we note that the expectation of
$\fluct$, defined as $\Expect(\fluct) = \int \fluct
\Prob(\fluct) \, \d \fluct$, is zero. This
is certainly necessary,
since the universe must contain as many underdense regions
as overdense ones, but it is a non-trivial restriction, since both the
$\fluct$ and $\fluct^3$ corrections to $\Prob(\fluct)$ do not separately
average to zero. The particular combination of coefficients in
\eqref{hubble:prob} is the unique correction [up to $\Or(\fluct^3)$,
containing only odd powers of $\fluct$] which
maintains $\Expect(\fluct)=0$.

Finally, we note that Eqs.~\eqref{hubble:prob}--\eqref{hubble:j} do not
explicitly involve the cut-off $\Lambda$, except as a limit of integration
in quantities such as $\Sigma^2_\Lambda$
and $\window_\Lambda$ which possess
a well-defined, finite limit at large $\Lambda$.
As a result, there is no obstruction to taking the $\Lambda \rightarrow \infty$
limit to remove the regulator entirely.

\subsection{When is perturbation theory valid?}
It is known from explicit calculation that the bispectrum is of order
$\ps_{\R}^2$ multiplied by the quantity, $\fnl$, which is predicted to be small
when slow-roll is valid. It is therefore reasonable to suppose that whenever
the window functions $\window$ are peaked around some probe wavenumber
$k_{\star}$, one has the order of magnitude relations 
$\Sigma^2 \sim \ps_{\star}$ and $\mathcal{J} \sim \ps^2_{\star}$,
where $\ps_{\star}$ represents the spectrum evaluated at $k =
k_{\star}$. Since the $\fluct^3$ correction dominates for $\fluct >
\sqrt{3} \Sigma$, this means that for $\fluct$ not too large, $\fluct
\ll \ps_{\star}^{-3/2}$, the perturbative correction we have
calculated will be small. As $\fluct$ increases, so that $\fluct \gg
\ps_{\star}^{-3/2}$, perturbation theory breaks down and the power
series in $\fluct$ needs resummation. In any case, at such large
values of $\fluct$, the calculation described above ought to be
supplemented by new physics which can be expected to become important
at high energy densities. The details of these corrections presumably
do not matter too much, because at any finite order, the fast-decaying
exponential piece suppresses any contributions from large values of
$\fluct$.   

  At some value of $\fluct$, corrections coming from the trispectrum
  can be expected to become comparable to those coming from the
  bispectrum that we have computed. We do not know preciselly
  which are the dominant contributions of the correction
  from the trispectrum. Such corrections to the non-Gaussian PDF are
  to be explored in the future.

\section{The probability density function for $\spect(k)$}
\label{sec:probrho}
The probability density function for $\spect(k)$ can be obtained by
a reasonably straightforward modification of the above argument,
taking account of the fact that the constraint, Eq.~\eqref{strength:local}
is now a functional constraint. This means that, when splitting the functional
measure $[\d \Rsm]$ into a product of $[\d \spect(k)]$ and the orthogonal
degrees of freedom $[\d \Rsm^\perp]$, the result after integrating out the
$\Rsm^\perp$ coordinates gives a functional probability density
in $[\d \spect(k)]$.
In particular, with the definition \eqref{sigman:psin}, the
$\diracd$-function in Eq.~\eqref{delta:zet} is now represented as 
\begin{equation}
  \int [\d z] \; \exp \left[ \imag \int_0^{\infty} \d k \; z(k) \left(
  \sum_{n=1}^\infty a^0_{0|n} k^3 \psi_n(k) - \frac{(2\pi)^3}{\sqrt{4\pi}}
  \spect(k) \right) \right] .
\end{equation}
In order to carry out this calculation, we write $z(k)$ formally as
\begin{equation}
  z(k) = \sum_{n=1}^\infty \frac{k^2}{\ps_{\R}(k) \window^2(k)}
  z_n \psi_n(k) .
\end{equation}
The integration measure $\int [\d z]$ becomes $\prod_n \breve{\mu}
\int_{-\infty}^ \infty \d z_n$, where, as before, $\breve{\mu}$
is a field-independent Jacobian representing the change of variables from
$z(k) \mapsto z_n$. Its value is not relevant to the present calculation.
In addition, we introduce a set of coefficients $\tilde{\spect}_n$ to describe
$\spect(k)$,
\begin{equation}
  \label{nonlocal:rhocoeff}
  \frac{\spect(k)}{k^3} = \sum_{n=1}^\infty \tilde{\spect}_n \psi_n(k) .
\end{equation}
The $\tilde{\spect}_n$ can be calculated using the rule
$\tilde{\spect}_n = \int_{0}^\Lambda \d k \; k^2 \ps_{\R}^{-1}(k) \window^{-2}(k)
\spect(k) \psi_n(k)$. Note that, in order to do so, we have made the implicit
assumption that $\spect(k)/k^3 \rightarrow 0$ as $k\rightarrow 0$, to ensure
that \eqref{nonlocal:rhocoeff} is compatible with the boundary conditions
for the $\psi_n(k)$. In other words, we make the ansatz of a
suppression of power in modes with low $k$.

With these choices, the $\diracd$-function  constraint becomes
\begin{equation}
  \prod_n \breve{\mu} \int_{-\infty}^\infty \d z_n \;
  \exp \left[ \imag \sum_{m=1}^\infty \left( a^0_{0|n} z_n -
  \frac{(2\pi)^3}{\sqrt{4\pi}} z_n \tilde{\spect}_n \right) \right] .
\end{equation}

\noindent In contrast to the nonlocal case of $\fluct$, where a single extra
integration over $z$ coupled to $\fluct$, we now have a situation where
a countably
infinite tower of integrations over $z_n$ couple to the the coefficients
$\tilde{\spect}_n$. In all other respects, however, the calculation is 
much the same as the nonlocal one, and can be carried out in the same way.
The shift of variables necessary to decouple $a^0_{0|n}$ and $z_n$ is
\begin{equation}
  a^0_{0|n} \mapsto a^0_{0|n} + \imag 2\pi^2 (2\pi)^3 z_n ;
\end{equation}
and the shift necessary to decouple the $z_n$ and $\tilde{\spect}_n$ is
\begin{equation}
  z_n \mapsto z_n - \frac{\imag \tilde{\spect}_n}{2\pi^2 \sqrt{4\pi}} \,.
\end{equation}

When only the two-point function is included, we obtain a Gaussian in the
$\tilde{\spect}_n$,
\begin{equation}
  \label{gauss:spectral}
  \Prob[\spect(k)] \propto \exp\left( - \frac{1}{2} \sum_n \tilde{\spect}_n^2
  \right) .
\end{equation}
The sum over the $\tilde{\spect}_n$ can be carried out using 
Eq.~\eqref{nonlocal:rhocoeff} and the
completeness and orthogonality relations for the $\psi_n(k)$:
\begin{equation}
  \sum_n \tilde{\spect}_n^2 = \int \d \ln k \; \frac{\spect^2(k)}
  {\ps_{\R}(k)\window^2(k)} .
\end{equation}
Using this expression, and integrating over all $\spect(k)$ which
give rise to a fluctuation of amplitude $\fluct$, one recovers
the Gaussian probability profile Eq.~\eqref{gauss:gauss} with
variance given by Eq.~\eqref{total:variance}. This serves
as a consistency check for Eqs.~\eqref{gauss:spectral} and \eqref{gauss:gauss}.

When the non-Gaussian correction $\correction^{(0)}$ is included,
one again generates a probability density of the form
\begin{equation}
  \Prob[\spect(k)] \propto (1 - K_0 - K_2) \exp \left( - \frac{1}{2}
  \sum_n \tilde{\spect}_n^2 \right) ,
\end{equation}

\noindent where $K_2$ has the same form as $\mathcal{J}_2$, and
therefore vanishes for the same reasons, and 
\begin{align}
  K_0 = \int &\frac{\d^3 k_1 \, \d^3 k_2 \, \d^3 k_3}{96\pi^2
  \prod_i \ps_{\R}(k_i) \window(k_i)}
  \diracd(\triangle) \bar{\ps_{\R}}^2 \\
&\qquad  \times \,\left( 3 \frac{\spect(k_1)}{k_1^3}
  \frac{\ps_{\R}(k_2)\window^2(k_2)} 
  {k_2^5} \diracd(\vect{k}_2 + \vect{k}_3) - \prod_i
  \frac{\spect(k_i)}{k_i^3} \right) .
 \notag
\end{align}

\noindent The first term contains a $\diracd$-function which squeezes
$k_1$ into the asymptotic past.
It formally vanishes by virtue of our assumption about the behaviour of
$\spect(k)$ near $k=0$, which is implicit in Eq.~\eqref{nonlocal:rhocoeff}.
As a result, the total probability density for the fluctuation spectrum
can be written as
\begin{equation}
  \label{fluct:prob}
  \Prob[\spect(k)] \propto (1 - K) \exp \left( - \frac{1}{2}
  \int \d \ln k \; \frac{\spect(k)^2}{\ps_{\R}(k)\window^2(k)} \right) ,
\end{equation}

\noindent where
\begin{equation}
  \label{fluct:k}
  K = - \int \frac{\d^3 k_1 \, \d^3 k_2 \, \d^3 k_3}{96\pi^2
  \prod_i \ps_{\R}(k_i) \window(k_i)}
  \diracd(\triangle) \bar{\ps_{\R}}^2 \prod_i \frac{\spect(k_i)}{k_i^3} .
\end{equation}

\noindent As before, one can show that this expression is
consistent with Eqs.~\eqref{hubble:prob}--\eqref{hubble:j}
by integrating over all $\spect(k)$ which
reproduce a total fluctuation of amplitude $\fluct$, after dropping
another term which is squeezed into the asymptotic past owing to the
presence of a $\diracd$-function. This is a non-trivial consistency
check of Eqs.~\eqref{fluct:prob}--\eqref{fluct:k}.

As in the local case, Eqs.~\eqref{fluct:prob}--\eqref{fluct:k} are entirely
independent of $\Lambda$ (except as a limit of integration), so the
regulator can be freely removed by setting $\Lambda = \infty$.

\section{Summary of results}
\label{sec:conclude}
In this chapter we have obtained the connection between
the $n$-point correlation functions of the primordial curvature perturbation,
evaluated at some time $t$,
 $\langle \R(\vect{k}_1) \cdots \R(\vect{k}_n) \rangle$, and
the PDF of fluctuations in the spatial configuration
of $\R$. We have obtained an explicit expression for
the PDF of a fluctuation of amplitude $\fluct$ when $\R$ is smoothed
over regions of order the horizon size. This is a probability density in the
conventional sense. In addition, we have obtained
an expression for the probability that $\fluct$ has a spectrum 
$\spect(k)$. This is given by $\int \d \ln k \, \spect(k) = \fluct$, although
mapping $\spect(k) \mapsto \fluct$ is many-to-one. This is a functional
probability density, and can potentially be used to identify features in
the fluctuation spectrum near some specific wavenumber
$k \simeq k_{\star}$. Our result is independent of statistical reasoning
based on the central limit theorem and provides a direct route to incorporate
non-Gaussian information from the correlators of the effective quantum field
theory of the inflaton into theories of structure formation.

Both these probabilities are Gaussian in the limit where
$\R$ only possesses a two-point connected correlation function. If
there are higher-order connected correlation functions, then $\R$
exhibits deviations from Gaussian statistics, which we have explicitly
calculated using determinations of the inflationary three-point
function during an epoch of slow-roll inflation. Our method can be extended
to incorporate corrections from higher connected $n$-point functions to
any finite order in $n$. We have not computed these higher corrections, since
we anticipate that their contribution is subdominant to the three-point
correction (which is already small).

Our argument is based on a formal decomposition of the spatial configuration
of the curvature perturbation in $\vect{k}$-space into spherical harmonics,
together with
harmonics along the radial $k$ direction. However, we have emphasised that
our results do not depend on the details of this construction, but require
only a minimal set of assumptions or conditions.
The first assumption is that the power spectrum $\ps_{\R}(k)$ goes
to zero sufficiently fast on large scales, specifically 
$\ps_{\R}(k)/k^3 \rightarrow 0$ as $k \rightarrow 0$.
(In addition, in the case of the fluctuation spectrum, we require
$\spect(k)/k^3 \rightarrow 0$ as $k \rightarrow 0$.) Such a
condition is certainly consistent with our understanding of large-scale
structure in the universe and, within the perturbative approach we are
using, we have argued that in fact it describes a self-consistency
condition which prevents perturbative fluctuations from destroying the
background FRW spacetime. Our second assumption is that the spatial
configuration $\Rsp$ can be smoothed to $\Rsm$ via a window function
$\window$ to obtain a configuration for which $\Rsm \rightarrow 0$ as
$k \rightarrow \infty$. In this case, it is fair to compare $\Rsm$ to
the primordial power spectrum. 

In addition to these fundamental assumptions, which relate to the behaviour
of real physical quantities, a large part of the calculation has relied on
an auxiliary technical construction. This construction is based on an
artificial compactification of momentum space, implemented by a hard cutoff
$\Lambda$. There is an associated boundary condition on $\Rsm$ at
$k = \Lambda$ which discretises the harmonics (partial waves)
in $k$. However, in both the non-local (total fluctuation $\fluct$) and
local (fluctuation spectrum $\spect(k)$) cases, the final probability
density is independent of both the details of the partial wave construction
and $\Lambda$ (except as a limit of integration).
It is also independent of the choice of the family of
window functions $\window_\Lambda(k;\kmax)$, and depends only on the limit
$\displaystyle\lim_{\Lambda \rightarrow \infty} \window_\Lambda(k;\kmax)$
$ = \window(k;\kmax)$.
Therefore the regulator can be removed by taking the limit
$\Lambda \rightarrow \infty$.
Moreover, the boundary condition at $k = \Lambda$
becomes irrelevant in this limit, which is a familiar result from the theory
of Sturm--Liouville operators. As a consistency check, one can integrate
$\Prob[\spect(k)]$ with the condition $\int \d \ln k \, \spect(k) = \fluct$
in order to obtain $\Prob(\fluct)$.

In Chapters \ref{chapterfour} and \ref{chapterfive} we present
applications of this method, and the PDF obtained, to improve the
estimation of the probability of PBH formation.

\clearemptydoublepage
\chapter{Probability of primordial black hole formation}
\label{chapterfour}
\section{Introduction}
\label{intro-ngpbhs}

Primordial Black Holes (PBHs) are a unique tool to probe
 inhomogeneities in the early universe. The probability of PBH
 formation is extensively studied because it is useful in
 constraining the amplitude of primordial inhomogeneities generated by
 inflation
 (e.g.,
   \cite{carr-lidsey-blue,liddle-green,sendouda,zaballa-green,bugaev-constraints}).   
 What makes PBHs a unique tool in cosmology is the range of 
scales that can be probed by their formation. The anisotropies probed by the CMB
 data cover the range of  wavenumbers $7\times  10^{-4}\leq k /
 {\rm Mpc}^{-1} \leq 0.021 $. Equivalently these modes enter the
 horizon when the cosmological horizon or Hubble mass is between 
 $10^{19}\lesssim M / {M_{\odot}} \lesssim 10^{23} $ while the
 overdensities forming galactic haloes have associated masses 
 $ 10^8 \lesssim  M / {M_{\odot}} \lesssim 10^{12}$.   The
 inhomogeneities forming PBHs are much  smaller and they can span the
 range of wavenumbers 
 $10^{3} \lesssim k / {\rm Mpc}^{-1} \lesssim 10^{16} $ which
 correspond to  $10^{-24}\lesssim M / {M_{\odot}} \lesssim 10^{6}$, a
 set of values that can change with the model of inflation and it's
 reheating scale. In any case, this is the largest range of scales
 probed by any single observable in the universe.   

Another advantage of studying PBH statistics  is that, in a radiation
background, the gravitational collapse of fluctuations takes place
shortly after horizon crossing. Consequently, PBH statistics do not
suffer the bias problem or the late-time nonlinear evolution that
significantly modifies the mass and statistics of other bound objects.  

The absence of direct detections of PBHs has prompted studies of
processes that could be influenced by the gravitational effects of PBHs
or their evaporations. Some of the processes and observations that limit the
abundance of PBHs are the following: 
\begin{enumerate}
  \item  If the number of PBHs is large enough, they
could constitute a significant fraction of the dark matter. The
current density of PBHs therefore cannot exceed the observed density
of dark matter, i.e., $\Omega_{\rm {PBH}}(M\geq 10^{15} \gr) \leq
\Omega_{\rm   DM} = 0.28 $ \citep{wmap2008-params}.

\item The Hawking radiation from PBHs \citep{hawking-pbhs} can be the
  source of the background radiation  at various wavelengths in our
  universe \citep{carr-76,page-hawking,bugaev-neutrinos} and cosmic
  rays \citep{bugaev-neutrinos}. As  mentioned in
  Chapter~\ref{chapterone}, PBHs of mass $M_{\rm evap}=  5\times
  10^{14}\gr$ should be evaporating today and observations of the
  gamma-ray 
 background imply  
 $\Omega_{\rm PBH}(M_{\rm evap}) \lesssim 5 \times 10^{-8}$
 \citep{page-hawking,carr-76,
macgibbon-carr,kim-macgibbon}. This is the tightest constraint on the
density of PBHs although  future observations of the $21\, {\rm cm}$
radiation might impose a tighter limit \citep{mack-wesley}. 
\item Black holes with mass $M < M_{\rm evap}$ have already
evaporated and the decay products should not spoil the well understood
chemical history of the universe. Indeed, limits on $\beta_{\rm PBH}(M)$
can be obtained in the mass range $10^9 < M / \gr < 10^{12}$ by
looking at the effects of hadrons and neutrinos emitted by PBHs on the
Big Bang nucleosynthesis of helium and deuterium
\citep{miyama,novikov-polnarev}.    
\end{enumerate}

\noindent A complete list of numerical bounds can be found in Table I,
as compiled by \cite{green-constraints}. All these bounds have been used
to probe early universe fluctuations
\citep{carr-lidsey-blue,liddle-green,sendouda,zaballa-green,bugaev-constraints}. They
can be translated into limits on the root-mean-square amplitude of
density or curvature perturbations $\R_{\rm RMS}$ on scales
inaccessible to the CMB.

Here we explore how the bounds to $\R_{\rm RMS}$ can be modified in
view of the consideration of a non-Gaussian probability
distribution. We use the PDF derived in Chapter 
\ref{chapterthree} and calculate the mass fraction of PBHs with the
aid of the Press-Schechter formalism.  
 The effects of non-Gaussian perturbations on PBHs have already been
 studied for specific models \citep{bullock-primack,ivanov, avelino}
 but a precise quantification of the non-Gaussian effects is still
 required. Indeed, it is only now, with a much better understanding of
 the effects of higher order perturbations, that we are able to
 describe the general effects on PBHs. This discussion is crucial in
 the light of recent claims that only  exotic extensions of the
 canonical slow-roll inflationary potentials can produce an appreciable
 number of PBHs \citep{siri,bugaev-pbhs} (see however
 \citep{peiris-pbhs} where it's argued that a large number of PBHs can
 be formed even within the slow-roll regime). Here we explore whether
 the consideration of non-Gaussian  perturbations in inflationary
 models could increase the mass fraction of PBHs significantly.   
 
\begin{flushleft}
  \text{Table I}
  Constraints on the mass fraction $\beta_{\rm PBH}(M)$ of the universe going
  into PBHs 
  \begin{align*}
    \renewcommand{\arraystretch}{1.5}
    \begin{array}{|l|c|l|}
      \hline
      \mathbf{CONSTRAINT} \,
      & \mathbf{MASS\, RANGE\, (\gr)} \,& \mathbf{NATURE\,}  \\
      \hline
      \hline
          { 1.25\times\ten{-8}
            \left(\frac{M}{\ten{11}\gr}\right)^{-1}} &  < \ten{11}  
          & \text{\small entropy of the universe}\\ 
	  { 4.1\times\ten{-3}\left(\frac{M}{\ten{9}\gr}\right)^{1/2}}
          &\ten{9}-\ten{11}&\text{\small pair-production at
            nucleosynthesis} \\ 
	  { 4.9\times\ten{-7}\left(\frac{M}{\ten{10}\gr}\right)^{3/2}}
          &  \ten{10} - \ten{11} & 
          \text{\small Deuterium destruction}  \\ 
               {
                 6.5\times\ten{-5}\left(\frac{M}{\ten{11}\gr}\right)^{7/2}}    
	       &  \ten{11} - \ten{13} & \text{\small Helium-4 spallation} \\
	       { \ten{-18} \left(\frac{M}{\ten{11}\gr}\right)^{-1}}
	       &  \ten{11}-\ten{13} & \text{\small CMB distortion} \\
	       { 3.1 \times 10^{-27}}
	       &\, 3.6 \times 10^{14} -  10^{15} &
	       \text{\small $\gamma$-rays from evaporating PBHs}\\
                    {\ten{-19}\left(\frac{M}{10^{15}\gr}\right)^{1/2}} 
	            &  > 10^{15} & {\textstyle\Omega_{\rm
                        PBH}(t_0)\leq 0.47} \\ 
	            \hline
    \end{array}
  \end{align*}
\end{flushleft}

\section{The non-Gaussian PDF}
\label{ngpdf-pbhs}

Let us introduce the elements of the non-Gaussian
distribution of probabilities for the curvature perturbation field
$\R$.  We first describe how the Gaussian PDF is
constructed in the context of the linear theory.  
The amplitude of the curvature perturbations $\R$ is derived by
solving the perturbed Einstein equations to linear
order. Statistically, the mean amplitude is written in terms of
the two-point correlation function as 
\begin{align}
  \langle \R_{\rm G}({\bf k}_1)\R_{\rm G}({\bf k}_2)\rangle =
  (2\pi)^3  \delta({\bf k}_1 + {\bf k}_2) |\R_{\rm RMS}(k)|^2,
  \label{eqn1.33}
\end{align}

\noindent where, as before, $\R_{\rm G}({\bf k})$ are Gaussian
perturbations in Fourier space.

The two-point correlator defines the dimensionless power spectrum
$\Pe(k)$ through the relation 
\begin{align}
  \label{eqn1.4}
  \langle \R_{\rm G}({\bf k}_1)\R_{\rm G}({\bf k}_2)\rangle =
  (2\pi)^3  \delta({\bf k}_1 + {\bf k}_2)\frac{2\pi^2}{k_1^3} \Pe(k_1).
\end{align}

 \noindent As discussed in Chapter \ref{chapterthree}, the perturbations are
 smoothed over a given mass scale $\km$. Here we choose a truncated Gaussian
 window function  
\begin{align}
  \label{eqn2.11}
  \window_{\rm M}(k) = \Theta(\kx - k)\exp{\left(-\frac{k^2}{2\km^2}\right)},
\end{align}

\noindent were $\Theta$ represents the Heaviside function and  the
fiducial scale $\kx$ is introduced to avoid ultraviolet divergences. The
smoothing scale $\km$ is defined by
\beq
	\km = 2\pi H_{\rm M} = M / 2,
\eeq

\noindent where $M$ is the Hubble mass at the time the scale $\km$
enters the horizon.  

The variance of the smoothed field is related to the power spectrum by
\begin{equation}
  \Sigmar^2(M) = \int \, \frac{dk}{k} \window_{\rm M}^2(k) \Pe(k).
  \label{eqn2.22}
\end{equation}

\noindent The power spectrum encodes important information about
the underlying cosmological model. For example, in the case of
perturbations deriving from the quantum fluctuations of a single
inflationary field $\phi$ with a potential  $V$ dominating the
cosmological dynamics, the explicit expression is \citep{stewart-lyth}  
\begin{align}\label{eqn1.5}
  \Pe(k) = \frac{H_*^4}{(2\pi)^2 \dot{\phi}_*^2 \mP^2}
  \approx \frac{V_*^3}{(d V / d\phi)_*^2\mP^2},
\end{align}

\noindent  Here an
asterisk denotes values at the time when the relevant perturbation
mode exits the cosmological horizon, $k = a_* H_* = a(t_*) H(t_*)$.

The tilt of the power spectrum is
parametrised with a second observable, the spectral index,
which is defined as 
\begin{align}
\label{eq1.55}
  n_s - 1 = \frac{d}{d \ln{k}}\ln{\Pe(k)}.
\end{align}

\noindent If $n_s < 1 $,  the
root-mean-square amplitude $\R_{\rm RMS}$ increases on larger scales,
corresponding to a red spectrum. Conversely, $n_s > 1$ indicates
larger power on smaller scales and corresponds to a blue spectrum.

The power spectrum and the tilt are derived directly from linear
perturbations as reviewed in Section \ref{inflation} of Chapter
\ref{chaptertwo}. In observations of the CMB, it is possible to
determine with great accuracy the numerical values of the
power spectrum and its tilt on scales larger than the horizon at the
time of last scattering, that is 
($k \leq k_{\mathrm{ls}} = 1.7 \times \ten{-3} {\rm Mpc}^{-1}$). On
such scales, the five-year results of WMAP, combined with the galaxy
counts, give $\Pe(k_{\mathrm{ls}}) = 2.4\times \ten{-9}$ and 
$n_s =  0.95 \pm 0.1$ \citep{wmap2008-params}.  

In linear perturbation theory one makes use of the central
limit theorem to construct the  
PDF. To first order, the perturbation modes are independent of
each other. If we assume the field of linear perturbations $\bar\R$ has 
zero spatial average,  then the central limit theorem indicates that
the PDF of $\bar\R$ is a normal distribution which depends only on the
variance $\Sigmar^2$, 
\begin{align}
\mathbb{P}_{\rm G}(\bar\R) = \frac{1}{\sqrt{2\pi}{\Sigmar}}
\exp\left(-\frac{\bar\R^2}{2\Sigmar^2(M)}\right).
\label{eqn2.0}
\end{align}

\noindent A successful linear theory of structure formation will
predict this probability distribution and match the numerical values
at the relevant observational scales. Higher order correlations of the
perturbation field $\R$ offer an exciting way to distinguish between
cosmological models with common properties at linear order. 
As discussed in Chapter \ref{chapterone} and Chapter \ref{chaptertwo},
the deviations from Gaussianity are described to lowest order by the
nonlinear parameter $\fnl$. This parameter appears in the expansion
(e.g. \cite{rodriguez-ngI}) 
\begin{align}
  \label{eqn1.3}
  \R(k) = \R_{\rm G}(k) - \frac{3}{5} \fnl (\R_{\rm G}\star\R_{\rm G}(k) -
  \langle \R_{\rm G}^2 \rangle),
\end{align}

\noindent where a star denotes the convolution of two copies of the
field. The interaction of Fourier modes does not admit the use of the
central limit theorem and the non-Gaussian probability distribution
must be constructed by other means. 
 
In Chapter \ref{chapterthree} we have provided a method to calculate
the correction to the Gaussian PDF, and to derive a new PDF which
includes the linear order contribution from the 3-point function. 
Such a correlator can be derived through a second order expansion of the
perturbations in the Einstein equations
\citep{italians03,rigopoulos-shellard,seery-kg}. Alternatively, an
explicit expression for the three-point correlator can be obtained
from the third-order quantum perturbations  to the Einstein-Hilbert
action. Pioneering works using this method come from  
\cite{maldacena} and \cite{seery-lidsey,seery-lidsey-a}.
Here we use the expression derived by \cite{rodriguez-ngI} for the
correlator in Fourier space. At tree-level this reduces to
\begin{align}
  \langle \R(k_1)\R(k_2)\R(k_3)\rangle
  =&   - ( 2 \pi )^3  \delta \left( \sum_i {\bf k}_i  \right)  4  \pi^4
  \frac{6}{5}  \fnl  \left[ \frac{\Pe(k_1) \Pe(k_2)}{k_1^3 k_2^3}
    + \perms \right]. \label{eqn1.6}
\end{align}

\noindent Current observations provide numerical bounds for $\fnl$
through the three-point correlation of the temperature fluctuation
  modes.
The WMAP satellite gives the
constraints $-151<\fnl^{\rm equil.}<253$ \citep{wmap2008-params} for
an equilateral triangulation of the bispectrum and 
$-4<\fnl^{\rm local}<80$ \citep{smith09} for a local triangulation
(the local and equilateral triangulations have been defined in
Sec.~\ref{intro-ng}). Both of these values are determined at the
$95\%$ confidence level and consider an invariant value at all scales
probed by the CMB. 

In the following, the basic components of the non-Gaussian PDF derived
in Chapter \ref{chapterthree} are presented. 
 The amplitude of the perturbation is characterised by its value at
 the centre of the configuration 
\beq
\vartheta_{0} \equiv {\bar \R}(\bx = 0).
\label{R:zero}
\eeq

\noindent This specification is necessary to construct an explicit
expression of the PDF. The parameter $\vartheta_0$ is particularly
useful to discriminate the relevant inhomogeneities forming PBHs
\citep{shibata-sasaki,green-liddle}. 
The non-Gaussian probability distribution function for a perturbation with
central amplitude $\vartheta_0$, derived in Eq.~\eqref{hubble:prob},
is 
\begin{equation}
  {\mathbb P}_{\rm NG}(\vartheta_0)    =\frac{1}{\sqrt{2\pi}\Sigmar}
   \left[1 + \left(\frac{\vartheta_0^3}{\Sigmar^3}
    -  \frac{ 3\,
      \vartheta_0}{\Sigmar}\right)\frac{\mathcal{J}}{\Sigmar^3}
    \right]\exp\left(-{\frac{\vartheta_0^2}{2\Sigmar^2}}\right),\label{eqn2.2}
\end{equation}

\noindent where the factor $\mathcal{J}$ encodes the non-Gaussian
contribution to the PDF:
\begin{align}
  \mathcal{J}  &= \frac{1}{6  }\int
  \,\frac{d{\bf k}_1 \,d{\bf k}_2 \,d{\bf k}_3}{(2\pi)^9}
  \window_{\rm M}(k_1)\window_{\rm M}(k_2)\window_{\rm M}(k_3) \langle\R({\bf k}_1)
  \R({\bf k}_2)\R({\bf k}_3)\rangle,
  \label{eqn2.4} \\
  &=   - \frac{1}{5}\int \,\frac{\,d{\bf k}_1
    \,d{\bf k}_2 \,d{\bf k}_3    }{(4  \pi)^2
    \prod_i     {\window}_{\rm M}^{-1}({ k}_i)}
  \delta\left(\sum_i {\bf k}_i\right) \; \fnl
  \left[\frac{\Pe(k_1)\Pe(k_2)}{k_1k_2} + \perms \right]. \label{eqn2.3}      
\end{align}

\noindent This last equation is valid at tree level in the
expansion of $\langle\R\R\R\rangle$. It is justified
as long as the loop contributions to the three-point function,
generated by the convolution of $\R$-modes, are
sub-dominant. This requirement is met when the second order
contribution to $\R$ in Eq.~\eqref{eqn1.3} does not exceed the linear
contribution. This is requirement is met if we demand that 
\beq
\fnl\leqslant
1/\sqrt{\Pe(k)}.
\eeq 

The complete derivation of the PDF in Eq. \eqref{eqn2.2} was already
provided in Chapter \ref{chapterthree}. Here it is sufficient to say
that the time-dependence of this probability is eliminated when the
averaging scale is $\km \leq a(t)H(t)$ providing the growing mode of
the perturbation $\R$ is constant on superhorizon scales. This is true
in particular for perturbations $\R$ considered in the radiation era,
when the PBHs considered here are formed (see Chapter
\ref{chapterone}).    

In order to adapt the PDF in Eq. \eqref{eqn2.3} to the computation of
PBH formation probabilities, this expression 
is integrated between the limits  $\kmi$ and $\kx$ defined 
to cover the relevant perturbation modes for PBH formation. PBHs are
formed long before today, so in the large-box
(small wavenumber) limit of integral \eqref{eqn2.3}, the present
Hubble horizon $\kmi = H_0$ is a reasonable lower
limit for PBH formation
\citep{lyth-axions}. At the other end of the spectrum, the smallest PBHs
have  the size  of the Hubble horizon  at the end  of inflation. A
suitable upper limit in this case is the wavenumber associated with
the comoving horizon at the end of inflation, $\kx = a(t_{\rm end})
H_{\rm end}$.  It is important to mention that, even though the integral in
Eq. \eqref{eqn2.3} should include all $k$-modes, finite limits are imposed
to avoid logarithmic divergences. Due to the window function factors
$\window_{\rm M}(k)$, the dominant part of the integral is independent
of the choice of integration limits as long as they remain finite. 

The integral \eqref{eqn2.3} is considered only at the limit of equilateral
configurations of the three-point correlator, that is,
considering correlations for which  $k_1 = k_2  = k_3 $. This is not merely a
computational simplification. In the integral, each perturbation
mode has a filter factor $\window_{\rm M}(k)$ which, upon integration, picks
dominant contributions from the smoothing scale $\km$
common to all perturbation modes. In this case $\mathcal{J}$ can be
written in the suggestive way: 
\begin{equation}
  \mathcal{J}  =   -  \frac{1}{8}  \int_{\kmi}^{\kx}\,\frac{dk}{k}
  \left[\window_{\rm M}(k)\Pe(k)\right]^2 \left( \frac{6}{5}\fnl\right).
  \label{eqn2.6}
\end{equation}

With the complete non-Gaussian PDF at hand, it is possible to
characterise its effects on the probability of PBH formation. 
In the next section, $\mathcal{J}$ is computed numerically
for inflationary perturbations generated in a single-field
slow-roll inflationary epoch. The results in this case are shown to be
consistent with previous works on non-Gaussian computations of the
probability of PBH formation. In Section \ref{ngbeta-pbhs}, the non-Gaussian
PDF is generated for the case of constant $\fnl$. This will be used to
test the magnitude of the effects of non-Gaussianity on the
probability of PBH formation. 

\section{Non-Gaussian modifications to the probability of PBH
  formation} \label{ngprob-pbhs}

The simplest models of structure formation within the inflationary
paradigm are those where a single scalar field drives
the accelerated expansion of the spacetime and its quantum
fluctuations evolve into the observed structure
in subsequent stages of the universe. 
Although small in magnitude, the non-Gaussianity of the fluctuations
generated in this simple model provide a qualitative hint to the
consequences that non-Gaussianity has for the probability of PBH
formation. 

In fact, for single-field inflationary models, the effects
of non-Gaussianity on PBHs have been explored in the past but with
inconclusive results.  \cite{bullock-primack} studied the
probability of formation of PBHs numerically for non-Gaussian
perturbations with a blue spectrum ($n_s > 0 $). 
The motivation for this was that any inflationary
model with a constant tilt and a normalisation consistent with the 
perturbations at the CMB scale must have a blue spectrum to produce a
significant number of PBHs \citep{carr-lidsey-blue,green-constraints}. Their
analysis is based on the stochastic generation of perturbations on
superhorizon scales, together with a Langevin equation for
computing the PDF. For all the cases tested, the non-Gaussian PDF
is skewed towards small fluctuations. In consequence, the probability
of PBH formation, which integrates the high amplitude tail, is
suppressed with respect to the Gaussian case. An example of the kind
of potential studied by \cite{bullock-primack} is
\begin{equation}\label{eqn3.4}
  V_1(\phi) = V_0\left\{\begin{array}{lcl}
  1 + \arctan\left(\frac{\phi}{\mP}\right),
  & \quad & \phi > 0, \\
  1 + (4\mbox{x}10^{33})\left(\frac{\phi}{\mP}\right)^{21},
  & \quad& \phi < 0.
  \end{array}\right.
\end{equation}

\noindent where $V_0$ is the amplitude of the potential at $\phi =
0$. This potential features a plateau for $\phi < 0$. This produces an
increase in the power of matter fluctuations corresponding to the
production of PBHs of mass $10^{32} {\rm gr}$.

Another way of generating large perturbations in the inflationary
scenario is to consider localised features in the potential dominating the
dynamics regardless of the tilt of the spectrum. As one can see from
Eq.~\eqref{eqn1.5}, an abrupt change in the potential would generate a
spike in the spectrum of perturbations. This is valid as long as we
avoid a 'flat' or 'static' potential in which $d V / d\phi = 0$. In
such case $\dot\phi = 0$ and Eq.~\eqref{eqn1.5} is invalid (for a
treatment of this particular case, also known as 'ultra-slow roll'
inflation, see \cite{kinney05}). 

The description of a model of inflation with large amplitude in the
power spectrum is incomplete if we do not take on account the effects
of nonlinear fluctuations. The effects of non-Gaussianity for an
inflationary model producing features in an otherwise red spectrum
($n_s < 0$) were explored by \cite{ivanov}, using the toy model 
\begin{equation} \label{eqn3.5}
  V_2(\phi) = \left\{\begin{array}{lcl}
  \lambda \frac{\phi^4}{4}\quad & \mbox{for} & \phi < \phi_1, \\
  A(\phi_2 - \phi)  +\lambda \frac{\phi_2^4}{4} & \mbox{for} &
  \phi_2 > \phi >\phi_1, \\
  \tilde{\lambda} \frac{\phi^4}{4} \quad& \mbox{for} & \phi > \phi_2.
  \end{array}\right.
\end{equation}

\noindent where $\lambda$ and $\tilde{\lambda}$ are coupling constants.
Through a stochastic computation of the PDF, Ivanov found that
the non-Gaussian PDF is skewed towards large perturbations. This
result goes in the opposite direction to that of \cite{bullock-primack}.

To understand this difference and generalise the effects of
non-Gaussianity, it is convenient to look at the fractional difference
of the Gaussian and non-Gaussian PDFs: 
\begin{equation}
  \label{eqn3.7}
  \frac{\mathbb{P}_{\rm NG} -\mathbb{P}_{\rm G}}{\mathbb{P}_{\rm G}} =
  \left[\left( \frac{\vartheta_0^3}{\Sigmar^3} -
    3 \frac{\vartheta_0}{\Sigmar}\right)\frac{\mathcal{J}}{\Sigmar^3}\right].
\end{equation}

\noindent Both \cite{bullock-primack} and \cite{ivanov} use
perturbations generated in a piecewise slow-roll inflationary
potential for which inflation is controlled by keeping the slow-roll
parameters, defined in Eq.~\eqref{sr:def}, smaller than one. Here  the
slow-roll approximation is used to explore the qualitative effects of 
Eq. \eqref{eqn3.7}. 

To linear order,
there is a straightforward expression for the spectral index in
terms of these parameters \citep{stewart-lyth},
\begin{align}
  \label{eqn3.71}
  n_s - 1  =  2(\etasr - 3 \epsilonsr).
\end{align}

\noindent On the other hand, by using a first order expansion in slow-roll
parameters, \cite{maldacena} provides an expression for the
nonlinear factor $\fnl$ in terms of these parameters
 \citep{maldacena}:
\begin{align}
  \label{eqn3.8}
  \fnl = &\frac{5}{12}\left( n_s + \mathcal{F}(k) n_t\right) =
  \frac{5}{6}\left( \etasr - 3 \epsilonsr + 2 \mathcal{F}(k) \epsilonsr \right),
\end{align}

\noindent where $n_t = 2\epsilonsr$ is the scalar-tensor perturbation
tilt and $\mathcal{F}(k)$ is a number depending on the triangulation
used. For the case of equilateral configurations, when $\mathcal{F}
=  5/6$,
\begin{align}
  \fnl = & \frac{5}{6}\left(\etasr - \frac{4}{3}\epsilonsr \right)_{\rm eq}.
\end{align}

\noindent This last expression is used to evaluate the integral 
\eqref{eqn2.6} for $\mathcal{J}$. The non-Gaussian effect on the PDF is illustrated in
Fig.~\ref{fig4.1} for the potentials given by Eqs. \eqref{eqn3.4} and
\eqref{eqn3.5} in terms of the fractional difference
\eqref{eqn3.7}. This difference represents the skewness of the
non-Gaussian PDF. The non-Gaussian contribution encoded in the factor
$\mathcal{J}$ is the integral of $\fnl$ over all scales relevant for PBH
formation. Consequently the sign of $\fnl$ is what determines
the enhancement or suppression of the probability
for large amplitudes $\vartheta_0$ in the non-Gaussian PDF. For the two
cases illustrated, the scalar tilt $n_s$ dominates over the tensor
tilt $n_t$, so that the sign of $\fnl$ coincides
with that of $n_s$. This result is illustrated in Fig. \ref{fig4.1}. 

\begin{figure}[h!]
  \begin{center}
    \psfrag{A}{\Large$\frac{\mathbb{P}_{\rm NG}(\vartheta_0) - \mathbb{P}_{\rm
	  G}(\vartheta_0)}{\mathbb{P}_{\rm G}(\vartheta_0)}$}
    \psfrag{B}{\Large$\vartheta_0$}
    \includegraphics[totalheight=0.35\textheight]{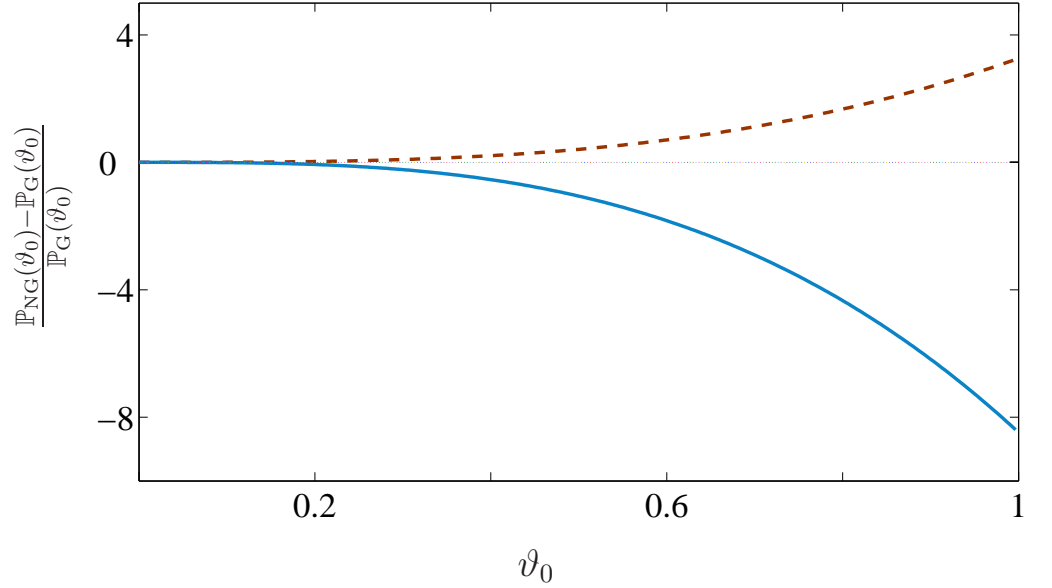}
    \caption{ {\small The fractional departure from the
        Gaussian PDF is plotted for two types of non-Gaussian distributions
	$\mathbb{P}_{\rm NG}$, as defined in Eq. \eqref{eqn2.2}. For
	the potential in Eq. \eqref{eqn3.4}, $\fnl > 0$ and the
	departure is plotted with a solid line. For the potential in
	Eq. \eqref{eqn3.5}, $\fnl < 0$ and the departure is shown by a
	dashed line.}}\label{fig4.1}   
  \end{center}
\end{figure}

\section{Constraints on non-Gaussian perturbations of PBH
  range}\label{ngbeta-pbhs}

A standard practise in calculating the PBH mass fraction is to use the
Press-Schechter formalism \citep{press-schechter}. As described in
Chapter \ref{chapterone}, this
involves integrating the probability of PBH formation over the
relevant matter perturbation amplitudes, $\delta$, measured at horizon epoch
\citep{carr} and gives  
\begin{align}
  \beta_{\rm PBH}(\gsim M) = 2 \int_{\delta_{\rm th}}^{\infty}
  \mathbb{P}(\delta_\rho(M))\,d \delta_\rho(M). \label{eqn4.2}
  \end{align}
  For the large ratio $\delta_{\rm th} / \Sigma_{\rho}$ this can be
  approximated as 
 \begin{align}
  \beta_{\rm PBH}(\gsim M) \approx \frac{\Sigma_{\rho}(M)}{\delta_{\rm th}}
  \text{exp}\left[-\frac{\delta_{\rm th}^2}{2 \Sigma_{\rho}^2(M)}\right],
  \label{eqn4.22}
\end{align}

\noindent where $\Sigma^2_{\delta}(M)$ is the variance corresponding
to the mass scale $M$ and $\delta_{\rm th}$ is the
threshold amplitude of the perturbation necessary to form a PBH.
When the relevant amplitudes of a smoothed perturbation are integrated,
$\beta_{\mathrm{PBH}}$ represents the mass fraction of PBHs with $M
\geq w^{3 / 2} M_{H} \approx w^{3 / 2} \km / (2\pi)$
\citep{carr}, where ${w}$ is the equation of state at the time
of formation. Note that the approximation \eqref{eqn4.22} is valid only
for a Gaussian PDF. 

The integral $\beta_{\mathrm{PBH}}$ establishes a direct relation between the
mass fraction of PBHs and the variance of perturbations. The set of
observational constraints on the abundance of PBHs is listed
in Table I and has been used to place a bound to the mean amplitude
$\delta$ in a variety of cosmological models (e.g.
\cite{carr-lidsey-blue,green-constraints,clancy-liddle,sendouda}). The
Press-Schechter formula has also been tested against other
estimations of the probability of PBH formation, such as peaks theory
\citep{green-liddle}. 

The threshold value $\delta_{\rm th}$ used in Eq. \eqref{eqn4.2}
has been modified with the improvement of gravitational collapse studies
\citep{carr, niemeyer-jedamzik-a, shibata-sasaki, hawke-stewart}. A
more appropriate approach has been noted  recently, where simulations
have addressed the problem using curvature fluctuations
\citep{shibata-sasaki,musco,musco-polnarev}. The corresponding
threshold value of the curvature perturbation can be deduced from the
relation \citep{liddle-lyth} 
\begin{align}
  \delta_{k}(t) = \frac{2(1+w)}{5 + 3w}
  \left(\frac{k}{aH}\right)^2 \R_k,\label{eqn4.3}
\end{align}

\noindent which at horizon-crossing during the radiation-dominated era
gives,  $\R_{\rm th} = 1.01$ for $\delta_{\rm th} = 0.3$. This value
has been also confirmed in the numerical simulations of
\cite{shibata-sasaki}, \cite{green-liddle} and \cite{musco}. We will
make use of it throughout.     

The threshold value $\R_{\rm th}$ indicates the minimum amplitude of an
inhomogeneity required to form a PBH.  Consequently, the probability
of PBH formation is best described by a nonlinear treatment and this
is the major motivation for our analysis.  
In the following we adapt the Press-Schechter formalism to derive the
non-Gaussian abundance of PBHs. The use of the Press-Schechter
integral for distributions of the curvature perturbation is not new. 
 \cite{zaballa-green} use it to estimate the PBH formation from the
 curvature perturbations which never exit the cosmological horizon. We
 apply the integral formula in Eq. \eqref{eqn4.2} to the non-Gaussian 
probability distribution \eqref{eqn2.2}. The result of the integral is
the sum of incomplete Gamma functions $\Gamma_{\rm inc}$ and an
exponential: 
\begin{align}
  \begin{split}
    \beta(M) = &\int_{\vartheta_{\rm th}}^\infty \, \mathbb{P}_{\rm
      NG} (\vartheta_0) \d\vartheta_0  =  
     \frac{1}{\sqrt{4\pi}} \Gamma_{\rm inc} \left(1/2,
    \frac{\vartheta_{\rm th}^2}{2\Sigmar^2(M)}\right) \label{eqn4.31} \\ &- 
      \frac{1}{\sqrt{2\pi}}\frac{\mathcal{J}}{\Sigmar^3(M)}
    \left[2\,\Gamma_{\rm inc}\left(2,\frac{\vartheta_{\rm
	  th}^2}{2\Sigmar^2(M)} \right) - {3} \exp{\left(-
    \frac{\vartheta_{\rm th}^2}{2
	  \Sigmar^2(M)}\right)}  \right].
  \end{split}
\end{align}

\noindent The Taylor series expansion of these functions
around the limit $ \Sigmar / \vartheta_{0} = 0$ gives
\begin{align}
\begin{split}
  \beta(M) \approx \frac{\Sigmar(M)}{\vartheta_{\rm th}\sqrt{2\pi}}
  & \exp \left[-\frac{1}{2}
    \frac{\vartheta_{\rm th}^2}{\Sigmar^2(M)}\right] \label{eqn4.4}
  \\ 
  &\times \left\{ 1 -  2\left(\frac{ \Sigmar}{\vartheta_{\rm th}}\right)^2  +
  \frac{\mathcal{J}}{ \Sigmar^3} \left[ \left(\frac{ \Sigmar}{\vartheta_{\rm
      th}}\right)^{-2} - 1 \right] \right\}.    
\end{split}
\end{align}

For the mass fraction shown in Eq. \eqref{eqn4.4}, the
observational limits of Table I could in principle constrain the
values of the variance $\Sigmar^2$ and of $\fnl$. However,
when the mean amplitude of perturbations, $\Sigmar$, is normalised to
the value at CMB scales, the obtained limits for $\fnl$ are of
order $ 10^4$. This is inconsistent with the analysis
presented here because in such r\'egime higher order contributions are
expected to dominate non-Gaussianity. In fact, the expansion in
Eq. \eqref{eqn1.3} shows that when  
\begin{align}
  \left|\fnl \right| \leq \frac{5}{3}\frac{1}{\R_{\rm RMS}} = \frac{5}{3
  \Sigmar},  
  \label{eqn4.44}
\end{align}

\noindent the quadratic term of Eq. \eqref{eqn1.3} dominates over
the linear term, and in the computation of the three-point function
Eq. \eqref{eqn1.6}, the loop contributions to the correlators become
dominant. For the values of $\R_{\rm RMS}$ required to form a
significant number of PBHs, the limit on  $|\fnl|$ is of order $10$. The
computation of non-Gaussianities in this case goes beyond the scope of
the present work.( For discussions on the loop corrections to the
correlation functions, see \cite{weinberg-qcI},
\cite{zaballa-rodriguez}, \cite{byrnes-diagrams} and
\cite{seery-loops}.)

It is interesting to look at the values allowed for
$\fnl$ from WMAP and test the modifications that large
non-Gaussianities bring to the amplitude of $\R$ at the PBH scale. 
 Fig. \ref{fig4.2} presents the set of bounds on the initial mass
fraction of PBHs listed in Table I. The corresponding limits to the
variance of the curvature $\Sigmar$ are shown in Fig. \ref{fig4.3} for
the Gaussian and non-Gaussian cases. Independently of the model of
cosmological perturbations adopted, one can use the observational
limits on $\fnl$ to modify the bounds for $\Sigmar$ on small
wavelengths. For the non-Gaussian case we choose to plot the central
value of the present limits to $\fnl^{equil.} = 51$ \citep{wmap2008-params}
and the limit value $\fnl = 5/( 3 \Sigmar)\approx - 66 $ mentioned in
Eq.~\eqref{eqn4.44}. The tightest constraints 
on $\Sigmar$ come from perturbations of initial mass $M  \approx
10^{15} \gr$. With the non-Gaussian modification the limit is
$\log{(\Sigmar)} \leq - 1.2$, compared to the Gaussian case
$\log{(\Sigmar)}\leq -1.15$. As shown in Fig. \ref{fig4.3}, the
modification to $\Sigmar$ cannot be much larger if instead the limit
value of Eq. \eqref{eqn4.44} is used.  

\begin{figure}[h!]
  \begin{center}
    \psfrag{A}{\Large$\log{(\beta_{\rm PBH})}$}
    \psfrag{B}{\Large$\log{\left(\frac{M}{1 \gr}\right)}$}
    \includegraphics[totalheight=.35\textheight]{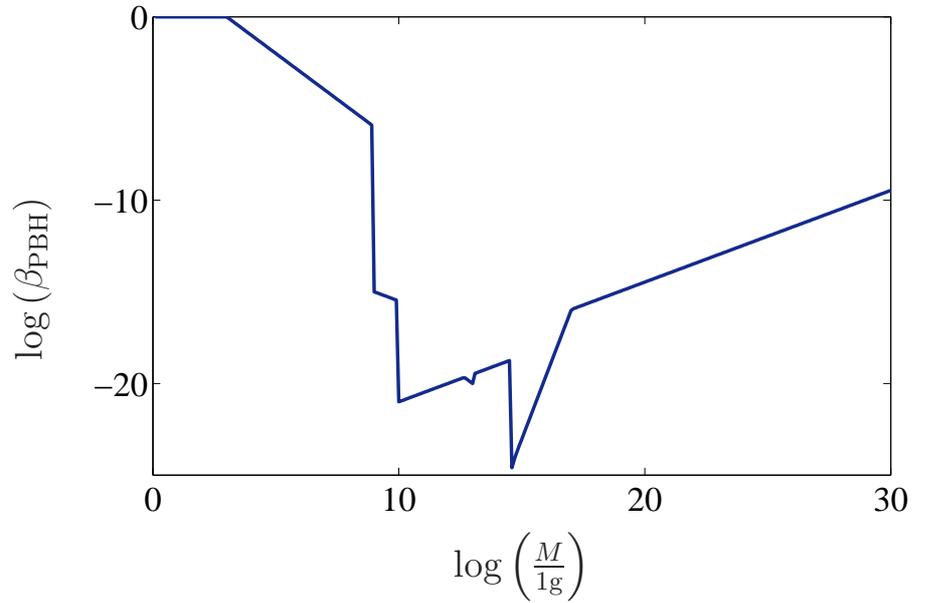}
    \caption{\small{The constraints on $\beta_{\rm PBH}$ in Table I
        are plotted together 
	with the smallest value considered for each
	mass.}}\label{fig4.2}
  \end{center}
\end{figure}

\begin{figure}[h!]
  \begin{center}
    \psfrag{F}{\Large$\log{(\Sigmar)}$}
    \psfrag{E}{\Large$\log{\left(\frac{M}{1 \gr}\right)}$}
    \includegraphics[totalheight=0.35\textheight]{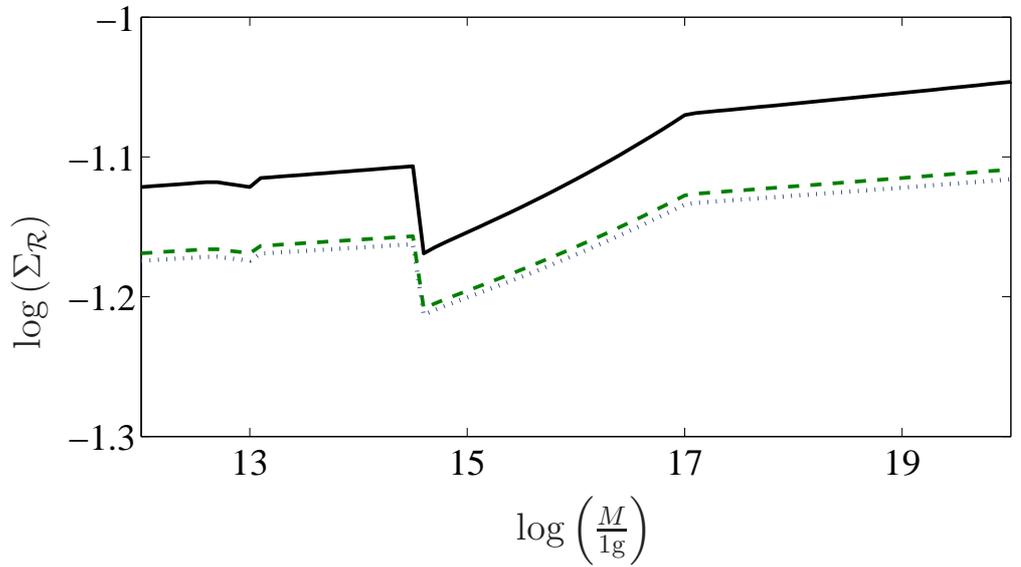}
    \caption{\small A subset of the constraints on $\Sigmar$ from
      overproduction of PBHs 
      is plotted for a Gaussian and non-Gaussian correspondence
      between $\beta$ and $\Sigmar$, Eqs. \eqref{eqn4.22} 
      and \eqref{eqn4.4} respectively. The dashed line assumes a
      constant $\fnl = 51$ and the dotted line a value $\fnl = - 1
      / \Sigmar^2 \approx - 66$. The solid line represents the
      constraints for in the Gaussian case}\label{fig4.3}. 
  \end{center}
\end{figure}

\section{Closing remarks}
\label{closing-ngpbhs}

The present chapter shows, to lowest order in the contribution of the
bispectrum, the effects of non-Gaussian perturbations on PBHs
formation.  Using curvature perturbations with a non-vanishing
three-point correlation, an explicit form of the non-Gaussian PDF is
presented, which features a direct contribution from the non-Gaussian
parameter $\fnl$. Furthermore, it is shown how the sign of this
parameter determines the enhancement or suppression of probability for
large-amplitude perturbations. Using the simple slow-roll expression
for $\fnl$ in the context of single field inflation, a previous
discrepancy in the literature regarding effects of non-Gaussianity on
the abundance of PBHs has been solved.

As a second application of the non-Gaussian PDF presented here is
to use the Press-Schechter formalism of structure formation to
determine the non-Gaussian effects on PBH abundance. In section
\ref{ngbeta-pbhs} it is shown how the PBH constraints on the amplitude
of perturbations can be modified when a non-Gaussian distribution is
considered. As an example, it is shown that the limit $\Sigmar(M =
10^{15}\gr) < 6.3 \times 10^{-2}$ is reached for the marginal value 
$\fnl = -66.35$ modifying the known bounds for $\R$ at the end of
inflation \citep{carr-lidsey-blue}.   
This limit is, however, much larger than the
observed amplitude at CMB scales, where  $\Sigmar \approx 4.8\times
10^{-5}$. The order of magnitude gap between the mean amplitude
observed in cosmological scales and that required for significant PBH
formation remains almost intact and, as a consequence, non-Gaussian
perturbations do not modify significantly the standard picture of
formation of PBHs.

\clearemptydoublepage
\chapter{Curvature profiles of large overdensities}
\label{chapterfive}

\section{Introduction}
As mentioned in Chapter \ref{chapterone}, previous
studies of PBH formation take the amplitude of the matter density
or curvature inhomogeneities as the only parameter determining the
probability density of PBH formation. Also the mass fraction in the
form of PBHs is usually calculated with the aid of the Press-Schechter
formula. Here we argue that this rough estimation is
incomplete and that a different approach should be taken to  evaluate
the threshold value $\delta_{\textrm{th}}$,  or the equivalent
curvature inhomogeneity $\R_{\textrm{th}}$, in the investigations of
PBH formation.     

From the first numerical simulations of PBH formation, it was evident
that  the process of PBH formation depends on the
pressure gradients in the collapsing configuration as well as
their amplitude. \cite{nadezhin-novikov} 
found that such pressure gradients can modify the value of
$\delta_{\textrm{th}}$ significantly. This has been confirmed in
more recent works, which describe the configuration in terms of the
curvature inhomogeneity $ \R(\br)$ (note that in this chapter we work
with spherical coordinates $\{\br\}$ and not any set of coordinates
$\{\bx\}$).   As we will show below, the Einstein equations relate the
curvature profiles directly with the internal pressure gradients. This
is the main motivation for considering the probability of curvature
configurations.  

We extend here the Press-Schechter formalism to consider a
two-parameter probability. We include here for the first time a
parameter related to the slope of curvature profile at the edge of the
configuration.  
We start by calculating the probability of finding a spherically
symmetric curvature configuration with a given radial profile. We can
justify the sphericity assumption using the argument of
\cite{zabotin-polnarev}: PBH
formation takes place only from nearly spherical configurations. 
In our analysis, we describe the radial profiles by
introducing two parameters: the central amplitude of the curvature
inhomogeneity $\R(\br = 0)$ and the central second radial derivative
$\R''(\br = 0)$. The introduction of these parameters is a first step
towards the full parametrisation of profiles in terms of all even
derivatives at the centre of configurations. (The odd derivatives are
all zero due to the assumed spherical symmetry.) 

The method presented to derive a multiple-parameter probability
enables us to compute the probability of any number of parameters
describing the curvature profile. However,  only
families of curvature profiles described by two parameters are currently
available, so we limit ourselves to a two-parametric
description. More accurate future codes will simulate PBHs formation
with a larger number of parameters. The number of parameters required
for the complete description of these profiles and their probability
distribution will be the same \footnote[10]{In the context of dark
  matter haloes, the profile of the initial inhomogeneity  is
  effectively irrelevant because galaxies are formed from pressureless
  configurations. The density profiles and shapes of virialised haloes
  result from the evolution of the initial high peaks and are not
  linked to the profile of initial configurations that we investigate here}.  

The central amplitude $\R(0)$ has been used in previous calculations of
gravitational collapse and the probability of PBH
formation, as illustrated in the previous chapter. Here we compute the
probability to find a given configuration as a function of the two
parameters $[\R(0),\R''(0)]$. We subsequently illustrate how this
two-parametric probability is used to correct the probability of PBH 
formation. For this purpose we use the results of the latest numerical
simulation of PBH formation \cite{musco-polnarev}.
Such an exercise shows
how the corrections to $\beta_{\mathrm{PBH}}$ are potentially
significant and they will be considered in more detail in future
studies of PBH formation.

\section{Probability of profile parameters of cosmological
  perturbations}\label{prob:construction}

Formally, the high amplitude inhomogeneous profiles describing
configurations which collapse into PBHs are not perturbations.
However, such regions are included in the statistics of random
primordial curvature perturbations in the sense that the statistics of random
fields can be used to estimate the probability of finding high-amplitude
inhomogeneities.
To describe such inhomogeneities, we consider the nonlinear
curvature field $\R(t,r)$, as first described by \cite{salopek-bond}.
The nonlinear curvature $\R(t,r)$, defined in terms of the metric
in the following equation, represents the relative expansion of a
given local patch of the universe with respect to its neighboring
patches. It is described by 
the metric 
\begin{align}
  \d s^2 = - N^2(t,\vect{r}) \, \d t^2 + a^2(t) \e{2\R(t,\vect{r})}
  \tilde{\gamma}_{ij} 
  (\d r^i + N^i(t,\vect{r}) \, \d t)( \d r^j + N^j(t,\vect{r}) \, \d
   t), \label{0.1}
\end{align}

\noindent where $a(t)$ and $\tilde\gamma$ are the usual scale factor
and the intrinsic metric of the spatial hypersurfaces. The
gauge-dependent functions $N$ and $N^i$ are the lapse function and
shift vector, respectively. These variables are determined by
algebraic constraint equations in terms of the matter density$~\rho$,
pressure$~p$ and metric variables $\R,~a$ and $\tilde{\gamma}_{ij}$. 

Here we consider the nonlinear configurations which
correspond to large $\R$ inside some restricted volume
and zero $\R$ outside, where the expansion of the universe follows the
background FRW solution.
There are several advantages of working with metric
\eqref{0.1}. First, as shown in Chapter \ref{chaptertwo}, $\R$ is
defined as a gauge-invariant combination of the metric and matter
variables \citep{wands-malik}. Second, with the aid of the gradient
expansion of the metric quantities, $\R(\br,t)$ appears in the
Einstein equations in a non-perturbative way 
\citep{starobinsky86,salopek-bond,deruelle-langlois,rigopoulos-shellard}. 
Third, $\R$ does not depend on time for  scales larger than the
cosmological  horizon, as proved by \cite{lyth-zeta} and
\cite{langlois-vernizzi-b}. In the present chapter we work in the
superhorizon r\'egime, where the field $\R(\br)$ can be assumed to be
time-independent. 

The primordial field of random perturbations we consider presents a Gaussian
probability distribution. The expressions for the PDF of the parameter
$\R(0)$  in Chapter \ref{chapterthree} are recovered here. For
convenience we use a different notation, replacing $\varrho$ in
Section \ref{sec:harmonic} with the amplitude $ \vartheta_0 \equiv
\R(\vect{0})$ and the variance with $\Sigma_{(2)} \equiv \Sigmar$. The
PDF for the central amplitude $\vartheta$ is identical to that in
Eq.\eqref{gauss:gauss}:  
\beq
	\mathbb{P}[\vartheta_0] =  \frac{1}{\sqrt{2\pi}\Sigma_{(2)}} 
        \exp\left[-\frac{\vartheta_0^2}{2\Sigma_{(2)}^2}\right].
	\label{pdf:r0}
\eeq

We now derive the density of the probability for the central second
derivative to have amplitude   
\begin{align}
  \vartheta_2  \equiv \R''(0) = \left[\frac{\partial^2}{\partial
  r^2}\R(r)\right]_{r = 0}. 
\end{align}

\noindent In order to compute the probability of a specific property of
$\R(\br)$, we integrate the original PDF, which encodes all the
information about the field, weighted with the Dirac 
$\delta$-functions of relevant arguments. Hereafter we assume $\R(0)$ and
$\R''(0)$ as statistically independent parameters. The validity of
this assumption is not explored here but is left for future
investigations.  Following this assumption, the
probability of having $ \R''( 0)= \vartheta_2$ is given by the integral
\begin{equation}
  \mathbb{P}(\vartheta_2) = \int[d\R]
  \mathbb{P}(\R)\, \delta\left[\R''(0) - \vartheta_2
  \right],
   \label{delta:integralddr}
\end{equation}

  \noindent where $[d\R]$ indicates integration over all possible
configurations $\R(\bk)$ in Fourier space. In order to compute this integral, we expand the smoothed curvature perturbation profile $\bar\R(\br)$ in terms of spherical harmonic functions:
\begin{align}
     \bar \R({\br}) = &{\int}
    \frac{d^3k}{(2\pi)^3}\bar\R({\bf k})\exp{(\textrm{i}{\bf k\cdot r})},
    \label{1.3a}
\end{align}

 \noindent with
\begin{align}
  \bar\R(\bk) = &\sum_{\ell = 0}^\infty\,
    \sum_{m = - \ell}^{\ell}\,  \sum_{n = 1}^\infty\,
    \R^m_{\ell|n}\, Y_{\ell m}(\theta,\phi)
    \psi_n(k). 
    \label{1.3b}
\end{align}

\noindent Here $Y_{\ell m}$ are the usual spherical harmonics on the
unit 2-sphere and $\psi_n(k)$ are a complete and orthogonal set of
functions in an arbitrary finite interval $0\,<\, k \,<
\,\Lambda$. (The explicit expression for $\psi(k)$ and the value of
$\Lambda$ are given by Eq.~\eqref{harmonic:psi} of Chapter
\ref{chapterthree}.)  
The coefficients in the expansion are generically complex,
so we separate the real and  imaginary parts by  introducing $\R^m_{\ell|n} =
a^m_{\ell|n} + \textrm{i}b^m_{\ell|n} $. The reality condition for the
curvature field,  ${\R}^{*}(\bk) = \R(-\bk) $, is met when
\bea
    \label{1.4}
    a^{-m}_{\ell|n} = (-1)^{\ell+m}a^m_{\ell|n}, \\
    b^{-m}_{\ell|n} = (-1)^{\ell+m+1}b^m_{\ell|n}.\label{1.41}
\eea

\noindent In particular, the $m = 0$
modes require $a^{0}_{\ell|n}$ and $b^{0}_{\ell|n}$  to be
zero for odd and even $\ell$, respectively. To integrate
\eqref{delta:integralddr} we use the Fourier expansion \eqref{1.3a} so
that 
\begin{align}
    \R''(0) = &\int \frac{d^3k}{(2\pi)^3} \R(k)
    (\textrm{i} k)^2 \exp{(\textrm{i}{\bf k\cdot r})}\vert_{r = 0}.
\end{align}

\noindent Furthermore, we use \eqref{1.3b} and the orthogonality of
the spherical harmonics  
\begin{align}
  \int\,  \,Y^m_\ell(\theta,\phi)\sin\theta\, d\theta d\phi =  \sqrt{4\pi}\,\delta^{m
  0}\,\delta_{\ell 0}, 
\end{align}

\noindent to obtain
\begin{align}
   \R''(0) = & - \sum_{n = 1}^\infty \left(a^0_{0|n} +
    \sqrt{\frac{4}{5}}a^0_{2|n}\right) \int \frac{dk}{\sqrt{\pi}(2\pi)^2}
    \psi_{n}(k) k^4 \equiv \vartheta_2.\label{1.6}
\end{align}

 To compute the probability \eqref{delta:integralddr} we proceed by
 integrating over all configurations in Fourier space. With the aid of
 the expansion \eqref{1.3b} we can express the measure of the integral
 in terms of the expansion coefficients satisfying the
 reality conditions \eqref{1.4} and \eqref{1.41}, as shown in
 Eq.~\eqref{harmonic:expand} this is 
\begin{align}
\begin{split}
     \int\,\Psi[\R]\,[\d \R] =
     \Bigg[ \prod_{\ell =0}^{\infty} \prod_{m=1}^{\ell}\prod_{n=1}^{\infty}&
       \mu \int_{-\infty}^{\infty}\,\Psi[\R]\,\d a^m_{\ell|n}
     \int_{-\infty}^\infty \,\Psi[\R]\, \d b^m_{\ell|n}
     \Bigg]\times\\
     \Bigg[&\prod_{\substack{p=0 }}^\infty \prod_{q = 1}^\infty
       \tilde{\mu} \int_{-\infty}^\infty\,\Psi[\R]\, \d a^0_{2p|q}
     \int_{-\infty}^\infty \,\Psi[\R]\,  \d b^0_{2p+1|q} \Bigg],\label{2.1}
     \end{split}
\end{align}

\noindent for any given functional $\Psi$ of $\R(\bk)$. The
constants $\mu$ and $\tilde{\mu}$ are weight factors to be included
in the final normalisation of the joint probability.

The Gaussian PDF we are restricted to is written in
terms of the spherical harmonic coefficients as (cf. Eq.~\eqref{gaussian:harmonics}) 
\begin{align}
\begin{split}
    \mathbb{P}[\R] = \exp\Bigg( - \frac{1}{2 \pi^2(2\pi)^3}
   &\sum_{\ell = 0}^\infty \sum_{m = 0}^{\ell}  \sum_{n = 1}^\infty
    \left[ |a^m_{\ell|n}|^2 + |b^m_{\ell|n}|^2 \right] \\
   & - \frac{1}{4 \pi^2(2\pi)^3}
    \sum_{p = 0}^\infty \sum_{q = 1}^\infty
   \left[ |a^0_{2p|q}|^2 + |b^0_{2p + 1|q}|^2 \right] \Bigg). \label{2.2}
   \end{split}
\end{align}

\noindent In order to obtain the probability  in
   Eq.~\eqref{delta:integralddr}, we use the standard representation
   of the Dirac 
$\delta$-function 
\begin{align}
  \delta(x) = \int^{\infty}_{-\infty} \,dz \,\exp[\textrm{i} z\,x].
\end{align}

\noindent  This allows us to write the $\delta$-function in Eq. \eqref{delta:integralddr} in
terms of the spherical harmonic coefficients as
\begin{align}
  \delta\left[\, \R''(0) - \vartheta_2\, \right] =  \int \,dz
    \exp\left[\textrm{i} z\left( \sum_{n = 1}^\infty \left(a^0_{0|n} +
    \sqrt{\frac{4}{5}}a^0_{2|n}\right) \int \frac{dk}{\sqrt{\pi}(2\pi)^2}
    \psi_{n}(k) k^4 +  \vartheta_2\right) \right]. \label{2.3}
\end{align}

\noindent We now have all the elements required to integrate
probability density of finding $\R''(0)$ with amplitude
$\vartheta_2$. Substituting expressions \eqref{2.2} and \eqref{2.3}
into Eq. \eqref{delta:integralddr}, we perform  the functional
integral with the aid of the decomposition \eqref{2.1}. For this case
we have 
\begin{align}
   \mathbb{P}(\vartheta_2) \propto \int \,[d \R]\int \,dz
   \,\mathbb{P}[\R] \exp\left[\mathrm{i}z \left(
   \frac{3(2\pi)^3\vartheta_2}{\sqrt{4\pi}} 
   +\sum_n  \Sigma_{n}^{(4)}\left(\sqrt{\frac{4}{5}} a^0_{2|n} + a^0_{0|n}\right)
   \right)\right], \label{proba:ddr}
 \end{align}

\noindent where we have simplified the expression by defining the factor 
\begin{align}
  \Sigma_n^{(4)} = \int \,dk \,k^4 \psi_n(k).  \label{sigma:four}
\end{align}

\noindent In the process of integration, we discard all the Gaussian integrals
because they contribute to the probability only with a multiplicative
constant which will be included in the final normalisation. On the
other hand, the Dirac
$\delta$-function contributes to the integral with exponential functions of $a_{0|n}^0$ and $a_{2|n}^0$. The integrals of these parameters are
computed by completing squares of the exponential arguments. 
First, we collect the terms of the integral with factors
of $a^0_{0|n}$, that is,  
   \begin{align}
     \exp\left[- \frac{1}{4\pi^2(2\pi)^3}\sum_{n=1}^\infty |a^0_{0|n}| +
     \mathrm{i}z \sum_{n=1}^\infty |a^0_{0|n}|\Sigma_n^{(4)}\right].
   \end{align}

  \noindent Completing the squares, this last expression becomes
    \begin{align}
  \exp\left[- \frac{1}{4\pi^2(2\pi)^3}\sum_{n=1}^\infty \left(|a^0_{0|n}| -
     \mathrm{i}(2\pi)^3 2\pi^2 z \Sigma_n^{(4)}\right)^2 - (2\pi)^3
     \pi^2 z^2 \Sigma_{(4)}^2\right]. \label{compsq:a0}    
  \end{align}

\noindent  In the same way we can complete the squares for the expansion factors
  $a^0_{2|n}$:
\begin{align}
     \exp&\left[-\frac{1}{4\pi^2(2\pi)^3}\sum_{n=1}^\infty |a^0_{2|n}| +
     \mathrm{i}z \frac{4}{5}\sum_{n=1}^\infty
     |a^0_{2|n}|\Sigma_n^{(4)}\right] = \label{compsq:a2} \\
     &\exp\left[-\frac{1}{4\pi^2(2\pi)^3}\sum_{n=1}^\infty
     \left(|a^0_{0|n}| - \imag (2\pi)^3 \frac{4\pi^2}{\sqrt{ 5}}z
     \Sigma_n^{(4)}\right)^2 - (2\pi)^3 \pi^2 \frac{4}{5} z^2
     \Sigma_{(4)}^2\right].\notag
   \end{align}

\noindent  Finally we can complete the squares for the terms containing
the variable $z$, these being  independent of $a^0_{0|n}$ and $a^0_{2|n}$:
\begin{align}
  \exp&\left[ -
    (2\pi)^3\pi^2\left(\frac{9}{5}\right)z^2\Sigma_{(4)}^{2}  +
    \mathrm{i}\frac{3 (2\pi)^3}{\sqrt{4\pi}}  \vartheta_2 z \right]  =
     \\ 
    &\exp \left[- (2\pi)^3 \pi^2
    \left(\frac{9}{5}\right)\Sigma_{(4)}^{2} \left(z - 
    \mathrm{i}\frac{5 }{12\sqrt{\pi^5}}
    \frac{\vartheta_2}{\Sigma_{(4)}^2} \right)^2 - \frac{5}{2}
    \frac{\vartheta_2}{\Sigma_{(4)}^{2}}\right],\notag
\end{align}

\noindent where for simplification we have written
\begin{align}
  \Sigma_{(4)}^2 \equiv  \sum_{n=1}^{\infty}\left(
  \Sigma_n^{(4)}\right)^2.   \label{sumsigma:four} 
\end{align}

\noindent  So by making the change of variables
\begin{align*}
  a^0_{0|n} \mapsto\,\, & a^0_{0|n}  + \mathrm{i} 2   \pi^2 (2\pi)^3
  \Sigma_n^{(4)} z,\\ 
  a^0_{2|n} \mapsto\,\, & a^0_{2|n}  +  \mathrm{i} \frac{4  \pi^2}{\sqrt{5}} 
  (2\pi)^3 \Sigma_n^{(4)} z\\
 \qquad z \mapsto\,\, & z + \mathrm{i}\frac{5}{12 \sqrt{\pi^5}}
  \frac{\vartheta_2}{\Sigma_{(4)}^{2}},
\end{align*}
we can perform all the integrals and eliminate the Gaussian ones
which contribute only up to an overall numerical factor subsequently
  absorbed by normalisation. The remaining factor expresses the
  probability of finding a perturbation $\R$ with a central second
  derivative of value $\vartheta_2$:
  \begin{align}
    \mathbb{P}\left[ \R''(\br = 0 ) = \vartheta_2 \right] \propto \exp\left(-
    \frac{5 \vartheta_2^2}{2 \Sigma_{(4)}^{2}}\right).
    \label{pdf:ddr}
  \end{align}

\noindent The quantity  $\Sigma_{(4)}^2$ represents the `variance' of
the PDF for $\R''(0)$. To evaluate this variance we integrate
Eq.~\eqref{sumsigma:four} and  use the property
\eqref{harmonic:orthonormal} in Chapter \ref{chapterthree} to
integrate the complete sum and obtain 
\begin{align}
  \Sigma_{(4)}^2 = \int_0^{\Lambda} d \ln{k}\,
  \window^2(k,k_H)\ps(k)\,k^4.
  \label{sigma4:kh}
\end{align}

The final probability density for the pair of parameters
$\R(0)$ and $\R''(0)$ is the product  of Eqs.~\eqref{pdf:r0} and \eqref{pdf:ddr}
\begin{align}
    \mathbb{P}\left(\,\R(0) = \vartheta_0,\, \R''(0) =
    \vartheta_2  \,\right) =  A
    \,\exp\Big(-\frac{\vartheta_0^2}{2\Sigma_{(2)}^2}
    -\frac{5\,\vartheta_2^2}{2\Sigma_{(4)}^2} \Big). \label{2.4}  
\end{align}

\noindent Here $\Sigma_{(2)}$ and $\Sigma_{(4)}$ are the dispersion
of the amplitude and the second derivative respectively, and $A$
is a normalisation factor obtained from the condition that the
integral of the joint PDF over all possible values of the two
independent parameters equals unity.  The final normalised joint
probability density is 
\begin{align}
    \mathbb{P}(\vartheta_0,\vartheta_2) =\, \frac{4\sqrt{12}}{2\pi}\,
    \Sigma_{(2)}^{-1}\Sigma_{(4)}^{-1}\,
    \exp\Big(-\frac{\vartheta_0^2}{2\Sigma_{(2)}^2}
    -\frac{5\,\vartheta_2^2}{2\Sigma_{(4)}^2} \Big). \label{2.5}
\end{align}

It is worth mentioning that the standard PDF containing only
amplitudes $\vartheta_0$, Eq.~\eqref{pdf:r0}, is recovered from
Eq.~\eqref{2.4} when we set all gradients in the Hubble
scale equal to zero, i.e. $\nabla \R|_{r = r_H} = 0$. The Fourier
transform of this expression demands $|k_H| \to \infty$. Using this in
Eq.~\eqref{sigma4:kh} means that $\Sigma_{(4)} \to \infty$ and the
argument $\vartheta_{2}$ goes to zero in the probability density
of Eqs.~\eqref{2.4} and \eqref{2.5}. 

\noindent According to the Press-Schechter formalism
\citep{press-schechter}, the PDF is integrated over all
perturbations which collapse to form  the astrophysical objects under
consideration. In this way we calculate the mass fraction of the
universe in the form of such objects. To apply this formalism and
calculate the probability of PBH formation and integrate the 
PDF \eqref{2.5}, we require the range of values  $\R(0)$ and $\R''(0)$
which correspond to PBH formation. In the next section we will obtain
this range with the help of the results of numerical computations
presented by \cite{musco-polnarev}.

\section{The link between perturbation parameters and
  the curvature profiles used in numerical calculations}\label{metrics}

\subsection{Initial conditions}

As demonstrated by the first numerical simulations of PBH formation
\citep{nadezhin-novikov}, whether or not an initial configuration with
given curvature profile leads to PBH formation predominantly
depends on two factors:

$\bullet$ The ratio of the size of the initial configuration $r_0$ to
the size of the extrapolated closed universe 
$r_{\mathsf{k}} = a(t)\int_0^1\, dr / \sqrt{1 - r^2}$,  which is a
measure of the strength of gravitational field  within the
configuration.

$\bullet$ The smoothness of the transition from the region of high curvature
   to the spatially flat FRW universe, which is characterised by the
   width of the transition region at the edge of the initial
   configuration and is inversely proportional to the pressure
   gradients there, strong pressure gradients inhibiting PBH formation
   (This is an argument beyond the Jeans' stability criterion and
   applies to configurations beyond the linear regime).

The numerical computations presented in \cite{musco-polnarev} (hereafter
 PM) give the time evolution of the configurations with initial
 curvature profiles  accounting for the 
above-mentioned factors. In that paper the initial conditions are obtained
with the help of a quasi-homogeneous asymptotic solution valid in the
limit $t \to 0$. This solution to the Einstein equations was first
introduced by  \cite{lifshitz-khalatnikov}; see also
\cite{zeldovich-novikov-II} and \cite{landau-lifshitz}. Following
\cite{nadezhin-novikov},  PM used this 
asymptotic solution to set self-consistent initial
conditions for curvature
 inhomogeneities, the initial curvature
 inhomogeneity being described  by the spherically symmetric curvature
 profile $\K(\hr)$. This sets the 
initial conditions for the process of black hole
formation. Asymptotically, the metric can be presented in terms of
$\K(\hr)$ as 
\begin{align}
  ds^2 = - d\hat{\eta}^2 + s^2(\hat{\eta})\left[ \frac{1}{1 -
      \K(\hr)\hr^2}d \hr^2 + \hr^2 \,\Big( d\theta^2 + \sin^2\theta
      \,d \phi^2 \Big)\right],  \label{4.2} 
\end{align}

\noindent where $\eta$
is the conformal time, $s(\eta)$ is the scale factor for this
metric. As we will show in the paragraph after
Eq. \eqref{angular:elem}, this is
identical to the usual scale factor $a(\eta)$ of a flat Friedmann
universe, only here we use a different notation to distinguish
between metrics. Also, we write $\hr$ for the radial coordinate to
distinguish it from the coordinate of the metric \eqref{0.1}. 

An advantage of working with
this metric is that it contains the curvature profile $\K(\hr) $
explicitly. We choose a set of coordinates with the origin at the 
centre of spherical symmetry and fix $\K(0) = 1$. The condition that
$\K(\hr)$ is a local inhomogeneity requires that $\K(\hr)
=0$ for  radii $\hr$ larger than the scale $\hr_0$ where
the metric matches the homogeneous FRW background.

In PM the profiles $\K(\hr)$ are presented in two forms,
one of which is characterised by two independent parameters $\alpha$ and
$\Delta$ as
\begin{align}
  \K(\hr)= \left[1 + \alpha \frac{\hr^2}{2 \Delta^2}\right] \exp\left(
  -\frac{\hr^2}{2 \Delta^2}\right). \label{K:profile}
\end{align}

\noindent The parameter $\Delta$ describes the width of the Gaussian profile, while $\alpha$ parametrises linear deviations from this profile. The results of the numerical simulations in PM indicate that PBHs are formed in the region of the parameter space $\left[\alpha,\Delta  \right]$ shown in Fig. \ref{fig1}a.

\begin{figure}[h!]
 \centering
 \includegraphics[width = 0.5\textwidth]{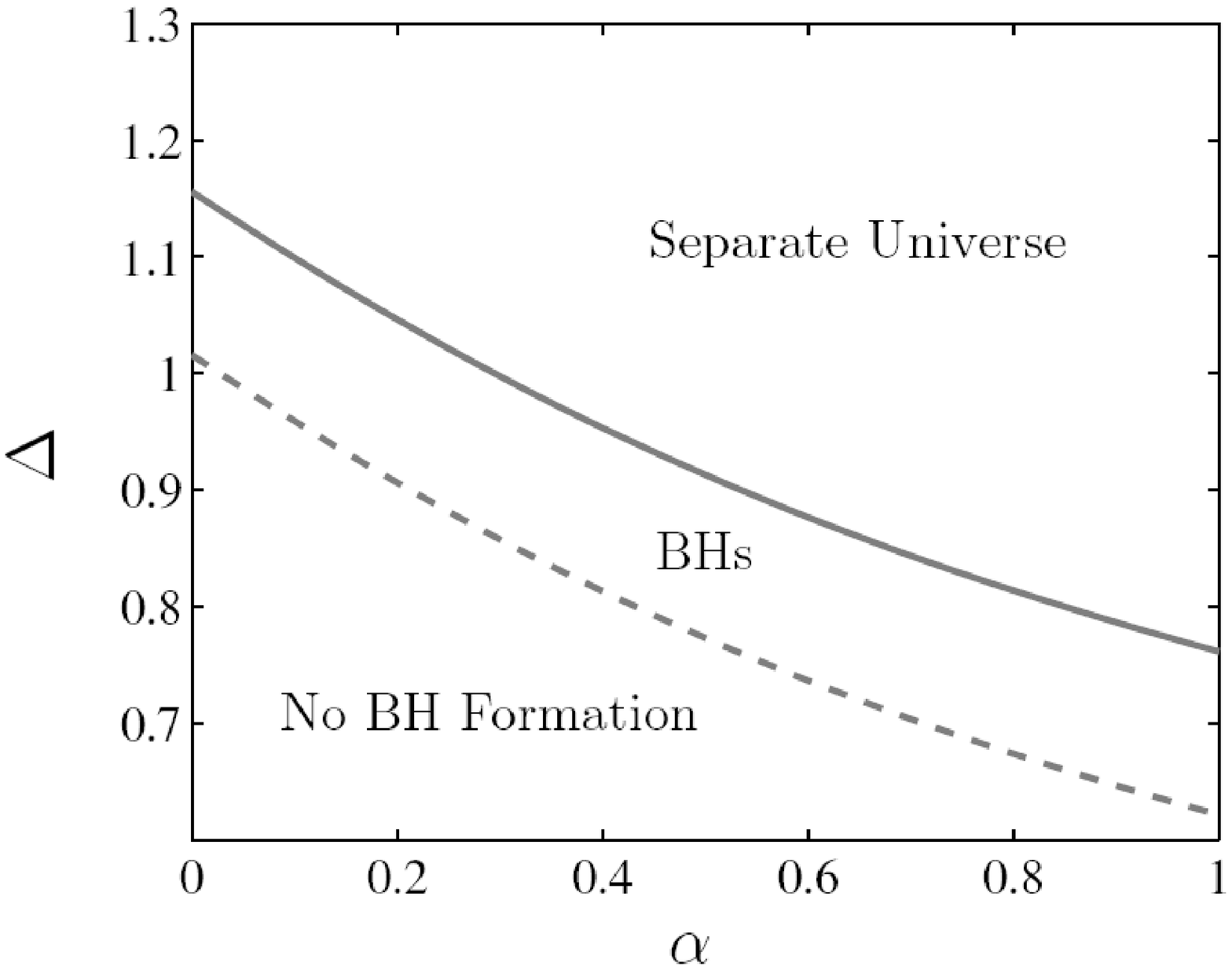}
 \hfill
 \includegraphics[width = 0.5\textwidth]{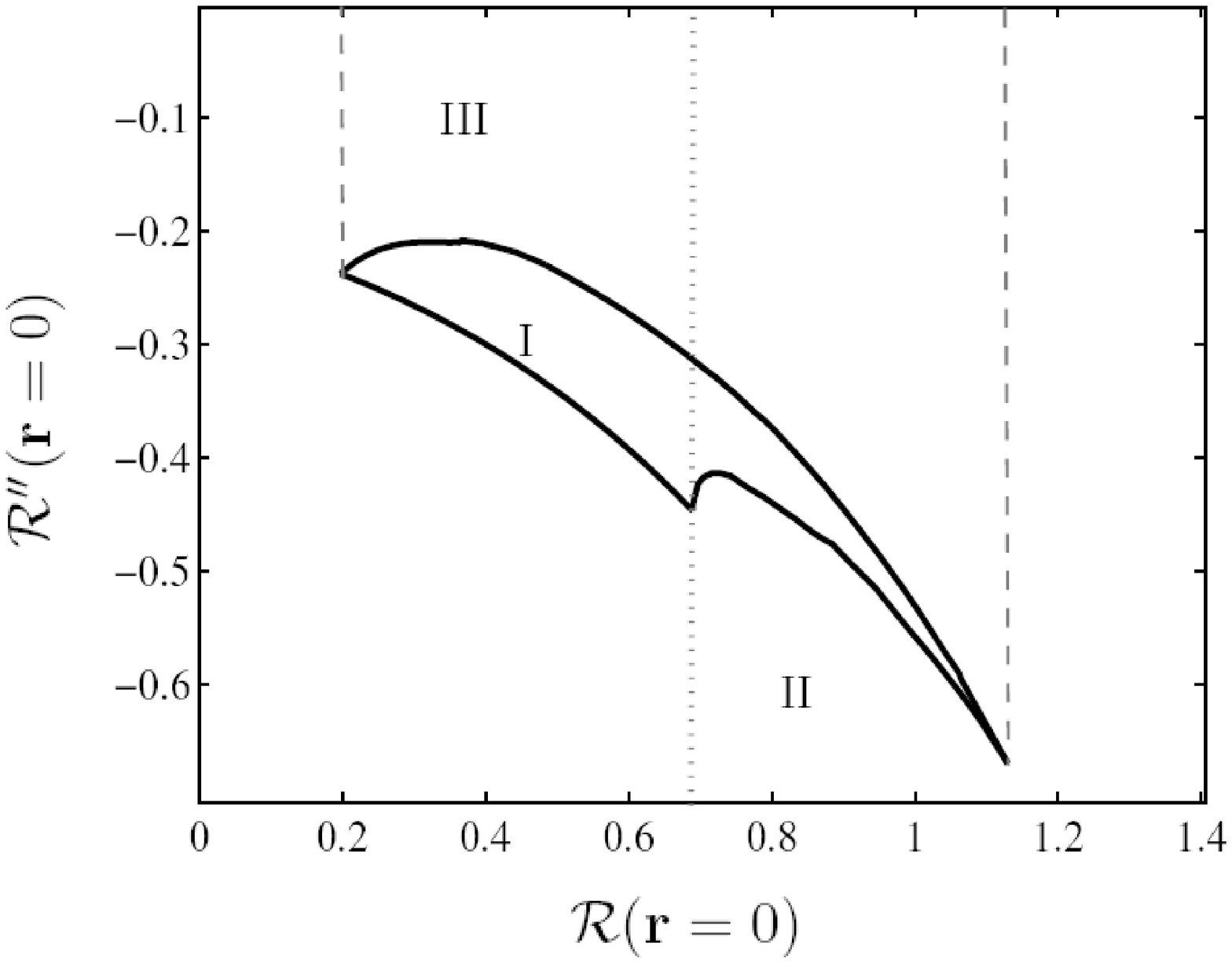}
 \caption{{\small (a) The top plot shows the parameter values for initial
 configurations which collapse to form black holes according to
 \cite{musco-polnarev}. (b) In the $\left[\R(0),\R''(0)\right]$ plane three
 regions of integration are considered to compute the probability of
 PBH formation. Area I is the region enclosed by the solid curves and 
 corresponds to the area denoted by BH in Fig. 1a. Area II is the region
 to the right of the grey dotted line, representing the area of
 integration considered in previous studies where only the amplitude
 is taken into account. Area III is the region above the solid line
 and between the dashed lines. This contains those configurations which have
 a smooth profile in the centre and present the amplitudes $\R(0)$ that
 are found to form PBHs in \citep{musco-polnarev}. The complete
 description of the physical characteristics of profiles with values in this
 region is given in Section \ref{metrics}. }} \label{fig1} 
\end{figure}

\subsection{Physical criteria for the identification of parameters }

We proceed to find the correspondence between the two sets of
parameters, $[\R(0),\R''(0)]$ and $[\alpha,\Delta]$, both of which
describe the initial curvature profiles. First let us note that the
sets of coordinates $\{t,r\}$ and $\{\hat\eta,\hat{r}\}$ are those of
the metrics \eqref{0.1} and \eqref{4.2}, respectively. Thus we require
a relationship between these set of coordinates too. Assuming that the
size of the configuration, $r_0$, is much larger than the Hubble
radius, $r_{H} = H^{-1}$,  we can use the gradient expansion of the
functions in metrics \eqref{0.1} and \eqref{4.2}. In this case, the 
time derivative of any function $f(t,r)$ is of order $f / t \sim
Hf$ and significantly exceeds the spatial gradient which is of
order $f / r_0$. Hence the small parameter in the gradient expansion
is
\begin{align}
  \varepsilon  \,\equiv \,\frac{r_H}{r_0}  = \frac{k}{a H}, \label{epsilon:def}
\end{align}

\noindent where $k$ is the wave-number corresponding to the scale of
the configuration.

For the metric \eqref{0.1}, using the coordinate freedom to set $N^i =
0$ and ignoring any tensor contributions (i.e. $\tilde\gamma_{ij} =
\delta_{ij}$), the expansion of the Einstein equation
$G^{~0}_{0}~=~8\pi G T^{~0}_0$ to order $\varepsilon^2$ can be written
as,\footnote[11]{For the complete second order expansion
  of the metric quantities, see 
  \cite{lyth-zeta} and \cite{langlois-zeta}.} 
\begin{align}
  \frac{1}{2}\left(\frac{6\dot{a}^2}{a^2}  + ^{(3)}\Rsp -
  \frac{4\dot{a}^2}{a^2} (N - 1) \right) + \mathcal{O}(\varepsilon^{4})  =
  8\pi G\, (\rho_0 + \delta \rho) + \mathcal{O}(\varepsilon^{4}),
  \label{einstein:00}
\end{align}

\noindent where $^{(3)}\Rsp$ is the spatial curvature, or the Ricci
scalar for the spatial metric $g_{ij}$. To zero order in $\varepsilon$, we have
\begin{align}
   \frac{3 \dot{a}^2}{a^2} = 8\pi G\,\rho_0, \label{einstein:hom}
\end{align}

\noindent which corresponds to the homogeneous part of Eq.
\eqref{einstein:00}. 
The time-slicing can be taken to be the uniform expansion gauge in which
\begin{align}
  N - 1 =  - \frac{1 + 3 w}{1 + w} \delta_{\rho} + 
  \mathcal{O}(\varepsilon^4),\label{N:delta}
\end{align}

\noindent where $w$ is the equation of state
\citep{shibata-asada,shibata-sasaki,tanaka-sasaki}. Using 
\eqref{einstein:00},\eqref{einstein:hom} and \eqref{N:delta}, we find
the equivalence between the spatial curvature and the matter overdensity:
\begin{align}
  ^{(3)}\Rsp = \frac{8 \pi G}{3}\delta \rho  \,\left(\frac{7 +
  3w}{3 + 3w}\right).  \label{einstein:inhom} 
\end{align}

\noindent In consequence, the gradients if this quantity relates to
the pressure gradient: 
\begin{align}
  \nabla {^{(3)}\Rsp} = \frac{8 \pi G}{3} \,\frac{7 +
  3w}{3(w + 1)} \nabla\left(\delta \rho\right)  = \frac{8 \pi G}{3} 
  \,\left(\frac{7 + 3w}{3w\left(w +1\right)}\right) \nabla p,
  \label{einstein:grad} 
\end{align}

\noindent where $\nabla = (g_{rr})^{-1/2} \d / \d r$. Hence, subject
to the two physical conditions at the edge of the configuration listed
at the beginning of Section \ref{metrics}, we relate the profiles
$\R(r)$ and $\K(\hr)$ by equating the spatial curvature and its
gradient for the metrics \eqref{0.1} and \eqref{4.2}. That is,  
\begin{align}
   ^{(3)}{\Rsp} =  - \left[2 \R''(r) + \left(\R'(r)\right)^2
  \right] \exp(-2 \R(r)) = 3\K(\hr) + \hr \K'(\hr), \label{3curv:eq}
\end{align}

\noindent and
\begin{align}
  \frac{1}{\sqrt{\mathrm{g}_{rr}}}\frac{\d }{\d r} &\left( ^{(3)}\Rsp
  \right) = \notag\\
   - \left[ \R'\R'' + \R''' \right]& \exp(-3 \R(r)) =
   \left[\frac{1 - \K \hr^2}{\hr^2}\right]^{1/2}
   \left(2 \hr \K'(\hr) + \frac{1}{2}\hr^2 \K''(\hr)\right). \label{3curv:der}
\end{align}

\noindent By definition, the 3-curvature must vanish at the edge of
  the configuration, so Eq. \eqref{3curv:eq} implies 
\beq
   2 \R''(r_0) + \left(\R'(r_0)\right)^2 = 0 \label{3curv:r-0}
\eeq

\noindent and
\beq
   3\K(\hr_0) + \hr_0 \K'(\hr_0) = 0. \label{3curv:hr-0}
\eeq

\noindent Thus the gradient relation
\eqref{3curv:der} can be written as
\begin{align}
    \left[\R'(r_0)^3 - 2 \R'''(r_0) \right]& \exp(-3 \R(r_0)) =
   \left[\frac{1 - \K \hr_0^2}{\hr_0^2}\right]^{1/2}
   [-12 \K(\hr_0) + \hr_0^2 \K''(\hr_0)]. \label{3curv:der-0}
\end{align}

\noindent This establishes a relation between $\R(r)$ and $\K(\hr)$
at the edge points $r_0$ and $\hr_0$. The configuration $\K(\hr)$ is
parametrised by $[\alpha, \Delta]$, as shown in Eq.
\eqref{K:profile}. As follows from condition \eqref{3curv:hr-0}, the radius $r_0$ can be written in terms of those parameters as
\begin{align}
  \hr_0^2 =\left( \frac{5 \alpha  -  2 + \sqrt{(5\alpha - 2)^2 -
      24\alpha}}{2\alpha} \right)\Delta^2. \label{rhat:0}
\end{align}

\noindent  Then we use two more equations obtained from the conformal
transformation of coordinates at zero order in
$\varepsilon$:
\begin{align}
  a^2(\tau) \e{2\R(r)}\,dr^2 = s^2(\eta)\frac{d\hr^2}{{1 -
      \K(\hr)\hr^2}}
  \label{radial:elem}
\end{align}
\noindent and
\begin{align}
    a^2(\tau)\, \e{2\R(r)}\, r^2\, d\Omega^2 = s^2(\eta)\, \hr^2\,
    d\Omega^2.
    \label{angular:elem}
\end{align}

\noindent Asymptotically, in the limit $[r,\hr] \to \infty$, the
homogeneous Einstein equations are identical in both metrics,
therefore, the homogeneous scale factors $a(\tau)$ and $s(\eta)$ can be
identified. Thus we find a relation between
the radial coordinates,
\begin{align}
 \e{\R(r)}\, r =  \hr,
  \label{r:hr}
\end{align}

\noindent and an integral relation between the configurations,
\begin{align}
   \int_0^r \e{\R(x)}\,dx =  \int_0^{\hr} \frac{dx}{\sqrt{1 -
   \K(x)x^2}}.
   \label{radial:integral}
\end{align}

\noindent One can verify that Eqs.
\eqref{3curv:eq}, \eqref{3curv:der} and \eqref{radial:integral} are
not independent. For example, Eq. \eqref{3curv:der} follows from
\eqref{3curv:eq} and \eqref{radial:integral}. 

{{In the previous section we have developed a method to account
    for the probability of any set of
    parameters describing the curvature profile. For simplicity we
  have chosen the pair $[\R(0),\R''(0)]$. We now illustrate how to relate
    $[\R(0),\R''(0)]$ and $[\alpha, \Delta]$ by considering the parabolic
    profile}
  \begin{align}
 \R(r) =\R(0) + \frac{1}{2}\R''(0)\,r^2. \label{param:expansion}
  \end{align}

  \noindent {This parametrisation meets the minimal requirement of
    covering the $[\alpha, \Delta]$  parameter space  in Fig. \ref{fig1} (a).}}

\noindent Eqs. \eqref{3curv:r-0}, \eqref{radial:integral} and \eqref{r:hr}
are now reduced to the following system of algebraic equations:
\begin{align}
  &r_0^2 = - \frac{2}{\R''(0)}, \label{x:0}\\
  \R(0) = 2 \log&\left(\frac{2}{\textrm{erf}(1)}\left[\pi \exp(1) \hr_0
    \right]^{-1/2} \int_{0}^{\hr_{0}} \frac{dx}{\left(1  -
    \K(x)x^2\right)^{1/2}}\right), \label{R:centre}\\
  \R''&(0) = - 2 \frac{\exp(2 \R(0) - 2)}{\hr_0^2},\label{ddR:centre}
\end{align}
where $\hr_0$ is given in terms of $[\alpha, \Delta]$ by Eq.
\eqref{rhat:0}.

\subsection{Parameter values leading to PBH formation}
\label{3c}

The numerical computations of PM,
which used the parametrisation \eqref{K:profile}, show that PBHs are formed in
the [$\alpha$, $\Delta$] region shown in
Fig.~\ref{fig1}(a). Eqs. \eqref{R:centre} and
\eqref{ddR:centre} map this region to Area I in the [$\R(0)$, $\R''(0)$] plane shown in Fig.~\ref{fig1}b. The
Jacobian of the transformation corresponding to this mapping is
non-vanishing, which guarantees a one-to-one correspondence of the
`BH' region in Fig.~\ref{fig1}a with Area I in
Fig.~\ref{fig1}b. Each point here corresponds to a parabolic profile
which leads to PBH formation. 

For each one of these parabolic profiles, there is a family of
non-parabolic profiles with the same central amplitude $\R(0)$, the
same configuration size $r_0$, and the same behaviour near
the edge, as shown in Fig. \ref{fig11}. In this
figure, the profiles lying below the parabola correspond to larger
absolute magnitudes of $\R''(0)$ and do not form PBHs because they
have lower average  gravitational field strength and
higher average pressure gradient. The non-parabolic
profiles which lie above the parabolic one (with smaller absolute
magnitude $\R''(0)$) should also collapse to form PBHs because they
correspond to higher average gravitational field strength and
lower pressure gradient.

\begin{figure}[h!]
  \begin{center}
    \includegraphics[totalheight=0.40\textheight]{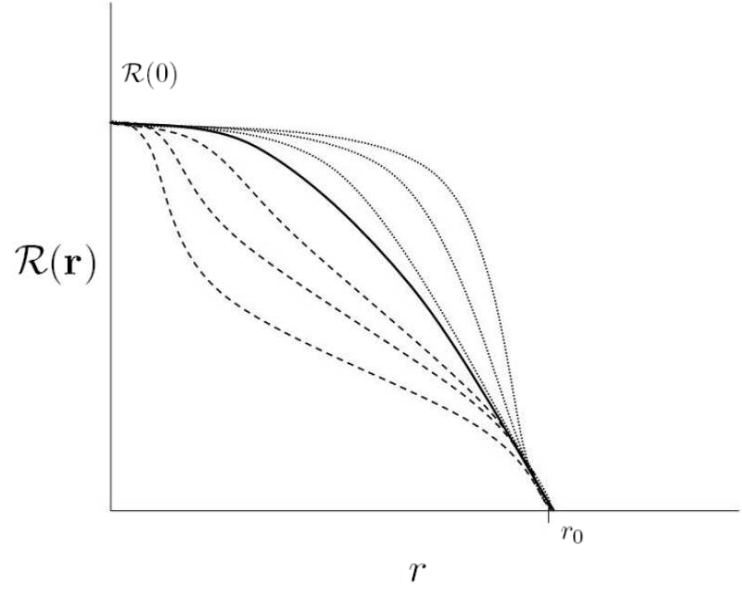}
    \caption{{\small The curvature profile for three different families of
      configurations with common central amplitude $\R(0)=1$. The
    configurations shown by the dashed lines have values of $\R''(0) $
    larger in absolute magnitude than the parabolic one shown in
    black. The configurations shown by the dotted lines have values
    of $\R''(0)$ smaller than the parabolic one. All profiles satisfy
    conditions \eqref{3curv:r-0}  and 
    \eqref{3curv:der-0}.}}\label{fig11} 
  \end{center}
\end{figure}

In the  parameter space  $[\R(0),\, \R''(0)]$, this last set of
profiles  corresponds to Area III in Fig.~\ref{fig1}b. This region
will be included in the calculation of the probability of PBH
formation in the next section.

 \section{Two-parametric probability of PBH formation}
    \label{prob:comparison}

To calculate the probability of PBH
formation, which is equivalent to the mass fraction of the universe
going into PBHs of given mass, it is customary to use the standard
Press-Schechter formalism \citep{press-schechter}. This has been widely used in
previous calculations of the one parametric probability of PBH
formation \citep{carr, carr-lidsey-blue,liddle-green, carr-recent,siri,zaballa-green}.
When the probability depends on a single amplitude
parameter, this method  reduces to  the integration of the
corresponding PDF over the relevant perturbation amplitudes. The 
final integral is equivalent to the mass fraction of PBHs of
mass \citep{carr}
\beq
M \sim w^{3 / 2} M_{H} \approx w^{3 / 2} k_M /
(2\pi)
\eeq

\noindent with the equation of state ${w}$ measured at their formation
time. Here we extend the standard Press-Schechter formalism to include
for the first time an additional parameter accounting for the radial
pressure in the initial configuration. When the
$\left[\R'(0),\,\R''(0)\right]$ area is a square $[\R_1<\R(0)<\R_2$,
  $\R''_1<\R''(0)<\R''_2]$, the integrated two-parametric probability is 
\begin{align}
\begin{split}
\beta_{\rm PBH}(M) = 2 \,\int_{\R_{\rm 1}}^{\R_{\rm
     2}} &\,d\vartheta_0 \int_{\R''_{\rm 1}}^{\R''_{\rm 2}}  \, d \vartheta_2 \,
  \mathbb{P}(\vartheta_0,\vartheta_2) = \\
  &\frac{1}{2}\left[{\rm erf}\left(\frac{\R_{\rm
      2}}{\sqrt{2}\Sigma_{(2)}(M)}\right)-  {\rm
     erf}\left(\frac{\R_{\rm
     1}}{\sqrt{2}\Sigma_{(2)}(M)}\right)\right] \times\\
 & \qquad \qquad\left[{\rm erf}\left(\frac{\R''_{\rm 2}}{\sqrt{2}\Sigma_{(4)}(M)}\right)
    - {\rm erf}\left(\frac{\R''_{\rm 1}}{\sqrt{2}\Sigma_{(4)}(M)}\right)
    \right].
  \label{proba:integral}
  \end{split}
\end{align}

\noindent{{We use this result to integrate numerically over a
    mesh of small squares covering each of the areas of the plane
    $\left[\R(0),\,\R''(0)\right]$ }}shown in Fig. \ref{fig1}b. The
    results of this integration for two different power-law spectra
    $\ps_{\R}(k) \propto k^{n-1}$ are shown in Fig.~\ref{fig2}.

From that figure we note that the probability function $\beta_{\rm
  PBH}(M)$ has a maximum at a value of $M_{\rm max}$ that changes with the 
of the spectral index. This can be easily derived by computing the
solution of  $d \beta_{\rm PBH} / d M = 0 $. We find that this
equation provides a formula for the value $M_{\rm max}$ which indeed
depends sensitively on the spectral index $n_s$. Assuming $n_s > 1$ we
have: 
\beq
  M_{\rm max} = M_{\rm eq} \frac{3 \gamma}{2}\frac{ P_{\rm
      eq}}{\R_{\rm th}^2}\, \exp\left(- \frac{2}{n_s - 1}\right),  
\label{mass:max}
\eeq

\noindent where $M_{\rm eq}$ and $P_{\rm eq}$ are the Hubble mass and the power
spectrum at the time of matter-radiation equivalence ($k_{\rm eq} =
8.9\times\ten{-2} \rm{Mpc}^{-1}$), and $\gamma$ is a factor of order 
unity that changes slightly with the value of $n_s$.  


\begin{figure}[h!]
  \begin{center}
    \includegraphics[width = 0.5\textwidth]{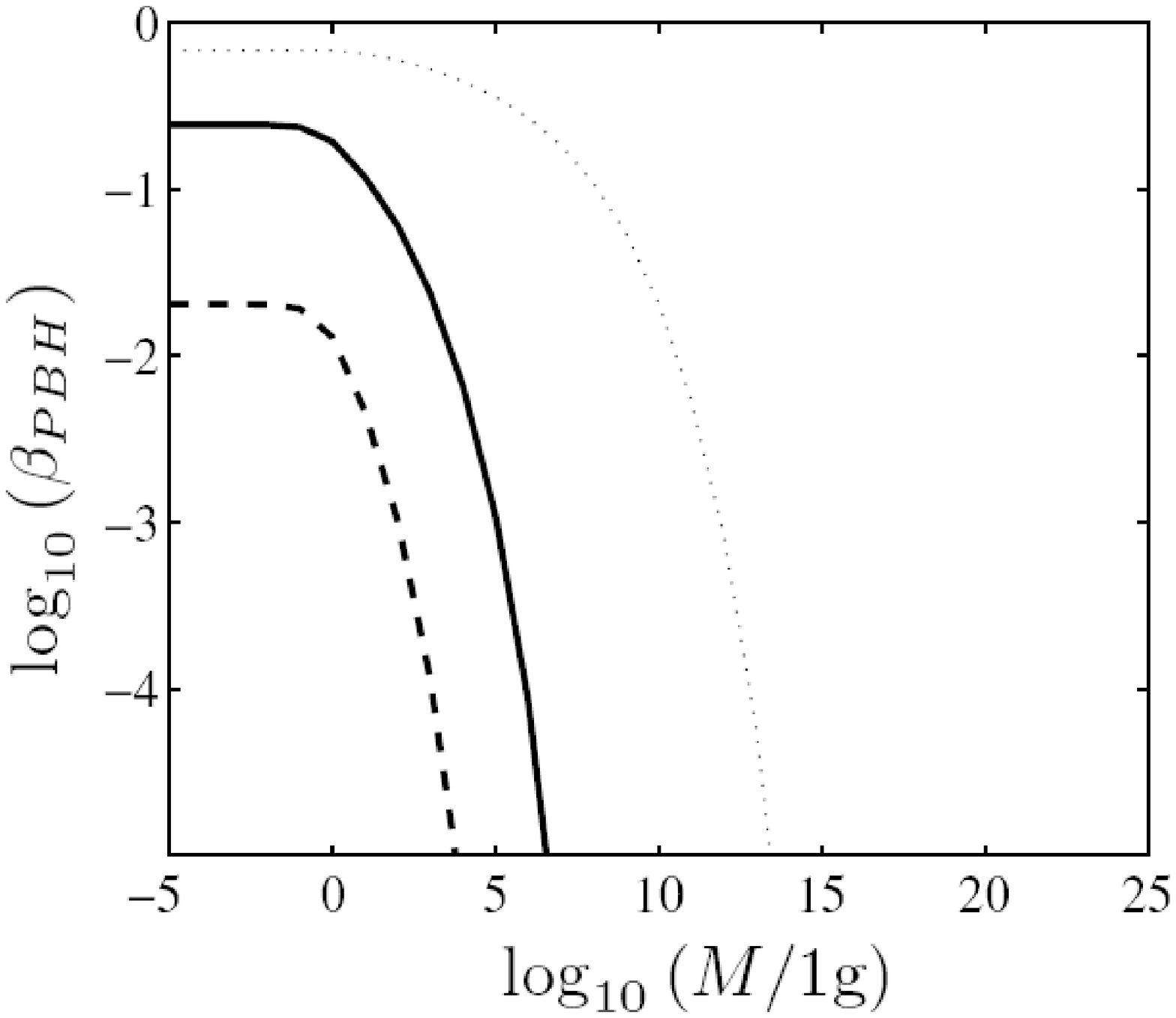}
    \hfill
    \includegraphics[width = 0.5\textwidth]{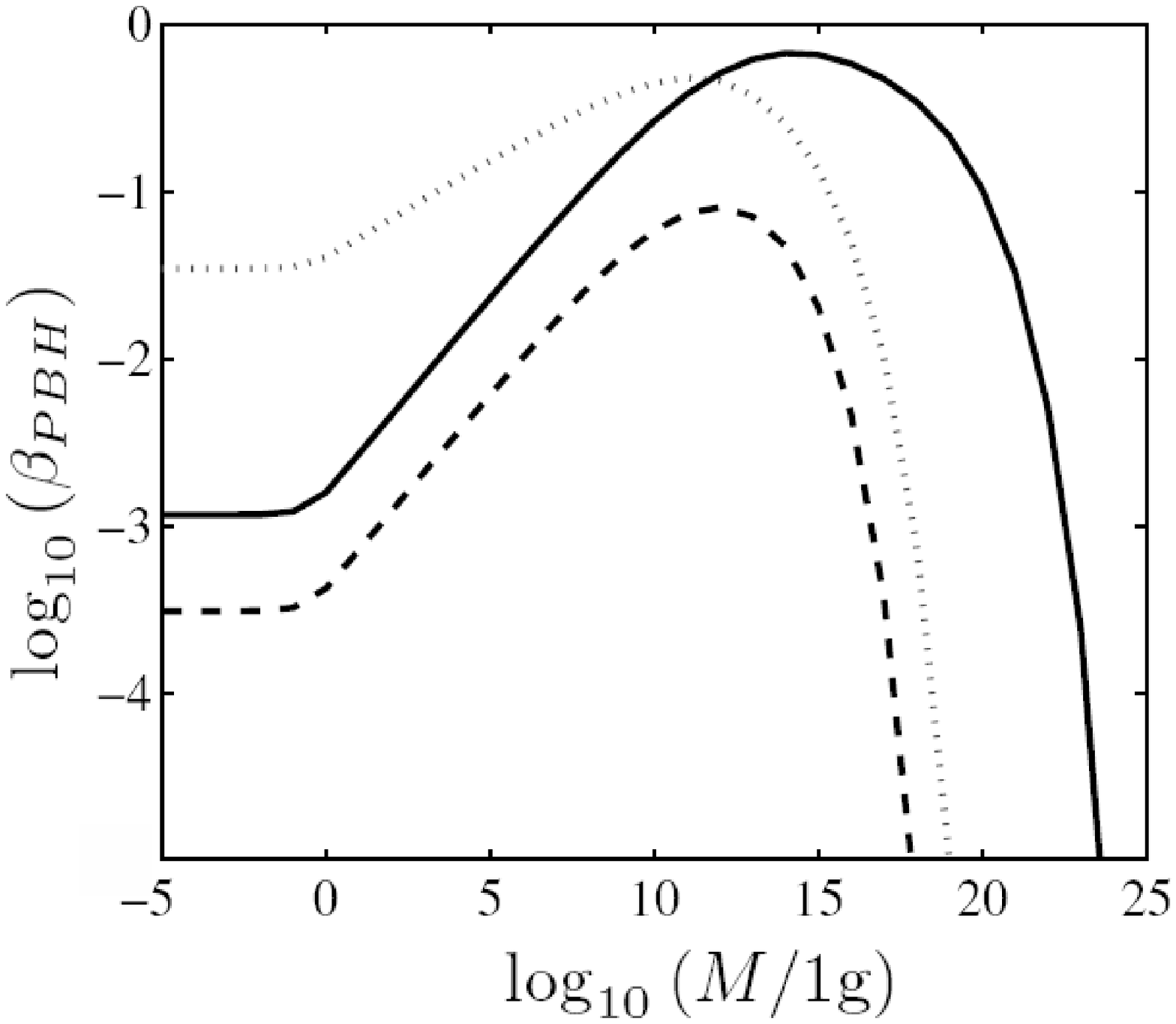}
    \caption{{\small The logarithmic probability of PBHs for two tilts in the
    power spectrum ($n_s = 1.23$ on the top figure, $n_s = 1.47$ on
    the bottom figure), integrated for the three different regions
    sketched in Fig. \ref{fig1}. The integrals over Areas I and II
    correspond to the dashed and solid lines, respectively. The
    probability integrated over Area III is represented by the dotted
    lines in both figures.}} 
    \label{fig2} 
  \end{center}
\end{figure}

We contrast the case of parabolic profiles described by
Eq. \eqref{param:expansion} with the non-parabolic set presented in
Fig. \ref{fig11} by plotting the probability $\beta_{\mathrm{PBH}}$
for different values of $\ps_{\R}$. This is presented in
Fig. \ref{fig3}. The figure shows that the probability of PBH
formation can be larger than the  one-parameter probability
computed in previous studies from the integration of Area II 
\citep{green-liddle}. This important result requires confirmation from more
detailed numerical simulations of PBH formation in this parameter
area. The uncertainty is explained by the fact that the two-parametric
calculation of the probability of PBH formation is still
incomplete. This should be complemented in the future by the
introduction of all relevant higher-order derivative parameters and the
higher-order correlations in the PDF.

\begin{figure}[h!]
  \begin{center}
    \includegraphics[width = 0.75\textwidth]{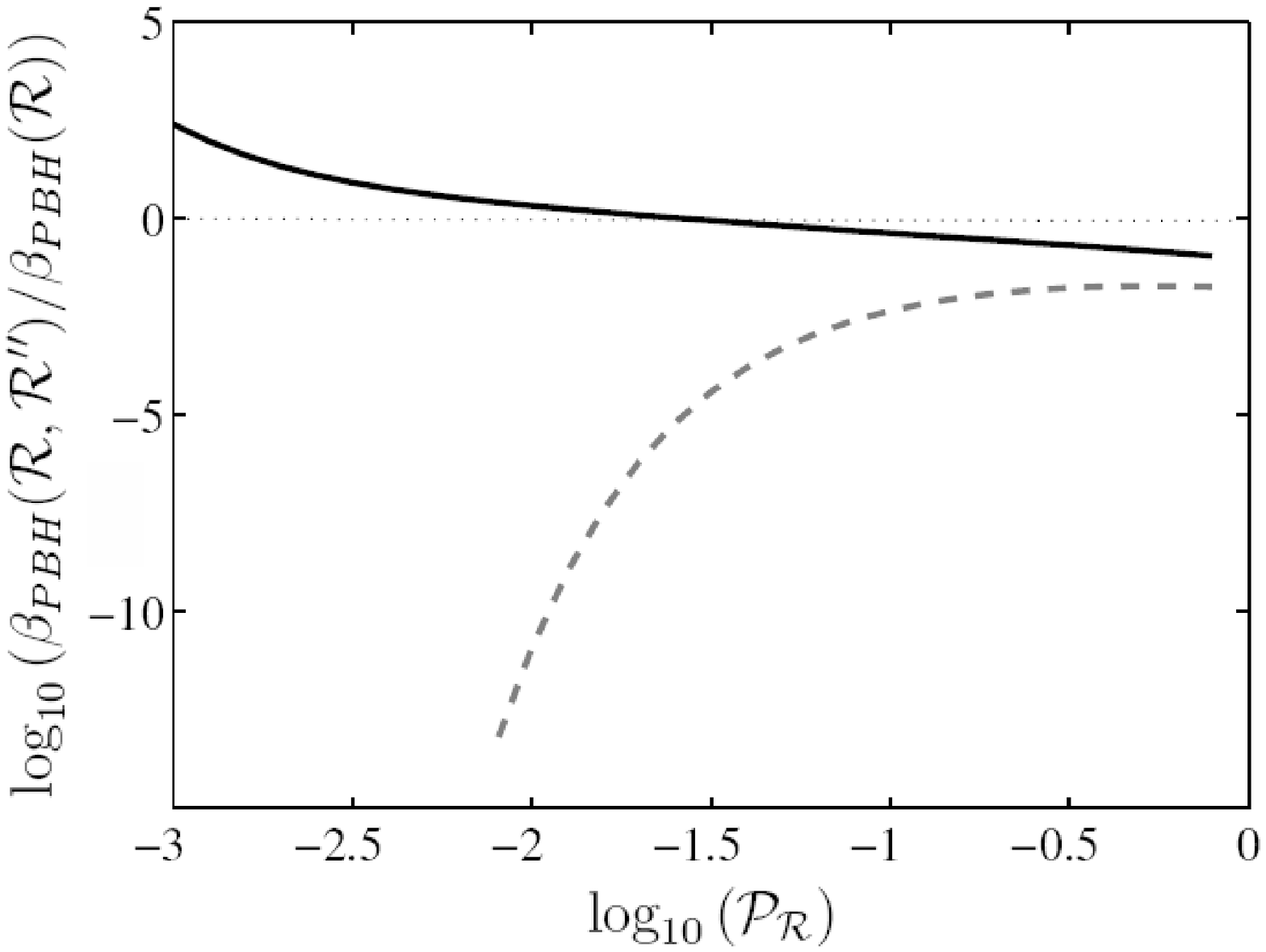}
    \caption{{\small The grey dashed line shows the ratio of the total
        probability $\beta_{\rm PBH}$ which results from integrating over
        Area I on the $[\R(0), \R''(0)]$ parameter space of Fig.~1b to
        the probability which results from the integrating over Area
        II. The black line is the ratio of the probability integrated
        over Area III to the probability integrated over Area II.}} 
    \label{fig3}
  \end{center}
\end{figure}

\section{Discussion}\label{conclusion}

We have developed a method for
calculating the two-parametric probability of PBH formation, taking
into account  the radial profiles of nonlinear curvature
cosmological inhomogeneities. This is the first step towards
calculating the $N$-parametric probability, which takes into account
the radial profiles more precisely than studies using the amplitude as
the only relevant parameter.  We have incorporated the derived
contribution to the total probability of PBH formation by considering
the range of values of $\R''(0)$ that will form PBHs, using the
results of the numerical computations presented by
\cite{musco-polnarev}. Finally, we have provided an example of the
consequences of this probability for the statistics of PBHs.

The results obtained show that, if we restrict ourselves to PBH
formation from parabolic profiles (as described in
Section \ref{metrics}), then the total PBH probability is orders of
magnitude below  previous estimates! On the other hand, if
non-parabolic configurations are also included (see Fig.~\ref{fig11}),
the total probability of PBH formation is higher than the
single-parametric probability estimated in previous works. In this
case, we can impose new bounds on the power
spectrum in the scales relevant for PBH formation. Analysing the
uncertainty of our results, we have demonstrated how much we still
have to understand about the formation and statistics of PBHs. The
physical arguments supporting our results should be verified by
numerical hydrodynamical simulations of PBH formation, which would
provide a valuable feedback to the initial motivation of this work. 

The main argument of this chapter is that the amplitude of
initial inhomogeneities is not the only parameter which determines the
probability of PBH formation. The ultimate solution of the problem
requires a greater set of parameters and a larger range of their
values to determine all high curvature configurations that form
PBHs. This is a huge task for future research. In the meantime,
we have a method to operate with the statistics of all these
parameters.  

\clearemptydoublepage
\chapter{Conclusions and future work}
\label{chaptersix}

In this thesis we have presented a study of large inhomogeneities in
the early universe.  
Such large concentrations of matter may collapse to form primordial
black holes (PBHs).   
The number of PBHs in our universe is calculated by integrating the
probability distribution function (PDF) of primordial inhomogeneities,
this encoding all the statistical information of primordial
inhomogeneities. The main objective of this thesis is to quantify the
probability of PBH formation in the context of nonlinear perturbation
theory. This represents a significant improvement in  the study of
large-amplitude inhomogeneities since, by definition, these are
nonlinear. \\

The statistics of inhomogeneities are the point of contact between
theory and observations. In theoretical studies the  statistics of
primordial fluctuations are studied in the framework  of cosmological
perturbation theory. Until recently, perturbation theory was
restricted to consider only linear departures from the homogeneous
background.  Linear perturbations are Gaussian due to the independence
the perturbation modes. This is an excellent approximation to describe
the structures  observed in the universe. Indeed, observationally,
only the variance, or second statistical moment, has been
measured. However, the detailed observations of large-scale structure
(LSS) and the cosmic microwave background (CMB) now allow us to test
for corrections to the linear approximation. This motivates the study
of extensions of linear perturbation theory. In particular, the
non-Gaussianity of curvature fluctuations has been a subject of
intense investigation.  

\section{Summary of results}

In Chapter \ref{chaptertwo} we presented a brief introduction to
cosmological perturbation theory within general relativity.  We
reviewed the basic results of this theory for the cosmological
inflation paradigm.  From the evolution equations, we identified the
conditions under which curvature  fluctuations can grow significantly
at superhorizon scales. As shown in
Eqs.~\eqref{zeta:evol}~and~\eqref{zeta2:evol}, these conditions are
mainly the presence of a non-adiabatic component in the matter field
fluctuations. This has motivated several previous studies of
non-Gaussianity resulting from the second-order perturbations of an
isocurvature (non-adiabatic) field $\chi$. The conditions for
inflation show that a non-adiabatic field is only a subdominant
component of the total matter during inflation.  The cosmological
model in which $\chi$ is responsible for the curvature perturbations
is called the curvaton model.  
\\

In Section \ref{non-gaussianity} we calculated non-Gaussian
correlators for some special cases of the curvaton model. The lowest
order non-Gaussian signature is a non-vanishing skewness or third
moment of the PDF. In perturbation theory this is equivalent to the
correlation of three copies of the curvature perturbation field. This
correlator itself is only present when we consider nonlinear
perturbations. In order to derive the three-point correlator we have
considered nonlinear field fluctuations $\delta \chi$. We have
calculated the second order perturbations of a single isocurvature
field during inflation and radiation domination. We have done this by
solving the Klein-Gordon equation of the perturbation $\delta \chi$ to
second order. For simplicity we consider only the matter fluctuations,
assuming a large contribution from the third derivative of the
potential $d^3 W(\chi) / d \chi^3$. We find that an effectively
massless field does not generate a large nonlinear contribution to
the perturbation $\delta \chi$. Conversely, a slightly massive field
allows an exponential growth of the nonlinear perturbation. With the
aid of a new method to compute non-Gaussian correlators, we derived
the field bispectrum $F(\bk_i)$ given by Eq. \eqref{ls_bi}.  We then
derived the curvature perturbation bispectrum $B(\bk_i)$, considering
a dominant contribution from the field bispectrum $F(\bk_i)$. Equation
\eqref{curvaton:fnl} expresses the the non-Gaussian parameter $\fnl$
in terms of the elements of the potential
$W(\chi)$. Chapter~\ref{chaptertwo} closed with a brief discussion of
the  observational limits to the curvaton.  
\\ 
 
 One of the main objectives of this thesis is to present the modified
 probability of structure formation from the non-Gaussian PDF. From
 the central limit theorem, we know that the non-Gaussian PDF produces
 non-trivial moments of order higher than two. To determine the shape
 of the distribution uniquely, one requires the \emph{a priori}
 knowledge of all moments. From studies of non-Gaussianity in
 perturbation theory, however, we only know the skewness (third
 moment) and kurtosis (fourth moment) of some models of structure
 formation. Finding a PDF which  encodes the contribution of only
 these two higher order moments is not trivial. 
In Chapter~\ref{chapterthree} we have constructed, in the
    context of quantum field theory, the general non-Gaussian PDF for
    curvature perturbations $\R$. Formally, this is a probability
    functional for the ensemble of realisations of $\R(\vect{x})$ at
    some specified time $t$. We refer to this probability as
    $\Prob_{t}[\R]$.  We first derived a mathematical expression of
    the above statement by writing $\Prob_{t}[\R]$ in terms  of the
    $n$-point correlation functions (see
    Eqs.~\eqref{genfunc:reconstruct}~and~\eqref{genfunc:prob}).  We
    then constructed an explicit expression for the PDF using only the
    first three statistical moments.  
  \\  
    
   We found that, in order to calculate the PDF, it is necessary to
   consider a regularised function $\Rsm(\bk)$. We must therefore consider a
   field sufficiently smooth on small scales, with a smoothing scale
   customarily set as the horizon scale at time $t$. We also
   require an upper limit for the $k$-numbers in order to avoid
   divergences in the integrations required to construct the PDF. This
   is achieved by artificially compactifying the momentum space over a
   scale $\Lambda~>~k$.  This regularisation is a common requirement
   of calculations in the theory of perturbations  and in the
   statistics of LSS.  
   The results of this chapter are important because the final
   expression is an explicit  functional probability
   $\Prob_{t}[\R(\vect{k})]$. This means that  the probability of any
   parameter appearing in the function $\R(\vect{k})$, or equivalently
   $\R(\vect{x})$, can be retrieved from this PDF. We rely on this
   property to study two important modifications of the probability of
   PBH formation in the subsequent Chapter~\ref{chapterfour} and
   Chapter~\ref{chapterfive}.          
\\    

In the last two chapters of this thesis  we revisit the calculation of
the probability of PBH formation, taking into account two important
effects which are characteristic of nonlinear inhomogeneities.  
In Chapter~\ref{chapterfour}  we calculate the probability of PBHs
using a non-Gaussian PDF we considered the non-Gaussian PDF of
curvature perturbations $\R$. The featured PDF includes a linear
contribution from the three-point correlation function as derived in
Chapter~\ref{chapterthree}. In Section~\ref{ngpdf-pbhs}, this PDF was adapted
to curvature configurations $\R(\vect{r})$ that give rise to
PBHs. As previous works show, the amplitude at the centre of the
curvature configuration $\R(\vect{r} = \vect{0})$ is a good parameter
to determine the formation of PBHs.  
Eq.~\eqref{eqn2.2} gives the non-Gaussian PDF for the mentioned
parameter. With the aid of this PDF we have reproduced qualitatively
the effects of the non-Gaussian contribution considered in two
previous works. We first identified the source of inconsistencies in
previous works studying non-Gaussian effects in the probability of PBH
formation. We showed that the fundamental difference in the
inflationary models considered by \cite{bullock-primack} and by
\cite{ivanov} is the spectral index.  In the first work, perturbations
involve a blue power spectrum (for which the spectral index
accomplishes $n_s-1>0$), while the power spectrum is red ($n_s-1<0$)
in the second.  Noting that, in single-field inflation, the
non-Gaussian parameter $\fnl$ is directly related to the spectral
index $n_s-1$, we have shown the source of the discrepancy.  The
main effect on the non-Gaussian PDF is the respective suppression and
enhancement of the probability for large 
values of $\R$. 
\\

 Chapter~\ref{chapterfour} also  presented the non-Gaussian
 modifications  to the probability of PBH formation (or the mass
 fraction of PBHs). In Section~\ref{ngbeta-pbhs} we have shown how the
 new PDF can modify the bounds to the variance of curvature fluctuations
 $\Sigmar$. This comes from the observational limits to the abundance
 of PBH for each mass scale.  Such modifications are illustrated in
 Fig.~\ref{fig3}. Note that in this figure we have used the maximum
 value of the parameter $\fnl$ allowed by perturbation theory. In the
 future, greater values could be considered by constructing PDFs with
 the techniques described here.  
\\

In Chapter~\ref{chapterfive} we have studied the probability of
configurations $\R(\bx)$ from another perspective: We compute the
probability of a parameter describing the curvature profile in
addition to the probability of $\R(\vect{0})$. Specifically, we
compute the probability of the second radial derivative at the centre
of the configuration, $\R''(\vect{0}) = d^2 \R / d r^2
|_{\vect{\vect{r} = 0}}$. As studied in that chapter, the
consideration of additional parameters describing curvature profiles is a
significant improvement in the study of gravitational collapse. In
other words, the choice of initial configurations collapsing to form
PBHs relies on two sets of parameters. Parameters of the profile
$\R(\vect{r})$ are required in addition to the amplitude parameters
customarily used. We used the results of the latest simulations of PBH
formation to integrate all the allowed configurations parametrised
with the pair $[\R(0), \R''(0)]$. The result shows, heuristically, how
the probability of PBH formation can be drastically changed by
considering curvature profiles in the PDF.  

\section{Future research}
\label{future}
     
The non-Gaussian signatures of cosmological inhomogeneities offer good
prospects for model discrimination. An example of this is given at the
end of Chapter~\ref{chaptertwo}, where we were able to limit special
cases of the curvaton model with the observed constraints on the
parameter $\fnl$. A number of extensions to this work are
possible. First, the curvaton model can be adapted, with pertinent
modifications, to describe models of modulated reheating
\citep{zaldarrion}. In such models, the auxiliary field modifies the
expansion in different patches of the universe during the reheating
process. The present work can be extended to cover such models by
modifying the scales of the mass and expectation values of the
auxiliary field $\chi$.  In this way one can search for feasible
models of modulated reheating which satisfy the observational limits
of non-Gaussianity.  
  \\
  
Another application of our study is to compute higher-order
correlations from the derived solutions to the nonlinear Klein-Gordon
equation. Future probes of non-Gaussianity could detect the
`trispectrum' of curvature perturbations, which is a higher order
discriminator between models of inflation. Computing the corresponding
four-point function is thus crucial for a characterisation of the
hypothetical detection of non-Gaussianity at this level. 
\\
    
An important complement of the work presented in
Chapter~\ref{chaptertwo} is the computation of the solutions to the 
Klein-Gordon equation allowing for metric perturbations. This
has been ignored here because we assumed that the field fluctuations
dominate over all other sources, as applies in the slow roll limit of
the Klein-Gordon equation. As mentioned in
Section~\ref{non-gaussianity}, however, the curvature perturbation 
contribution (often called backreaction) may entail important
corrections for $\delta \chi$ [see e.g. \cite{malik-kg}]. It 
is important to compute such contributions because, on the one hand,
they could be the dominating component in the growth of the
fluctuations and, on the other hand, the curvature back-reaction could
cancel large $\fnl$ values.  This could prompt reconsideration of
models previously excluded by observations. Which case applies is an
open question that should be addressed in the near future.  
\\

There is another set of problems where the methods of Chapter
\ref{chapterthree} find an important application. This is the
determination stochastic sources in the evolution equations of
classical fields. A functional probability is written in terms of
products of the $n$-point correlation functions with  $n$ copies of
the field configuration in Fourier space (see Eq.\eqref{genfunc:prob}
for the case of curvature perturbations). This can be used, in
particular, to derive and extend the stochastic equations of inflation
by \cite{starobinsky-yokoyama}. This seminal work presents a
Fokker-Planck equation for the probability of the configuration
$\phi(\vect{x})$ on small scales and for a single-field inflationary
field. A first connection between the stochastic framework and the
work presented here  has been given by \cite{seery-stochastic}.  In
that paper, a wave-functional similar to Eq.~\eqref{wave-func:R} is
considered and the governing Hamiltonian operator for the scalar field
$\phi$ is recovered from its action. The Fokker-Planck equation
suggested by Starobinsky can be deduced easily from the Schr\"odinger
equation for that wave-functional.  
Such a method can be extended to calculate PDFs of multi-scalar or
non-canonical models of inflation. The construction of the probability
distribution for configurations $\phi(\vect{x})$ and other possible
fields allows for the consideration of full non-Gaussian
distributions. This is clearly the way to go beyond approximations
like the one considered in Chapter~\ref{chapterthree}. 
\\
     
Regarding the probability of PBH formation, the results of
Chapter~\ref{chapterfive} cannot be conclusive because 
we do not have at hand the complete set of collapsing
configurations. Determining the set of all configurations collapsing
to form PBHs is a huge task to be explored elsewhere. We can assert,
however, that if PBHs are to be used as a tool for cosmology, the
curvature profile parameters have to be taken into account in the
derivation of the PDF. A less ambitious task is to have an estimate of
how severe the modifications to the single-parameter approximation
are. This would require the determination of more appropriate
parameters describing curvature profiles, a topic currently under
investigation.  
\\

From the results of Chapter~\ref{chapterfour} we are able to set
constraints on inflationary models. Specifically one can look at
models with an enhancement of the power spectrum at small scales.  In
such cases, the constraints from PBHs can be more or less stringent,
depending on the values and the sign of the  non-Gaussian parameter
$\fnl$. Here we have provided a tool for testing those models. Such
tool can also be improved as more constraints are derived from
observational tests to the abundance of PBHs.

\clearemptydoublepage
}
\end{onehalfspace}
\begin{onehalfspace}
\bibliographystyle{thesis}
  \bibliography{thesis}
\end{onehalfspace}
\end{document}